\documentclass[3p,12pt]{elsarticle}
\usepackage{booktabs}
\usepackage[fleqn]{amsmath}
\usepackage{cases}
\usepackage{multirow}
\usepackage{xcolor}
\usepackage[normalem]{ulem}
\usepackage{tabularx}
\usepackage{siunitx}
\usepackage{comment}
\usepackage{algorithm}
\usepackage{algpseudocode}
\usepackage{algorithmicx}
\usepackage{grffile}
\usepackage{rotating}
\usepackage{lineno}
\usepackage{hyperref}
\usepackage{graphicx,enumerate,subfigure}
\usepackage{amssymb}
\usepackage{bm,amsmath}
\usepackage{caption,color,xcolor}
\usepackage{subeqnarray}
\usepackage{multirow}
\usepackage{lscape}
\usepackage{rotating}
\usepackage{graphics}
\usepackage{float}
\usepackage{epstopdf,epsfig}
\usepackage{threeparttable}
\usepackage{soul}
\usepackage{enumitem}

\usepackage[capitalize]{cleveref}

\floatname{algorithm}{Algorithm}
\biboptions{numbers,sort&compress} 
\journal{XXX}

\newcounter{bla}

\begin{document}

\begin{frontmatter}
\title{Adaptive GSIS for rarefied gas flow simulations}

 \author{Yanbing Zhang}
 \author{Jianan Zeng}
 \author{Lei Wu\corref{mycorrespondingauthor}}
 \cortext[mycorrespondingauthor]{Corresponding authors:}
 \ead{wul@sustech.edu.cn}

\address{Department of Mechanics and Aerospace Engineering, Southern University of Science and Technology, Shenzhen 518055, China}

\begin{abstract}
The parallel solver of the general synthetic iterative scheme (GSIS), as recently developed by Zhang \textit{et. al.} in Comput. Fluids 281 (2024) 106374, is an efficient method to find the solution of the Boltzmann equation deterministically. 
However, it consumes a significant computational memory due to the discretization of molecular velocity space in hypersonic flows. In this paper, we address this issue by introducing the adaptive GSIS, where the Boltzmann equation is applied only in rarefied regions when the local Knudsen number exceeds a reference value, $\text{Kn}{ref}$. In contrast, the Navier-Stokes equations, with and without the high-order corrections to the constitutive relations, are applied in the continuum and rarefied regimes, respectively. Numerical results indicate that setting $\text{Kn}{ref}=0.01$ yields acceptable outcomes. With the adaptive GSIS, the computational memory and time can be significantly reduced in near-continuum flows, e.g. 24 and 7 times, respectively. in the simulation of rarefied gas flow passing the International Space Station.
\end{abstract}


\end{frontmatter}

\section{Introduction}

There has been a growing interest in rarefied gas flows due to advancements in space exploration \cite{ivanov1998computational,votta2013hypersonic}, extreme ultraviolet lithography \cite{EUV-plasma2021,Teng2023JCProd}, and vacuum systems for nuclear fusion \cite{Tantos2024NuclFusion}. These flows are characterized by thermodynamic non-equilibrium, which cannot be fully described by the traditional Navier-Stokes (NS) equations. Instead, the Boltzmann equation is employed to describe the system's state at the mesoscopic level.

The Boltzmann equation utilizes the velocity distribution function (VDF) to characterize the state of gaseous systems, which encompasses the streaming operator and the binary collision operator for the free motion and collision of gas molecules, respectively. Given that the VDF is defined in a high-dimensional phase space and the collision operator involves a five-dimensional nonlinear integral, numerical simulation of the Boltzmann equation presents a significant research challenge.

The direct simulation Monte Carlo (DSMC) is the predominant method for solving the Boltzmann equation stochastically~\cite{bird1994molecular}, particularly when the gas rarefaction effects are pronounced, i.e., when the Knudsen number (Kn, defined as the ratio of the mean free path of gas molecules to the characteristic flow length) is high. However, DSMC becomes computationally inefficient when the Knudsen number is low. This is because the separate treatment of the streaming and collision operators necessitates that the spatial cell size and time step in DSMC be smaller than the mean free path and mean collision time of gas molecules, respectively. 
To enhance simulation efficiency, a natural approach is to employ the NS-DSMC coupling method~\cite{sun2004hybrid}. This involves using the NS solver in regions where the local Knudsen number is below a reference value $\text{Kn}_{ref}$, and applying the DSMC method in other regions. Additionally, recent advancements have led to the development of asymptotic preserving schemes~\citep{Gorji2011JFM,LiuZhu2020JCP,Feng2023CPC,Fei2023JCP,Luo2024AiA}, which enables the use of larger cell sizes and time steps in simulating near-continuum flows.

The surge in computer power has significantly propelled the development of the discrete velocity method (DVM) for solving the Boltzmann equation deterministically. Notably, the design of asymptotic preserving schemes, including the unified gas-kinetic scheme~\citep{Xu2010unified,guo2013discrete,zhu2016implicit} and the general synthetic iterative scheme (GSIS)~\cite{Su2020Can,su2020fast,liu2024further,zhang2024efficient}, has substantially enhanced the computational capabilities. 
For example, in the simulation of cavity flow, the GSIS is faster than DSMC by 4 orders of magnitude when $\text{Kn}=0.005$~\cite{zhang2024efficient}; even in the simulation of hypersonic flows, the GSIS is 62 times faster than DSMC when $\text{Kn}=0.1$.
This is attributed to the GSIS's ability to achieve fast convergence and maintain asymptotic-preserving properties, so that steady-state solutions can be obtained within just a few dozen iterations across the entire spectrum of gas rarefactions, even when using coarse spatial grids.

However, the GSIS (and other DVMs) consumes a substantial amount of computer memory in hypersonic flow simulations due to the discretization of the molecular velocity space. For instance, the simulations of rarefied gas flow passing the X38 vehicle and the International Space Station require $1.3\ \text{TB}$ and $21.5\ \text{TB}$ of memory, respectively~\cite{zhang2024efficient}. There are several strategies to mitigate the memory cost. For examples, combining the discrete velocity method with the regularized Grad moment method (the macroscopic equations which are more accurate than the NS equations) to simulate rarefied gas flows can significantly reduce computational costs and memory consumption~\cite{yang2020hybrid}; Techniques such as structured-unstructured velocity space discretization~\cite{zhang4724172implicit} and Cartesian velocity space adaptive techniques~\cite{chen2024global} can effectively decrease the number of discrete velocities; Furthermore, employing an adaptive velocity space decomposition strategy to capture non-equilibrium parts of the flow field conserves memory consumption in near-continuum flows~\cite{xiao2020velocity, wei2024adaptive, long2024implicit}.

Given the fast convergence and asymptotic preserving properties of the GSIS, this paper introduces an adaptive GSIS (aGSIS) method, which combines the GSIS method in non-equilibrium regions with the NS equations in continuum regions. The reason for not using moment equations is their poor performance in hypersonic flow simulations~\cite{yang2020hybrid}. Although this type of DVM-NS coupling has been demonstrated~\cite{xiao2020velocity, wei2024adaptive, long2024implicit}, this paper builds upon the recent parallel solver for GSIS~\cite{zhang2024efficient} and provides a detailed parallel implementation of aGSIS, with a particular emphasis on achieving load balance. Eventually, this implementation is significantly more efficient than other deterministic solvers.


The remainder of the paper is organized as follows. In Section~\ref{sec:2}, we introduce the modified Boltzmann-Rykov equation valid from the continuum to free-molecular flow regimes and the macroscopic equations obtained by taking moments of the kinetic equation.
In Section~\ref{sec:gsis}, the numerical procedure and parallel computing strategy in solving aGSIS are introduced.
In Section~\ref{sec:num_example}, the accuracy and efficiency of aGSIS are assessed in several challenging cases. Finally, conclusions are given in Section~\ref{sec:conclusion}.
\section{Governing equations}\label{sec:2}


Without losing of generality, we consider the molecular gas with 3 translational degrees of freedom and $d_r$ internal degrees of freedom. Two VDFs, $f_0(t, \bm{x}, \bm{\xi})$ and $f_1(t, \bm{x}, \bm{\xi})$, are used to describe the translational and internal states of gas molecules, where $t$ is the time, $\bm{x}=(x_1,x_2,x_3)$ is the spatial coordinate, and $\bm{\xi}=(\xi_1,\xi_2,\xi_3)$ is the molecular velocity. The macroscopic quantities are obtained by taking moments of VDFs, e.g., the density $\rho$,  velocity $\bm{u}$, traceless stress $\bm{\sigma
}$, translational and rotational temperatures $T_t$ and $T_r$, translational and rotational heat fluxes $\bm{q}_t$ and $\bm{q}_r$, are
\begin{equation}\label{eq:getmoment}
    \begin{aligned}
        \left(\rho,~\rho\bm{u},~\bm{\sigma},~\frac{3}{2}\rho RT_t,~\bm{q}_{t}\right)&=\int\left(1,~\bm{\xi},~\bm{c}\bm{c}-\frac{c^2}{3}\mathrm{I},~\frac{c^2}{2},~\frac{c^2}{2}\bm{c}
        \right) f_0 \mathrm{d}\bm{\xi},\\
        \left(\frac{d_r}{2}\rho RT_r,~\bm{q}_{r}\right)&=\int\left(1,~\bm{c}\right)f_1\mathrm{d}\bm{\xi},
    \end{aligned}
\end{equation}
where $\bm{c}=\bm{\xi}-\bm{u}$ is the thermal velocity, $\mathrm{I}$ is the $3\times 3$ identity matrix, and $R$ is the gas constant; the translational pressure is $p_t=\rho R T_t$, while the total pressure is $p=\rho RT$, with the total temperature $T=(3T_t+d_rT_r)/(3+d_r)$ defined as the equilibrium temperature between the translational and internal modes.

The evolution of VDFs is governed by the following kinetic equations which take into account the proper relaxations of energy and heat-flux exchanges between translational and internal modes~\cite{LeiJFM2015,li2021uncertainty,Li2023JFM}:
\begin{equation}\label{general_model}
    \begin{aligned}
        &\frac{\partial f_0}{\partial t}+\bm{\xi}\cdot \nabla f_0 = \frac{g_{0t}-f_0}{\tau}+\frac{g_{0r}-g_{0t}}{Z_r\tau}, \\
        & \frac{\partial f_1}{\partial t}+\bm{\xi}\cdot \nabla f_1 = \frac{g_{1t}-f_1}{\tau}+\frac{g_{1r}-g_{1t}}{Z_r\tau}, 
    \end{aligned}
\end{equation}
where the reference distribution functions are given by:
\begin{equation}
    \begin{aligned}
        g_{0t}&= \rho\left(\frac{1}{2\pi RT_t}\right)^{3/2}\exp\left(-\frac{c^2}{2RT_t}\right)\left[1+\frac{2\bm{q}_{t}\cdot\bm{c}}{15RT_tp_t}\left(\frac{c^2}{2RT_t}-\frac{5}{2}\right)\right],\\
        g_{0r}&= \rho\left(\frac{1}{2\pi RT}\right)^{3/2}\exp\left(-\frac{c^2}{2RT}\right)\left[1+\frac{2\bm{q}_{0}\cdot\bm{c}}{15RTp}\left(\frac{c^2}{2RT}-\frac{5}{2}\right)\right],\\
        g_{1t}&=\frac{d_r}{2}RT_rg_{0t} + \left(\frac{1}{2\pi RT_t}\right)^{3/2}\frac{\bm{q}_{r}\cdot\bm{c}}{RT_t}\exp\left(-\frac{c^2}{2RT_t}\right), \\
        g_{1r}&=\frac{d_r}{2}RTg_{0r} + \left(\frac{1}{2\pi RT}\right)^{3/2}\frac{\bm{q}_{1}\cdot\bm{c}}{RT}\exp\left(-\frac{c^2}{2RT}\right),
    \end{aligned}
\end{equation}
with $\bm{q}_{0},~\bm{q}_{1}$ being linear combinations of translational and internal heat fluxes \cite{li2021uncertainty}:
\begin{equation}
    \begin{bmatrix} 
        \bm{q}_{0} \\ \bm{q}_{1} 
    \end{bmatrix}
    =
    \begin{bmatrix}		
        (2-3A_{tt})Z_r+1 & -3A_{tr}Z_r  \\		
        -A_{rt}Z_r & -A_{rr}Z_r+1 \\ 
    \end{bmatrix}
    \begin{bmatrix} 
    \bm{q}_{t} \\ \bm{q}_{r} 
    \end{bmatrix},
\end{equation}
where $\bm{A}=[A_{tt},A_{tr},A_{rt},A_{rr}]$ is determined by the relaxation rates of heat flux. Finally, the mean collision time $\tau=\mu/p_t$, where $\mu$ is the shear viscosity of the gas; $Z_r$ is the internal collision number, and $Z_r\tau$ characterizes the internal-translational energy exchange time.  The power-law intermolecular potential is considered, so that the viscosity can be expressed as 
\begin{equation}
\mu(T_t)=\mu(T_0)\left(\frac{T_t}{T_0}\right)^{\omega},
\end{equation}
with $\omega$ the viscosity index and $T_0$ the reference temperature.

Taking the velocity moments of the kinetic equation~\eqref{general_model}, the multi-temperature governing equations for the molecular gas with mass density $\rho$, flow velocity $\bm{u}$, translational temperature $T_t$, and internal temperature $T_r$ are obtained as:
\begin{equation}\label{eq:macroscopic_equation_2}
	\begin{aligned}
		\frac{\partial{\rho}}{\partial{t}} + \nabla\cdot\left(\rho\bm{u}\right)  &= 0, \\
		\frac{\partial}{\partial{t}}\left(\rho\bm{u}\right) + \nabla\cdot\left(\rho\bm{u}\bm{u}\right) + \nabla\cdot\bm{P} &= 0, \\
		\frac{\partial}{\partial{t}}\left(\rho e\right) + \nabla\cdot\left(\rho e\bm{u}\right) + \nabla\cdot\left(\bm{P}\cdot\bm{u}+\bm{q}_t+\bm{q}_r\right) &= 0, \\
        \frac{\partial}{\partial{t}}\left(\rho e_r\right) + \nabla\cdot\left(\rho e_r\bm{u}+\bm{q}_r\right) &= \frac{d_r\rho R}{2}\frac{T-T_r}{Z_r\tau},
	\end{aligned}
\end{equation} 
where $e_r=d_rRT_r/2$ and $e=(3RT_t+u^2)/2+e_r$ are the specific internal and total energies, respectively; the pressure tensor is given by $\bm{P} = p_t\mathrm{I} + \bm{\sigma}$. 

In the continuum flow regime, i.e., when the Knudsen number (defined as the ratio of the molecular mean free path $\lambda$ to the characteristic flow length $L$)
\begin{equation}
     \text{Kn}=\frac{\lambda}{L}\equiv
     \frac{\mu(T_0)}{p_0L}\sqrt{\frac{\pi R T_0}{2}},
 \end{equation}
is small, the constitutive relations are given by the Newton law of viscosity and the Fourier law of heat conduction:
\begin{equation}\label{eq:NSF_constitutive}
    \begin{aligned}
        \bm{\sigma}_{\text{NSF}} &= -\mu \left(\nabla\bm{u}+\nabla\bm{u}^{\mathrm{T}}-\frac{2}{3}\nabla\cdot\bm{u}\mathrm{I}\right),\\
        \bm{q}_{t,\text{NSF}} &= -\kappa_t\nabla T_t,\\
        \bm{q}_{r,\text{NSF}} &= -\kappa_r\nabla T_r.
    \end{aligned}
\end{equation}
where $\kappa_t$ and $\kappa_r$ are the transitional and internal thermal conductivities, respectively, and the superscript $\mathrm{T}$ is the matrix transpose. and thermal conductivities $\kappa_t,\kappa_r$ are given by
\begin{equation}\label{eq:mu_kappa}
	\begin{aligned}[b]
        \left[ 
            \begin{array}{ccc} 
              \kappa_t \\ \kappa_r 
            \end{array}
          \right]
          &= \frac{\mu R}{2}
          \left[ 
            \begin{array}{ccc} 
              A_{tt} & A_{tr} \\ A_{rt} & A_{rr} 
            \end{array}
          \right]^{-1}
          \left[ 
            \begin{array}{ccc} 
              5 \\ d_r 
            \end{array}
          \right].
	\end{aligned}
\end{equation}

\section{Numerical methods}\label{sec:gsis}

We adopt the finite volume scheme with second-order accuracy to solve the kinetic equations and macroscopic equations. We only show the major steps here, leaving the details in Ref.~\cite{liu2024further, zhang2024efficient}. 
We emphasis on the spatial adaptation and parallel implementation of GSIS.

\subsection{Original GSIS}

Since $f_0$ and $f_1$ share the same form, in the following they are represented by $f$ for clarity of presentation. Given the gas information at the $n$-th iteration step, the discretized velocity distribution at the next intermediate iteration step $n+1/2$ is calculated as
\begin{equation}
    \begin{aligned}[b]
        \frac{f_i^{n+1/2}-f_i^n}{\Delta t} + \frac{1}{V_i}\sum_{j\in N(i)} \xi_nf_{ij}^{n+1/2}S_{ij}=\frac{g^{n}_{i}-f^{n+1/2}_{i}}{\tau^{n}_i},
    \end{aligned}
\end{equation}
where $g = \left(1-\frac{1}{Z_r}\right)g_t+\frac{1}{Z_r}g_r$;
$\Delta t $ is the time step;
the subscripts $i,~j$ are the indices of the control cells, and the subscript $ij$ denotes the interface between the adjacent cells $i$ and $j$, with
$S_{ij}$ and $V_i$ being the area of interface $ij$ and the volume of cell $i$, respectively. $\xi_n=\bm{\xi}\cdot\bm{n}$ is the molecular velocity component along normal direction $\bm{n}=\bm{S}/|\bm{S}|$ pointing from cell $i$ to cell $j$; the sum of fluxes $\xi_n f_{ij}$ is taken over all the faces of a cell $N(i)$.

This type of iteration is slow when the gas flow is in the near-continuum flow regimes. To facilitate rapid convergence, the velocity distribution function should be modified to more closely resemble the final state. In the GSIS method, this is achieved through the following equation:
\begin{equation}\label{eq:updatef}
	f^{n+1} = f^{n+1/2} + [f_{eq}(\bm{W}^{n+1})-f_{eq}(\bm{W}^{n+1/2})],
\end{equation}
where $\bm{W}^{n+1}=\left[\rho, \rho\bm{u}, \rho e , \rho e_r\right]^{n+1}$ are the macroscopic conservative variables at the next iteration step.

The macroscopic quantities are solved from the macroscopic equations~\eqref{eq:macroscopic_equation_2}, where the constitutive relation should not only contain the Newton and Fourier laws of viscosity and heat conduction, but also contain the high-order rarefaction effects. In GSIS-II, the stress and heat fluxes are composed of the linear constitutive laws and the higher-order terms (HoTs) characterizing the rarefaction effects:
\begin{equation}\label{eq:full_constitutive}
    \begin{aligned}
        \bm{\sigma}^{n+1} &= \bm{\sigma}^{n+1}_{\text{NSF}}  + 
        \underbrace{ \int \left(\bm{c}\bm{c}-\frac{c^2}{3}\mathrm{I}\right)f^{n+1/2}_0 \mathrm{d}\bm{v} -\bm{\sigma}^{n+1/2}_{\text{NSF}} }_{\text{HoT}_{\bm{\sigma}}},
        \\
        \bm{q}^{n+1}_{t} &=  \bm{q}^{n+1}_{t,\text{NSF}} + \underbrace{ {\frac{1}{2}}\int \bm{c}c^2f^{n+1/2}_0 \mathrm{d}\bm{v} -\bm{q}^{n+1/2}_{t,\text{NSF}} }_{\text{HoT}_{\bm{q}_{t}}},
        \\
        \bm{q}^{n+1}_{r} &= \bm{q}^{n+1}_{r,\text{NSF}} + \underbrace { \int \bm{c}f^{n+1/2}_1 \mathrm{d}\bm{v}  -\bm{q}^{n+1/2}_{r,\text{NSF}} }_{\text{HoT}_{\bm{q}_{r}}}.
    \end{aligned}
\end{equation}
Substituting Eq.~\eqref{eq:full_constitutive} into Eq.~\eqref{eq:macroscopic_equation_2}, the traditional NSF equations with source terms coming from the high-order constitutive relations are obtained, which can be solved by the following traditional finite-volume method to obtain $\bm{W}_i^{n+1}$:
\begin{equation}\label{eq:macro}
    \frac{\bm{W}_i^{n+1}-\bm{W}_i^n}{\Delta t} + \frac{1}{V_i}\sum_{j\in N(i)}\bm{F}_{ij}^{n+1}S_{ij}=\bm{Q}^{n+1}_i,
\end{equation}
where the detailed expressions of the fluxes $\bm{F}$ and the source terms $\bm{Q}$ are given in the appendix in Ref.~\cite{Zeng2023CaF}.

\subsection{adaptive GSIS}

To reduce the computational memory required for high-speed problems in DVM, this paper adopts a similar idea to NS-DSMC coupling methods~\cite{sun2004hybrid, yang2017development, kim2024evaluation} by dividing the computation domain into equilibrium regions (solved by the NS equations) and non-equilibrium regions (solved by the Boltzmann equation). Compared with traditional NS-DSMC coupling methods, the deterministic approach eliminates statistical noise interference and is better suited for complex physical flow fields. Furthermore, solving the macroscopic synthetic equations across the entire computational domain enhances the exchange of flow information between the DVM and NS regions, thereby facilitating rapid convergence.

To enable adaptive distinction of different domains, we employ a widely used gradient-length local Knudsen number determined by both density $\rho$, speed $U$ and temperature $T$, which is defined as~\cite{boyd1995predicting}:
\begin{equation}\label{eq:Kn_Gll}
  \text{Kn}_{Gll}=\lambda \times \max \left(\frac{\left | \nabla \rho  \right | }{\rho}, \frac{\left | \nabla U  \right | }{U}, \frac{\left | \nabla T  \right | }{T} \right).
\end{equation}
where $\lambda$ is the local mean free path of the gas. In the next section, we will show that $\text{Kn}_{Gll}$ has the ability to discern non-equilibrium regions.
Here, we introduce a reference Knudsen number $\text{Kn}_{ref}$ to distinguish between equilibrium and non-equilibrium:
\begin{equation}\label{eq:updatef}
    \text{DVM\ Region} = \left\{\begin{matrix} 
      1,\ \text{Kn}_{Gll}  \ge   \text{Kn}_{ref}\\  
      0,\ \text{Kn}_{Gll}   <   \text{Kn}_{ref}
    \end{matrix}\right.
\end{equation}
where $\text{Kn}_{ref}$ is a case-dependent user input parameter.
In the computation, the kinetic solver is only used in regions where DVM Region is 1, while the macroscopic solver is used to solve for whole domain (the NS equation is applied when DVM Region=0, while the NS equations with high-order constitutive relations in applied when DVM Region=1, thus facilating rapid convergence in the GSIS framework).


Boundary conditions need to be supplied at the interface of DVM and NS solvers. This is easy for the NS region to receive the boundary flux from the DVM region. However, for the newly divided non-equilibrium domains, the original boundary conditions is no longer applicable. 
Here, the VDFs at the NS region
are constructed as that in the Grad moment equations~\cite{liu2024further}:
    \begin{equation} \label{eq:G13_VDF}
        \begin{aligned}
            f_{0}&=\left[1+\frac{\bm{\sigma}\cdot \bm{cc}}{2\rho T_t^2} + \frac{\bm{q}_t\cdot\bm{c}}{\rho T_t^2}\left(\frac{\bm{c}^2}{5T_t}-1\right)\right]f_{eq}, \\
            f_{1}&=\frac{d_r}{2} T_r \left[1+\frac{\bm{\sigma} \cdot \bm{c c}}{2 \rho T_t^2}+\frac{\bm{q}_t \cdot \bm{c}}{\rho T_t^2}\left(\frac{\bm{c}^2}{5 T_t}-1\right)\right]f_{eq}
            +\frac{\bm{q}_r  \cdot \bm{c}}{\rho T_t}f_{eq},
        \end{aligned}
    \end{equation}
    where $f_{eq}$ is the equilibrium distribution function:
    \begin{equation}
       f_{eq}= \rho\left(\frac{1}{2\pi RT_t}\right)^{3/2}\exp\left(-\frac{c^2}{2RT_t}\right).
    \end{equation}

Note that the initial non-equilibrium region is obtained by calculating the initial field with the macroscopic solver.
For cases where the Knudsen number lies within the slip flow and transition regimes, the initial non-equilibrium region might not adequately capture the nuances of non-equilibrium conditions. In such scenarios, it is essential to dynamically adjust the non-equilibrium calculation domain throughout the iterative process. However, for higher Knudsen numbers, the use of aGSIS is actually unnecessary, and the original GSIS algorithm can be employed directly.

\subsection{Parallel design of aGSIS}

For the macroscopic solver, the conventional approach is using spatial domain decomposition to partition the grids across processors and parallelizing, as detailed in Ref.~\cite{zhang2024efficient}, which will not be repeated here. 
Typically, all computational cores are configured with macroscopic solver, and these cores execute calculations via data exchange. However, when utilizing a substantial number of computational cores, the computational load of mesoscopic equations can significantly exceed that of macroscopic equations in certain scenarios. To avoid any adverse effects on the parallel efficiency of the macroscopic equations, only a subset of cores is designated for the macroscopic solver.
For the kinetic solver, the parallel design idea is also consistent with that in Ref.~\cite{zhang2024efficient}, but since this paper has adaptive partitioning for non-equilibrium regions, there are some differences.

For two-dimensional cases, the reduction of the distribution function allows for a decrease in the velocity space from three dimensions to two~\cite{Zeng2023CaF}, thereby significantly reducing computational memory requirements. For standard computing servers, employing the most efficient and straightforward parallel strategy for full velocity space parallelization is optimal. In the case of dynamically varying computational domains, spatial grid load balancing is unnecessary; instead, dynamic memory allocation and deallocation for the distribution function should be performed as required.

\begin{figure}
    \centering
    \subfigure[]{
    \label{fig:3DApollo_Ma5_Kn0d01_partition8}
        {\includegraphics[width=0.4\textwidth,trim={20 0 10 0},clip = true]{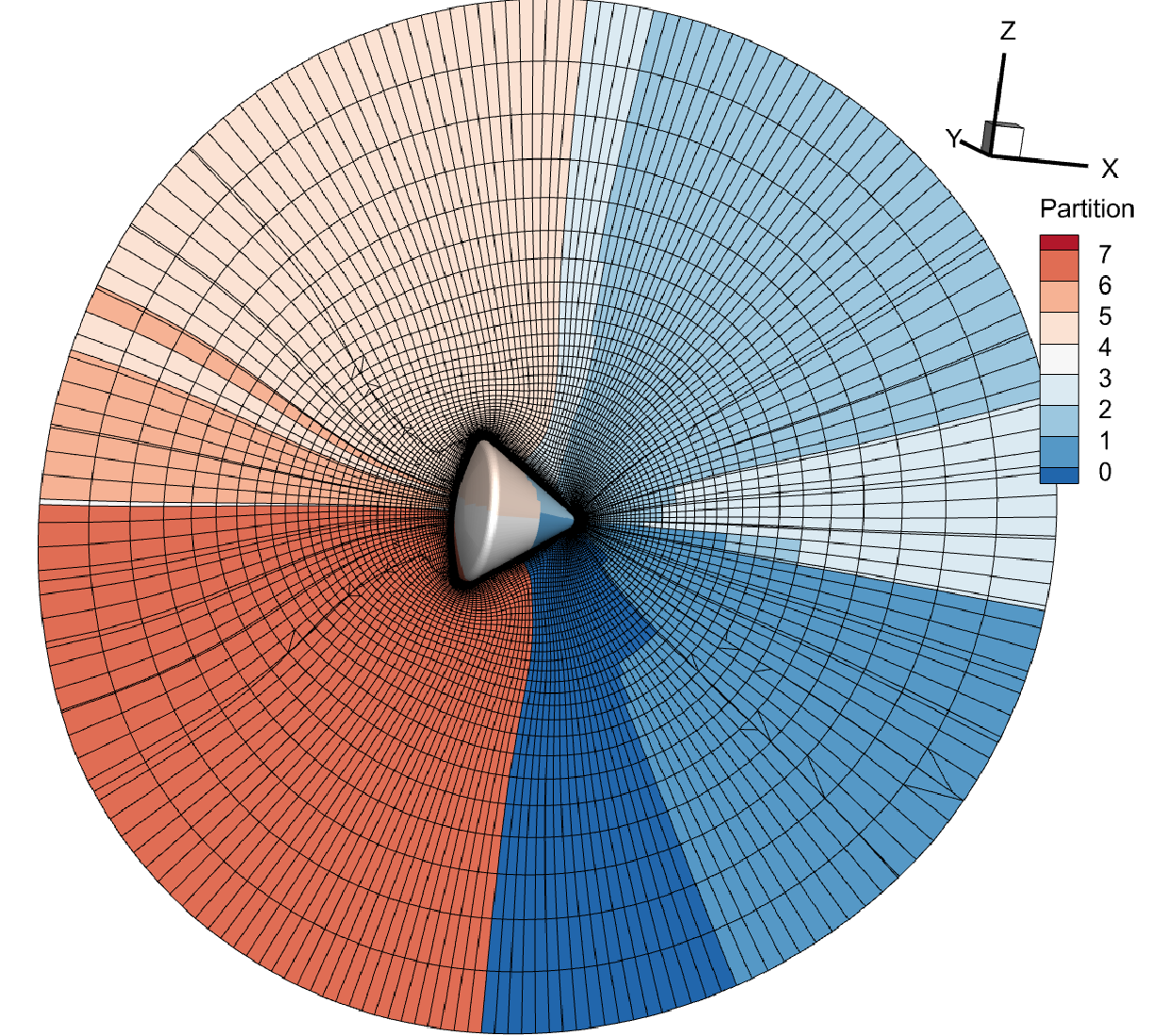}} } \\
    \subfigure[]{
    \label{fig:3DApollo_Ma5_Kn0d01_Knref0d01_dvmRegion}
        {\includegraphics[width=0.4\textwidth,clip = true]{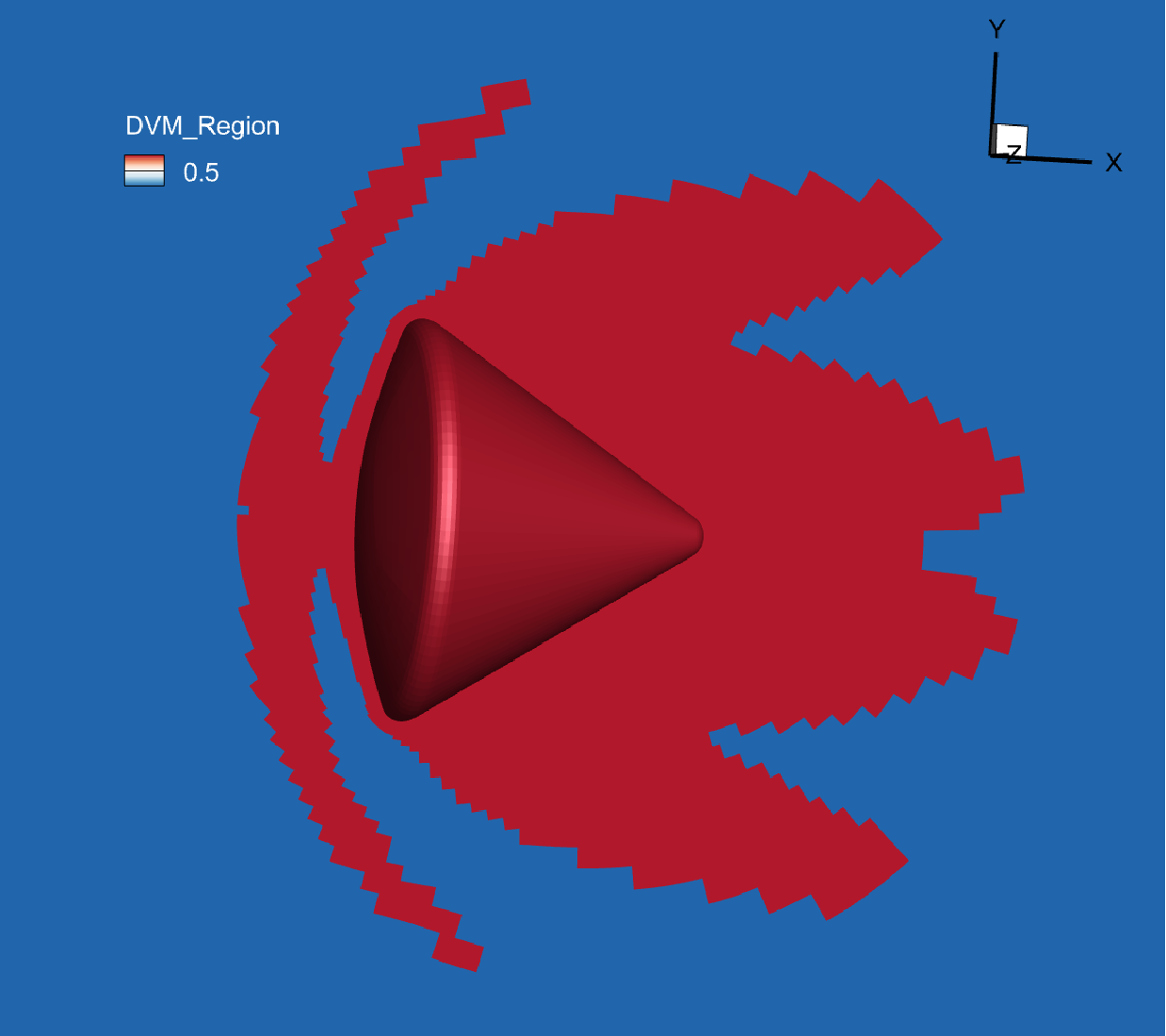}} }
    \subfigure[]{
    \label{fig:3DApollo_Ma5_Kn0d01_Knref0d01_partition8}
        {\includegraphics[width=0.4\textwidth,clip = true]{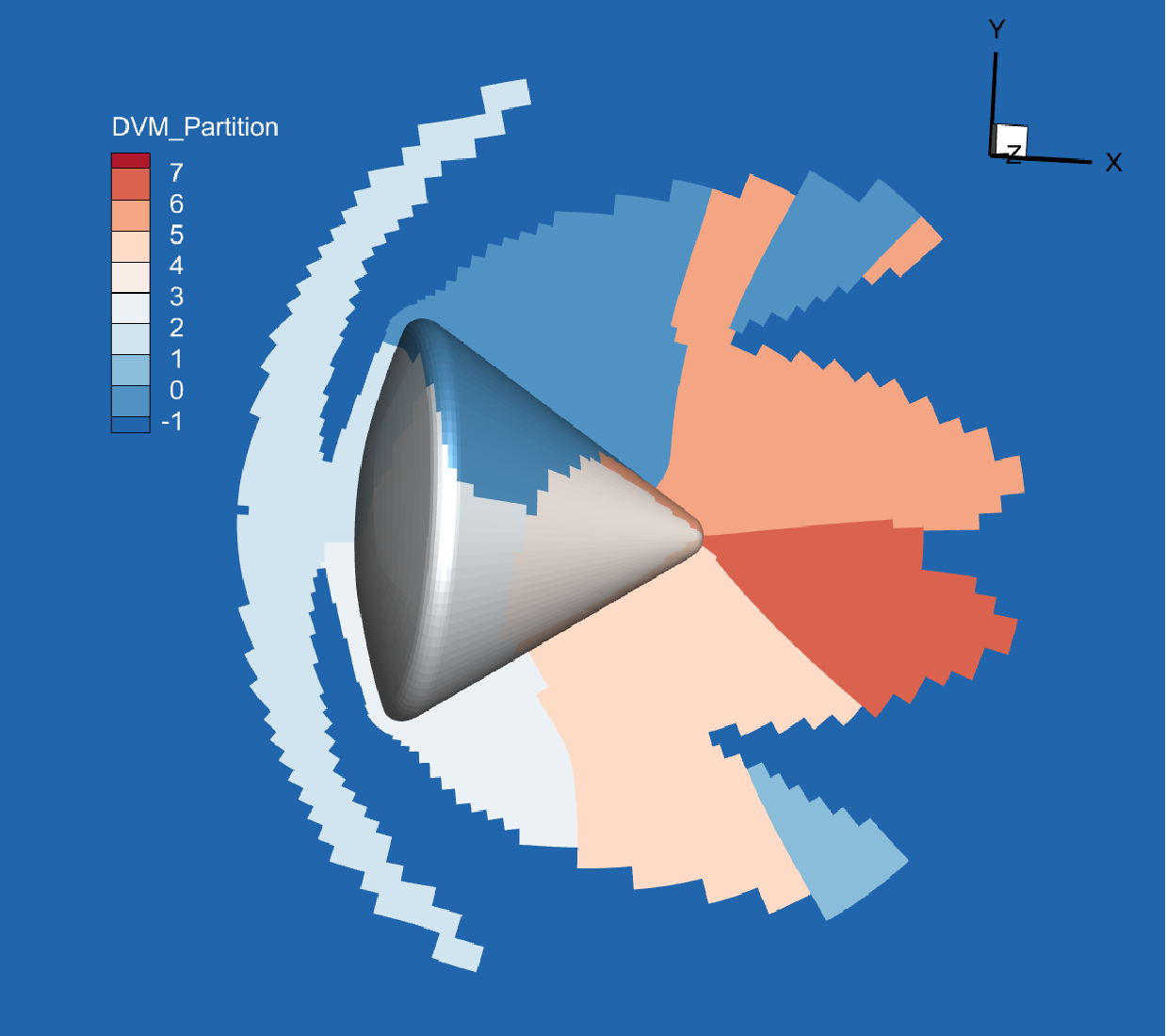}} }
    \caption{Spatial domain of the hypersonic flow around Apollo at $\text{Ma}=5$ and $\text{Kn}=0.01$. (a) Partition the whole spatial domain into 8 subregions. Note that only a two-dimensional slice is depicted within the three-dimensional domain, which may give the appearance of discontinuity in the partition. (b) The adaptive region of DVM at $\text{Kn}_{ref}=0.01$. (c) Repartition the DVM region into 8 subregions.
    }
    \label{fig:3DApollo_Ma5_Kn0d01_Knref0d01}
\end{figure}

For three-dimensional cases, owing to memory constraints, the kinetic solver in non-equilibrium regions continues to employ a two-level parallel strategy with a partitioned physical space-velocity space approach~\cite{zhang2024efficient}. However, because its computational domain differs from that of the macroscopic solver, individual load balancing is required for each. The macroscopic solver operates over the entire field without alteration throughout the computation, necessitating only an initial load balancing at the start. For example, as shown in Fig.~\ref{fig:3DApollo_Ma5_Kn0d01_partition8}, we use 8 cores for parallel computing and partition the entire computational domain into 8 regions, with each region having a roughly consistent number of grid cells. 
However, for kinetic solvers, the computational domain may change during adaptation. For each change, it is necessary to rebalance the computational load across the domain to ensure high parallel efficiency.
For example, as depicted in Fig.~\ref{fig:3DApollo_Ma5_Kn0d01_Knref0d01_dvmRegion}, the non-equilibrium regions are first identified (using NS data for the first time and GSIS data for subsequent times), with the red areas indicating these regions. Subsequently, as illustrated in Fig.~\ref{fig:3DApollo_Ma5_Kn0d01_Knref0d01_partition8}, these identified regions are further divided into 8 subregions.

In large-scale parallel computing, partitioning too many regions can reduce the parallel efficiency if the number of grid cells in non-equilibrium areas is limited. To maintain high parallel efficiency while ensuring that $N_x \times N_v = N_c$ (where $N_x$  denotes the number of spatial partitions, $N_v$  represents the number of velocity partitions, and $N_c$  is the total number of computing cores), it is advisable to adjust the ratio of $N_x$ to $N_v$  to achieve an optimal balance.
Here, we provide a reasonable empirical solution:
\begin{equation}\label{eq:adjust_nx_nv}
    N_x = 
    \begin{cases} 
        1, & \text{if } N_{\text{cell}} \le 20,000 \\
        \min_{\substack{x \in \mathbb{Z}^+ \\ x \in \left[ \left\lfloor \frac{N_{\text{cell}}}{1000} \right\rfloor,\ 2\left\lfloor \frac{N_{\text{cell}}}{1000} \right\rfloor \right]}} \left\{ N_c - x \left\lfloor \frac{N_c}{x} \right\rfloor \right\}, & \text{if } N_{\text{cell}} > 20,000
    \end{cases}
\end{equation}
where $N_{cell}$ represents the number of grid cells in the non-equilibrium regions. 


\subsection{Summary of algorithm}

Here, we outline the computational steps for the aGSIS method:
\begin{enumerate}[label=Step \arabic*: , itemjoin=\newline]
    \item Solve the macroscopic equations until convergence to obtain the initial field.
    
    \item Utilize the least squares method to compute the gradients of macroscopic quantities and determine the non-equilibrium computational regions based on Eq.~\eqref{eq:Kn_Gll}. For three-dimensional scenarios, it is imperative to conduct a novel load balancing procedure in the non-equilibrium regions. Furthermore, adjust the dimensions of $N_x$ and $N_v$ in accordance with the number of grid cells within these non-equilibrium regions as specified in Eq.~\eqref{eq:adjust_nx_nv}.
    
    \item Carry out five iterations of the GSIS method. After this process, compute the gradients of macroscopic quantities, re-identify the non-equilibrium regions, and perform load balancing on the new non-equilibrium regions as required. 
    
    \item Continue with the GSIS iterations until the flow field converges.
\end{enumerate}

\section{Computational Results and Discussion}\label{sec:num_example}

The performance of the aGSIS solver is evaluated in 2D planar micro nozzles, 2D circular cylinders, and 3D Apollo reentry modules. The simulated gas is nitrogen, which has a rotational degrees of freedom $d_r=2$, collision number $Z_r=2.667$ and viscosity index $\omega=0.74$. The thermal relaxation rates are \cite{LeiJFM2015}: $A_{tt}=0.786$, $A_{tr}=-0.201$, $A_{rt}=-0.059$ and $A_{rr}=0.842$, and hence the Eucken factors (corresponding to thermal conductivities) of translational and rotational degrees of freedom are $f_t=2.365$ and $f_r=1.435$, respectively.

The convergence criterion of the simulations is that the volume-weighted relative error $\varepsilon$ between two consecutive iterations 
\begin{equation}\label{convergence_critertion}
    \varepsilon=\max \left(\sqrt{\frac{\sum_{i}^{} \left ( W_i^{n}-W_i^{n-1} \right )^2 \varOmega _i }{\sum_{i}^{} \left ( W_i^{n-1} \right )^2 \varOmega _i } }  \right)
\end{equation}
is less than $10^{-6}$, where $W\in \{\rho,\bm{u},T_t\}$. This criterion is used to determine the wall clock time spent in the following simulations. 

\subsection{Supersonic flow in a planar micro nozzle} \label{subsec:nozzle}

\begin{figure}[!t]
    \centering
    \subfigure[Micro nozzle configuration]{
    \label{fig:2DNozzle_computationalDomain}
        {\includegraphics[width=0.45\textwidth,clip = true]{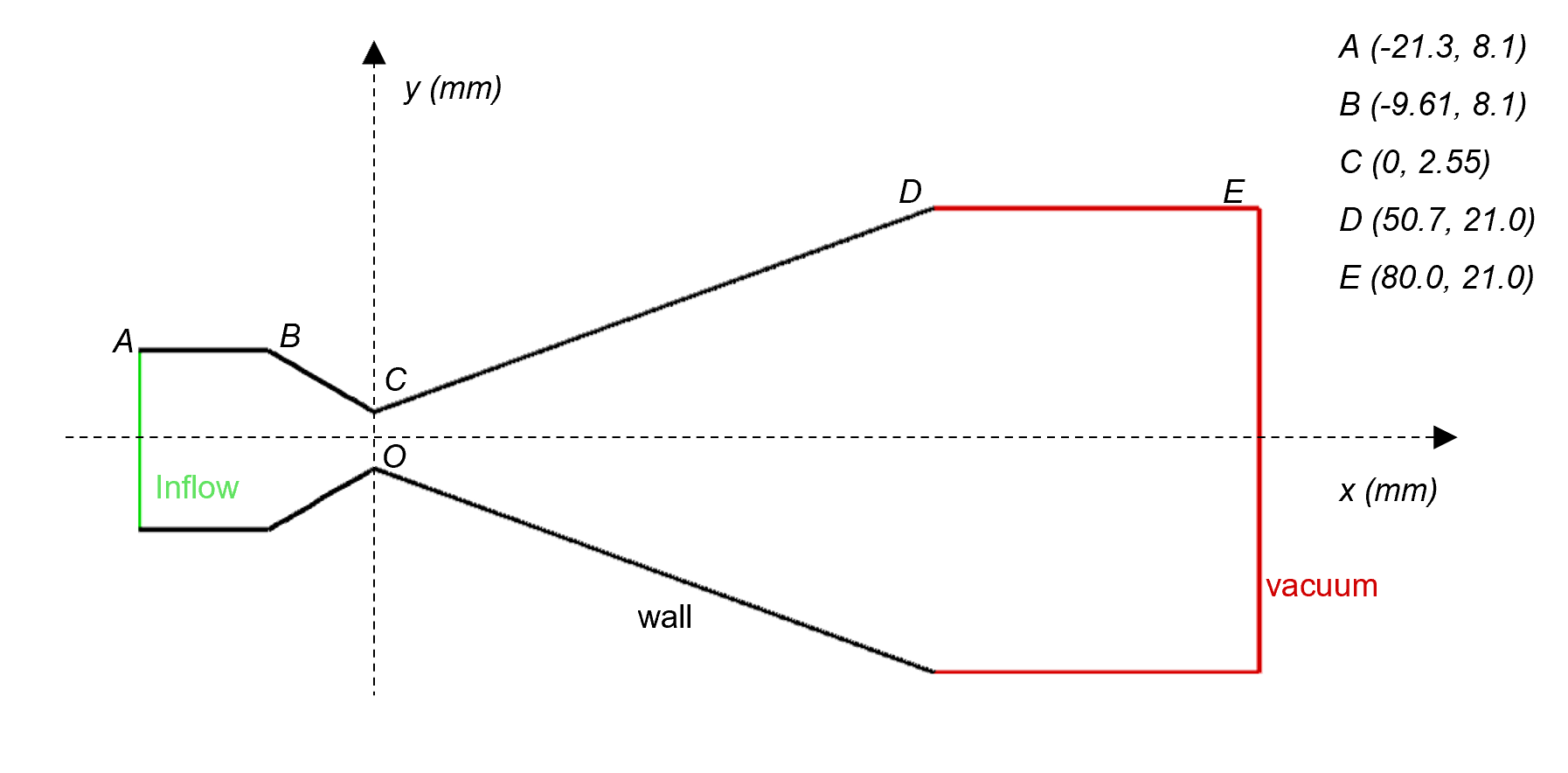}} }
    \subfigure[Spatial discretization]{
    \label{fig:2DNozzle_AGSIS_Kn0d01_config}
        {\includegraphics[width=0.45\textwidth,clip = true]{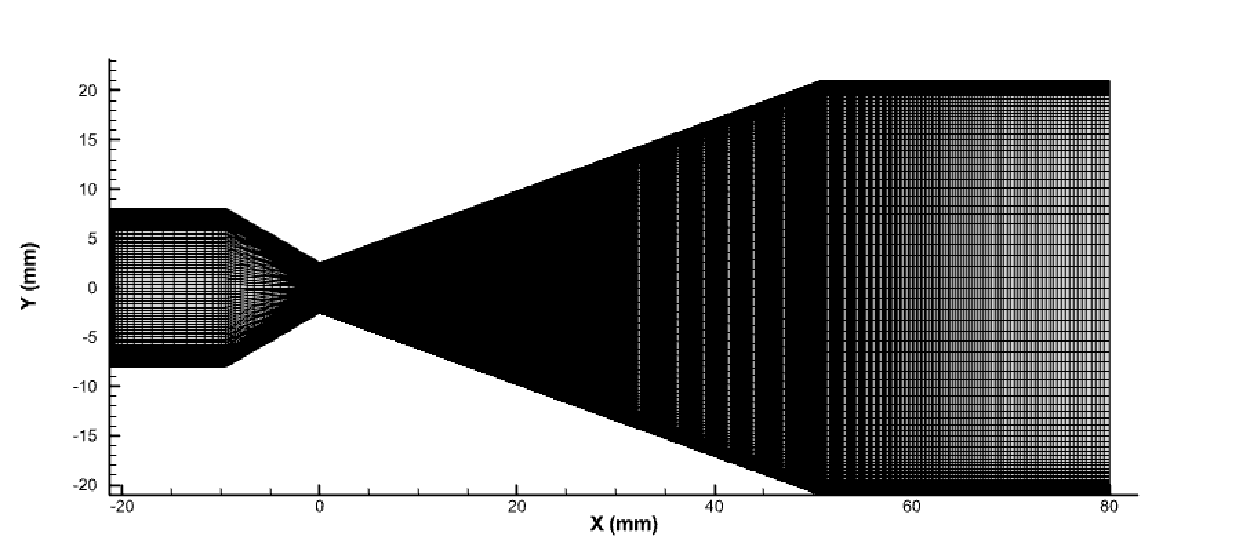}} }\\
 \subfigure[Adaptive region of DVM when $\text{Kn}_{ref}=0.01$]{
        {\includegraphics[width=0.45\textwidth,clip = true]{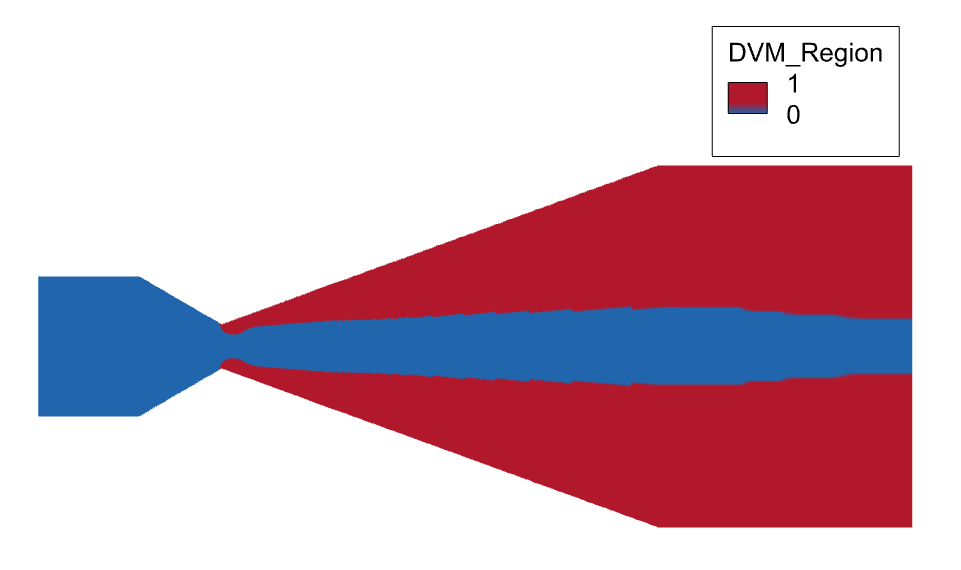}} }
    \caption{The computational domain for a micro nozzle and an example of computational domain adaptation. 
    }
    \label{fig:2DNozzle_computationalDomainIllustration}
\end{figure}

The flow in a planar micro nozzle presents a multi-scale challenge, transitioning gradually from a continuous flow in the inlet to a free molecular flow in the outlet, which can effectively test the capabilities of aGSIS. The geometry and dimensions of the nozzle are depicted in Fig.~\ref{fig:2DNozzle_computationalDomain}, which is based on the shape of the Rothe nozzle \cite{rothe1971electron, kim2024evaluation}. The ambient condition is a vacuum, and the left side boundary condition is set as an inflow with  density  $\rho_{in}=4.01 \times 10^{-3}\ \text{kg}/\text{m}^3$ and temperature $T_{in}=1000\ \text{K}$. 
When the throat diameter of 5.1 mm is chosen as the reference length, the Knudsen number corresponding to the inflow condition is $\text{Kn}$ = $4.9 \times 10^{-3}$.
An isothermal wall with $T_{w}=1000\ \text{K}$  and a fully diffuse gas-wall interaction model is employed. A non-uniform structured spatial grid is utilized, as illustrated in Fig.~\ref{fig:2DNozzle_AGSIS_Kn0d01_config}, with a total of 26,400 physical space grid points. The truncated velocity space $[-12\sqrt{RT_0},\ 12\sqrt{RT_0}] \times [-10\sqrt{RT_0},\ 10\sqrt{RT_0}]$ is uniformly orthogonally discretized into $72 \times 60$ points.


Figure \ref{fig:2DNozzle_computationalDomainIllustration}(c) displays the computational domain adaptation outcome at a reference Knudsen number $\text{Kn}_{ref}=0.01$. The red region indicates the DVM area, which encompasses a total of 17,319 elements, constituting 65.6\% of the domain.
The DVM computational domain undergoes dynamic adjustments twice. The initial adjustment occurs when the error in the NS solver reaches $10^{-6}$, leading to the formation of the initial DVM computational domain. After the GSIS solver has been run for 5 iterations, the DVM computational domain is updated a second time, based on the gradients of the computed macroscopic quantities. In a two-dimensional simulation, the DVM leverages velocity-space parallelization, eliminating the need for load balancing. Instead, memory can be dynamically allocated and deallocated as required. The simulation is performed using 8 cores of an AMD EPYC 7763 processor (2.45GHz), with a total computational CPU time of 1.5 core-hours.

\begin{figure}[!t]
    \centering
    {\includegraphics[width=0.45\textwidth,clip = true]{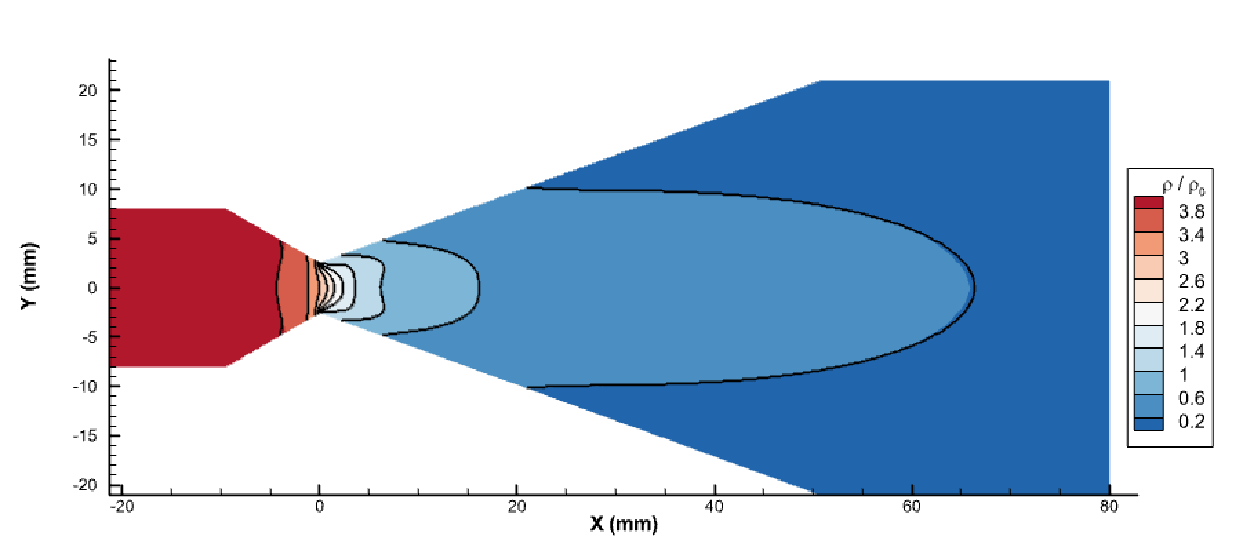}} 
    {\includegraphics[width=0.45\textwidth,clip = true]{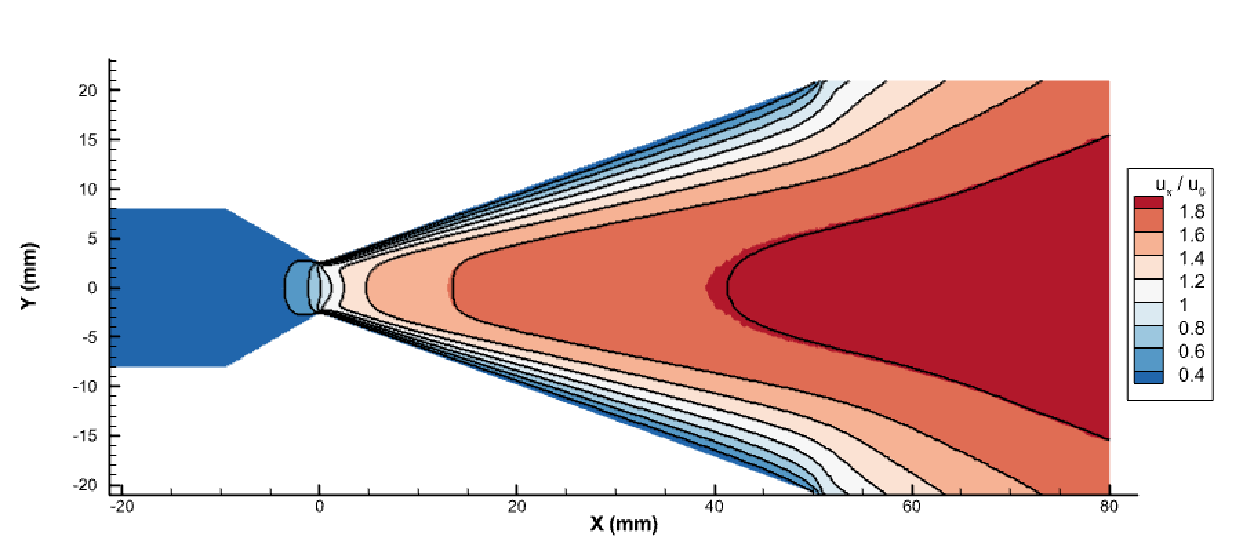}}
    \\
    {\includegraphics[width=0.45\textwidth,clip = true]{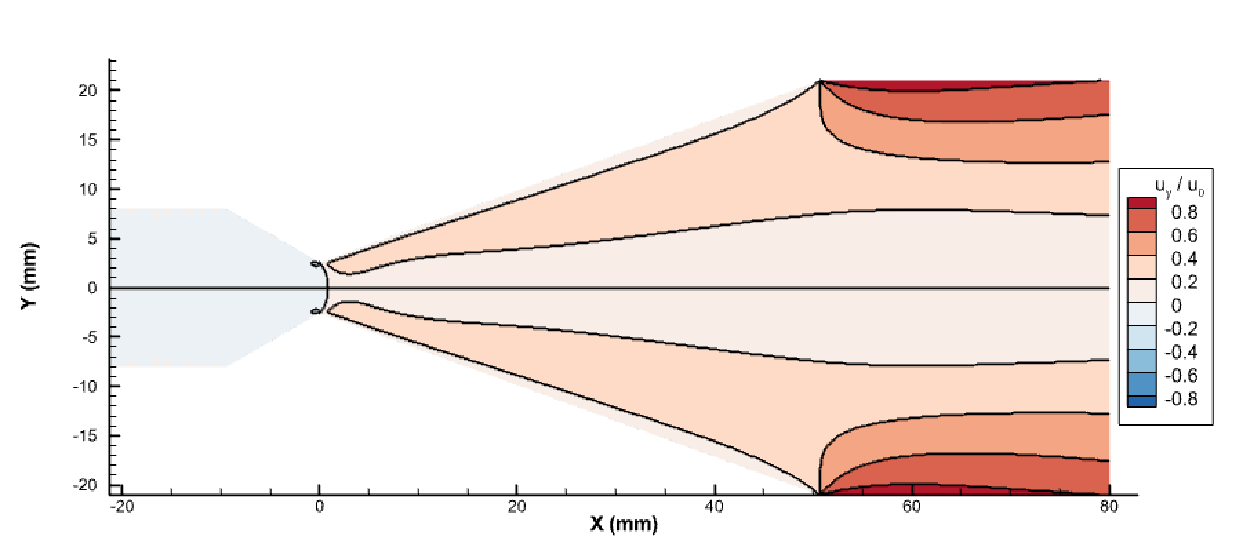}} 
    {\includegraphics[width=0.45\textwidth,clip = true]{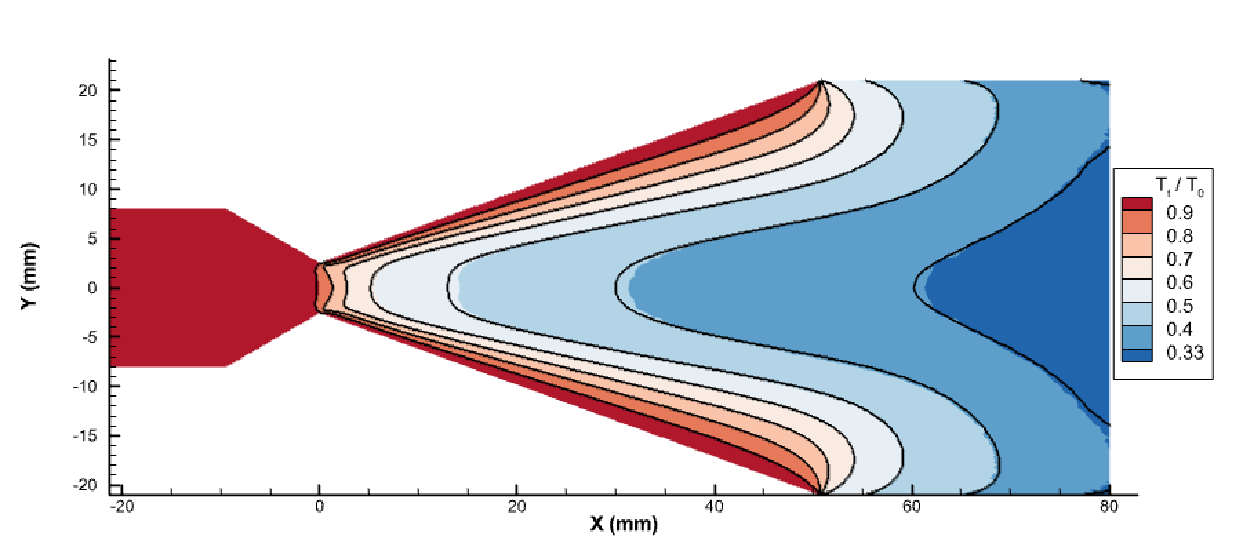}}
    \caption{A comparison of the macroscopic variables between DSMC (contours) and aGSIS (black lines).}
    \label{fig:2DNozzle_AGSIS_Kn0d01_cmp_dsmc}
\end{figure}

\begin{figure}[!t]
    \centering
    {\includegraphics[width=0.45\textwidth,clip = true]{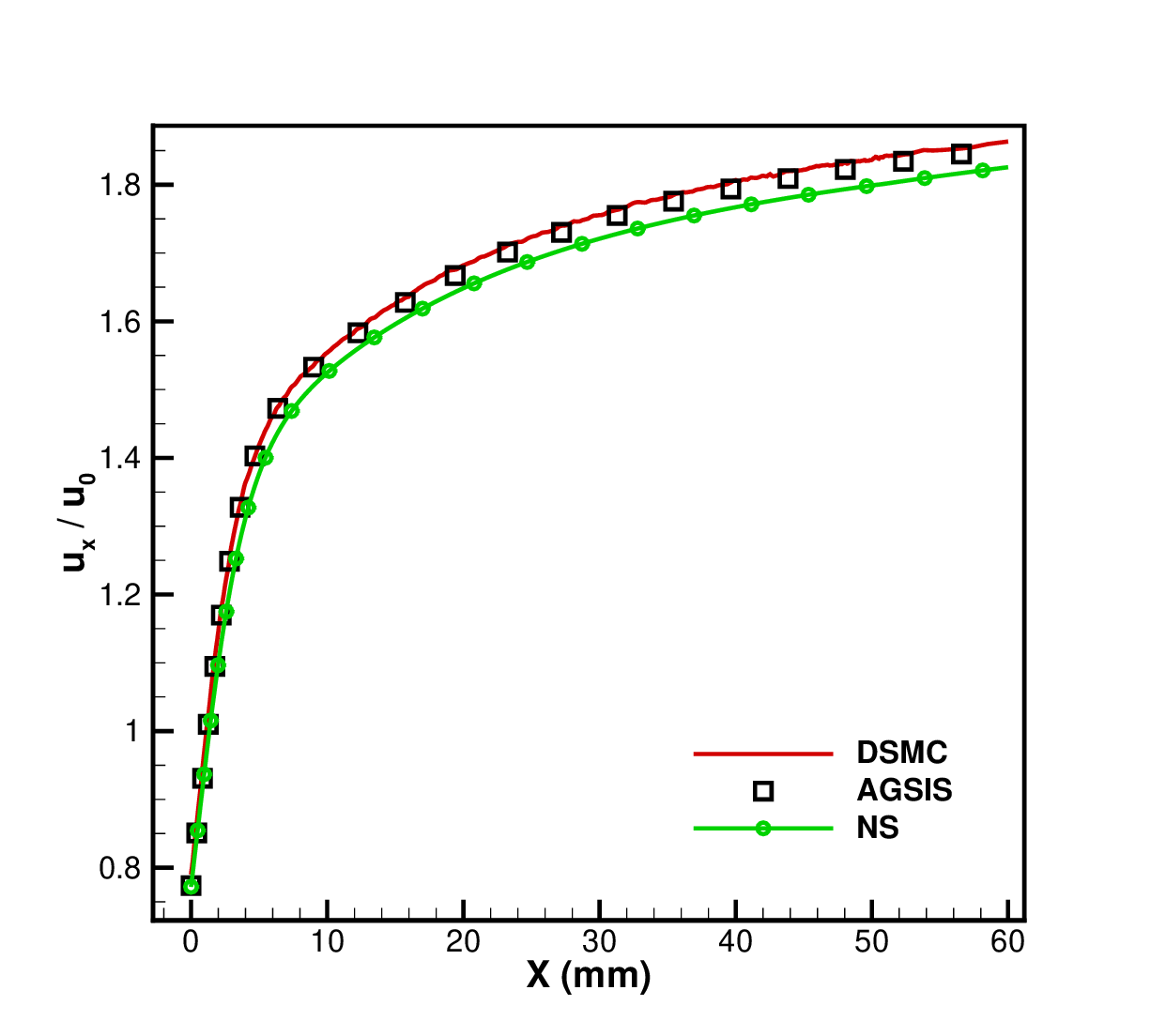}} 
    {\includegraphics[width=0.45\textwidth,clip = true]{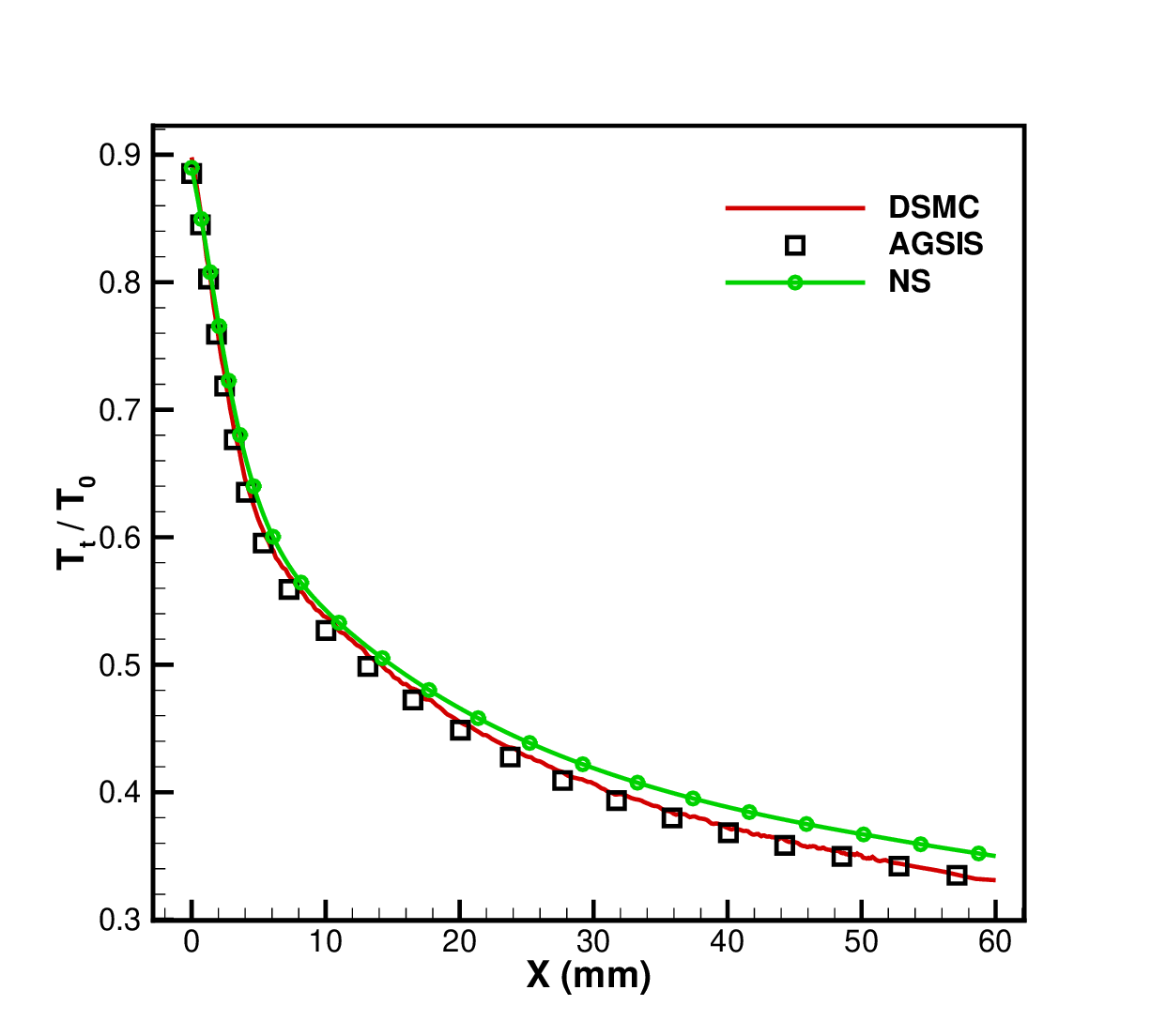}} 
    \\
    {\includegraphics[width=0.45\textwidth,clip = true]{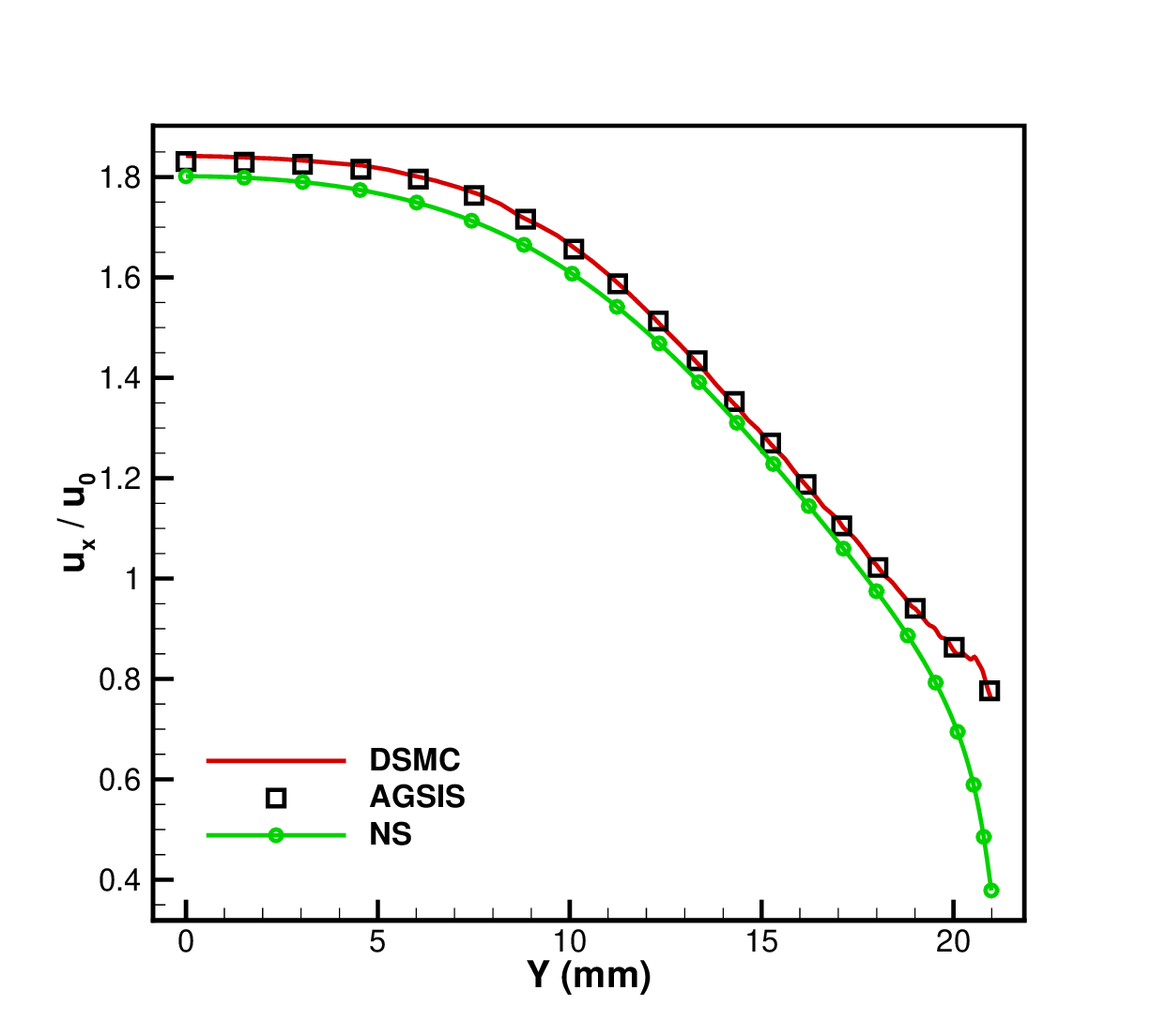}}
    {\includegraphics[width=0.45\textwidth,clip = true]{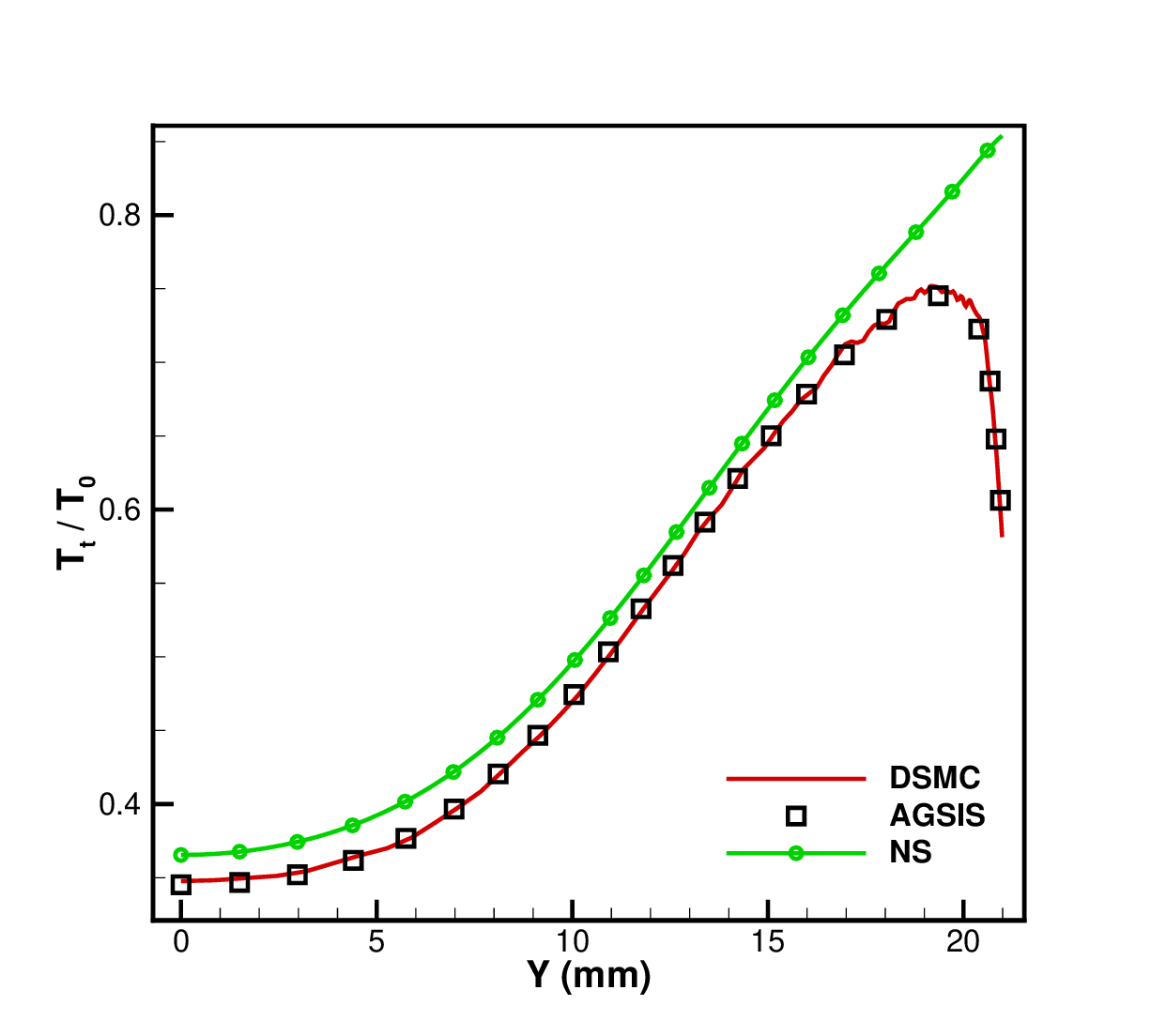}}
    \caption{Comparison of macroscopic variables along the central axis ($y=0\ \text{mm}$, top) and at the nozzle exit line ($x=51\ \text{mm}$, bottom).}
    \label{fig:2DNozzle_AGSIS_Kn0d01_cmp_gsis_dsmc_ns}
\end{figure}

Figure~\ref{fig:2DNozzle_AGSIS_Kn0d01_cmp_dsmc} presents a comparative analysis of the macroscopic variables between the GSIS and DSMC, where $\rho_0=\rho_{in}/4, T_0=T_{in}, u_0=\sqrt{\gamma RT_0}$. 
Meanwhile, Fig.~\ref{fig:2DNozzle_AGSIS_Kn0d01_cmp_gsis_dsmc_ns} compares the velocity and translational temperature along the nozzle axis ($y=0\ \text{mm}$) and in the vertical direction at the exit ($x=51\ \text{mm}$). It can be seen that the results of  aGSIS agree well with DSMC.  

The partitioning of the spatial domain is justified in the following manner. A comparison of macroscopic variables along the central axis reveals that the results obtained from the NS solver align closely with the accurate solution on the left side of the throat, exhibiting only a minor deviation from the precise results in the intermediate region extending from the throat to the exit. This implies that it is sufficient to utilize exclusively the NS solver within these areas. Notably, Fig.~\ref{fig:2DNozzle_computationalDomainIllustration} shows that the aGSIS also employs solely the NS solver in this region, thereby providing validation from an alternative perspective that partitioning the computational domain based on $\text{Kn}_{Gll}$ is indeed reliable.
However, from the throat to the outlet, NS predictions increasingly diverge from actual values as they near the exit. At the exit corner of the nozzle, due to strong flow expansion \cite{kim2024evaluation}, NS predictions deviate severely from the actual values, which is also the reason why we use a multi-scale solver in this region.
In practice, the flow of vacuum will reverse direction near the nozzle corner, impacting the spacecraft's surface and potentially leading to damage. Consequently, precise prediction of this region is essential for tackling practical challenges.

\subsection{Hypersonic flow around a circular cylinder}\label{2DCylinder}

\begin{figure}[!t]
    \centering
     \subfigure[]{\includegraphics[width=0.4\textwidth,trim={20 20 100 50},clip = true]{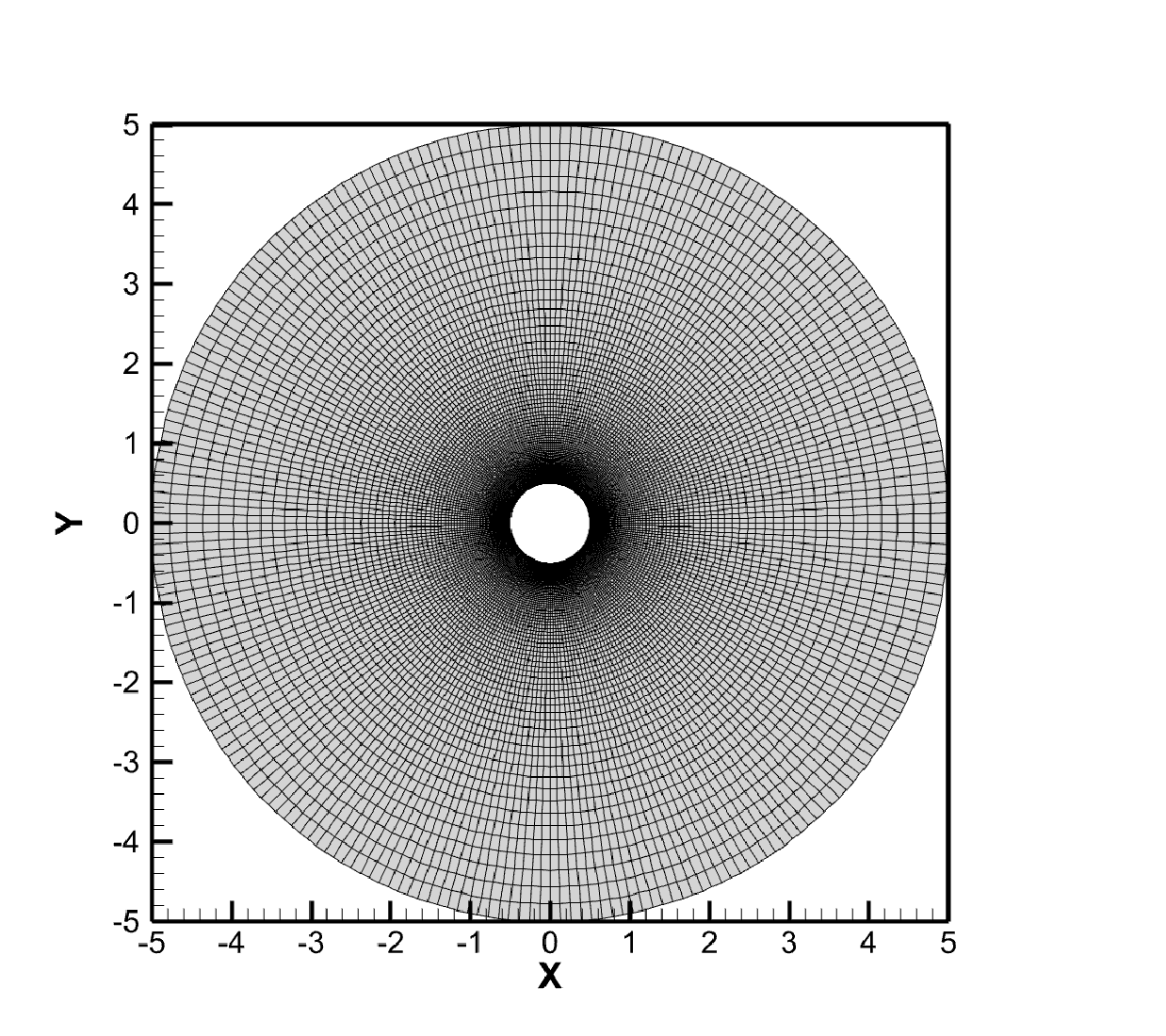}} 
    \subfigure[]{\includegraphics[width=0.4\textwidth,trim={50 50 100 50},clip = true]{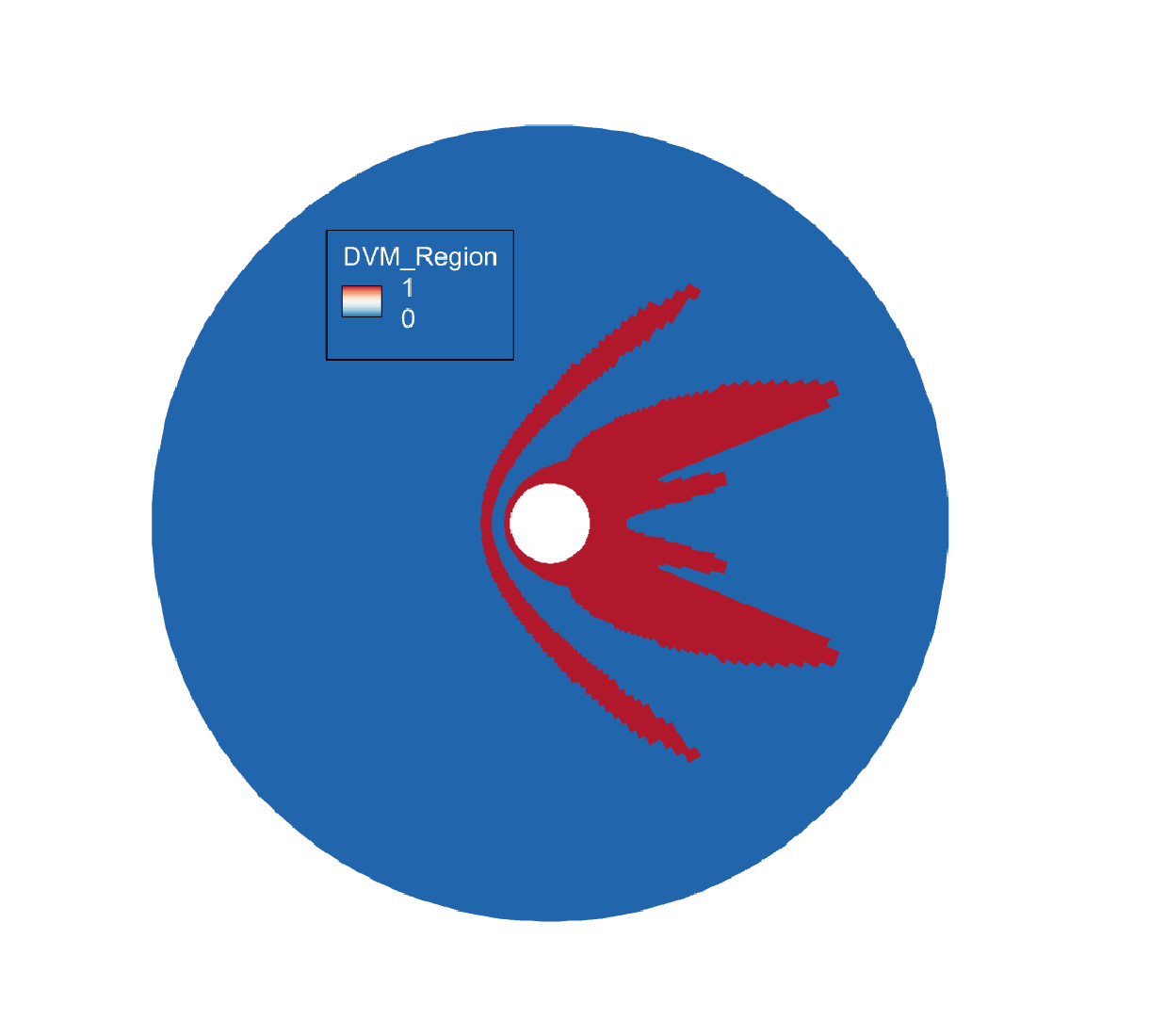}} \\
    \subfigure[]{
    	\label{fig:2DCylinder21120_Ma5_Kn0d01_DSMC_AGSIS_NS_cmp_rho_dvmRegion}
    	{\includegraphics[width=0.45\textwidth,trim={20 20 50 50},clip = true]{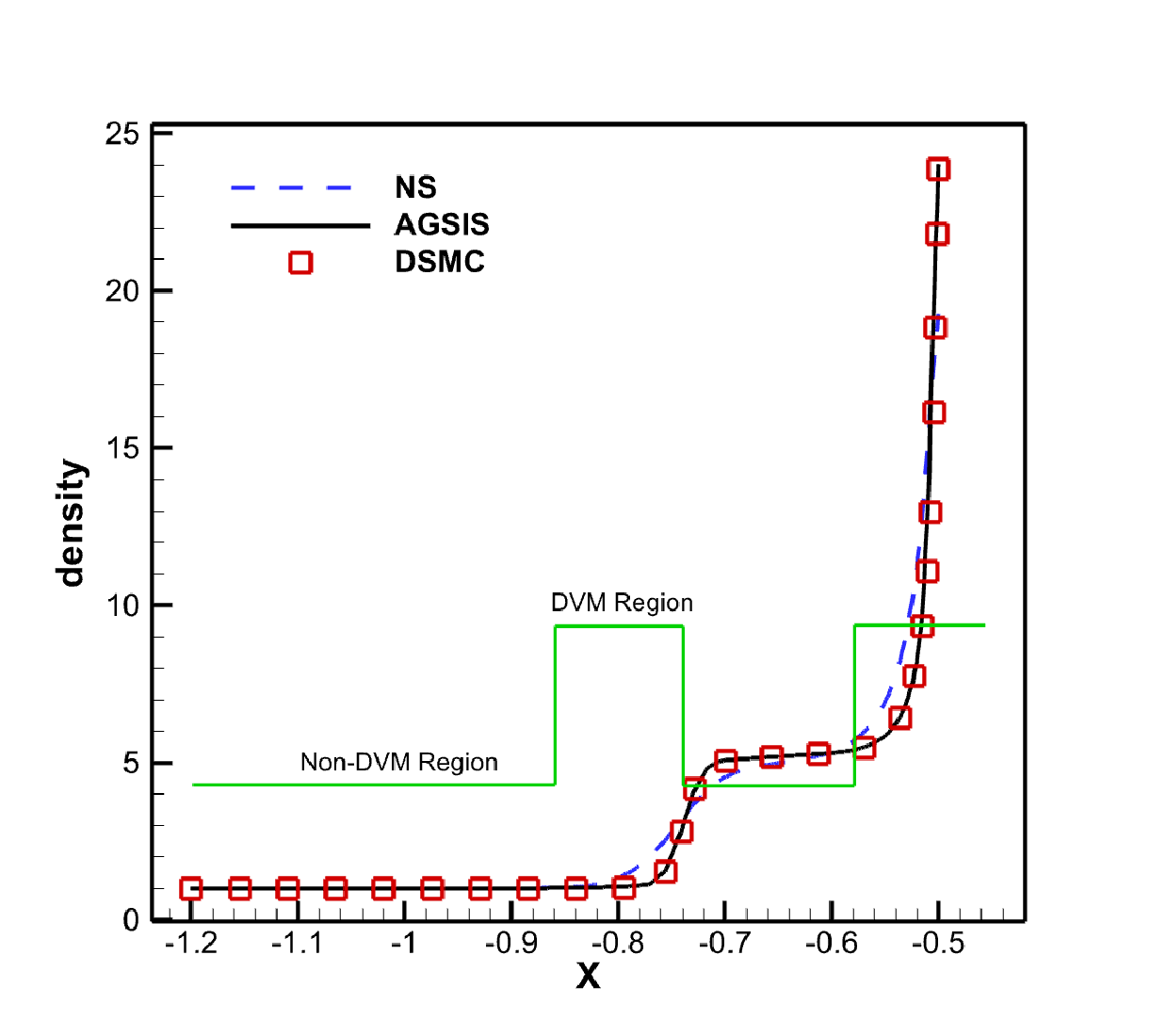}} }
    \caption{(a) The computational domain for hypersonic flow around a cylinder, which is discretized by $240 \times 88$ quadrilateral cells.
(b)  The spatial adaptation  in a hypersonic flow with Ma=5 and $\text{Kn} = 0.01$. When $\text{Kn}_{ref}=0.01$, the number of grid cells in the red DVM region is  10,752.
(c) Comparison of the density distributions  along the stagnation line for the hypersonic flow. 
}
    \label{fig:2DCylinder21120_Mesh}
\end{figure}

In this problem, a significant portion of the computational domain can be considered to be in a state of equilibrium, especially before the bow shock wave. For rarefied gas flows with a Knudsen number in the slip flow regime, there will also be a computational domain in the leeward region that is in a continuous flow state. Clearly, for these domains, we can treat the distribution function as a macroscopic quantity corresponding to an equilibrium state and perform simplifications. 

Hypersonic flows at Mach 5 with Knudsen numbers of 0.1 and 0.01 are considered. The inflow temperature is $T_0 = 273\ \text{K}$. The isothermal surface with $T_w=273\ \text{K}$ and fully diffuse gas–wall interaction is adopted. The reference length is taken as the cylinder diameter  $D=1\ \text{m}$. The mesh configuration is shown in Fig.~\ref{fig:2DCylinder21120_Mesh}(a), where the entire computational domain has a radius of  $5\ \text{m}$,  divided into $240\times88$ quadrilateral cells, and the first layer of grid cells adjacent to the wall has a thickness of  $0.005\ \text{m}$.

Figure~\ref{fig:2DCylinder21120_Mesh}(b) shows the computational domain of DVM at $\text{Kn} = 0.01$.
It is seen that, a significant portion of the free flow pertains to the computational grid that is not within the DVM computational domain, and accordingly, this computation domain only employs an NS solver for computation.  Also, there are regions near the rear of the cylinder and downstream where the changes in macroscopic quantities are relatively small. Interestingly, a discontinuity within the DVM computational domain in the upwind region of the cylinder. This occurs because the region behind the shock wave and adjacent to the wall exhibits high density and temperature, with minimal variations, thus residing in a continuous domain.

In aGSIS, the truncated velocity space  $[-14\sqrt{RT_0}, 14\sqrt{RT_0}] \times [-12\sqrt{RT_0}, 12\sqrt{RT_0}]$ is discretized uniformly into $70\times60$ cells. And this  discretization remains uniform when the Knudsen number varies. 
Figure~\ref{fig:2DCylinder21120_Mesh}(c) shows that the aGSIS and DSMC results are in good agreement, while the NS prediction shows a deviation, with the main discrepancy occurring in the DVM region corresponding to the green solid line. This also provides indirect evidence for the effectiveness of the parameter $\text{Kn}_{Gll}$ in capturing non-equilibrium regions and the applicability of aGSIS in such problems. 


\begin{figure}[!t]
    \centering
    {\includegraphics[width=0.32\textwidth,trim={0 0 20 20},clip = true]{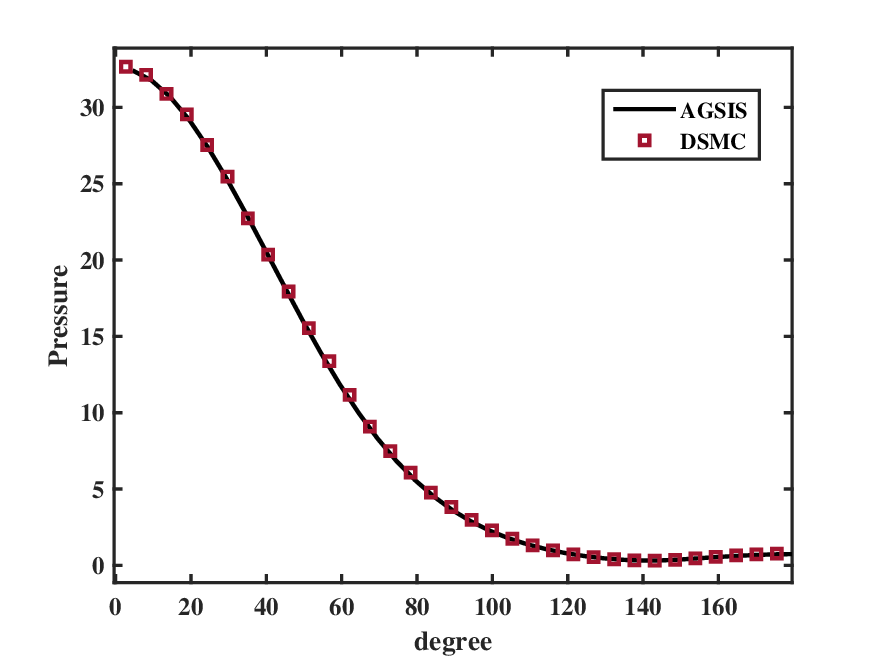}} 
    {\includegraphics[width=0.32\textwidth,trim={0 0 20 20},clip = true]{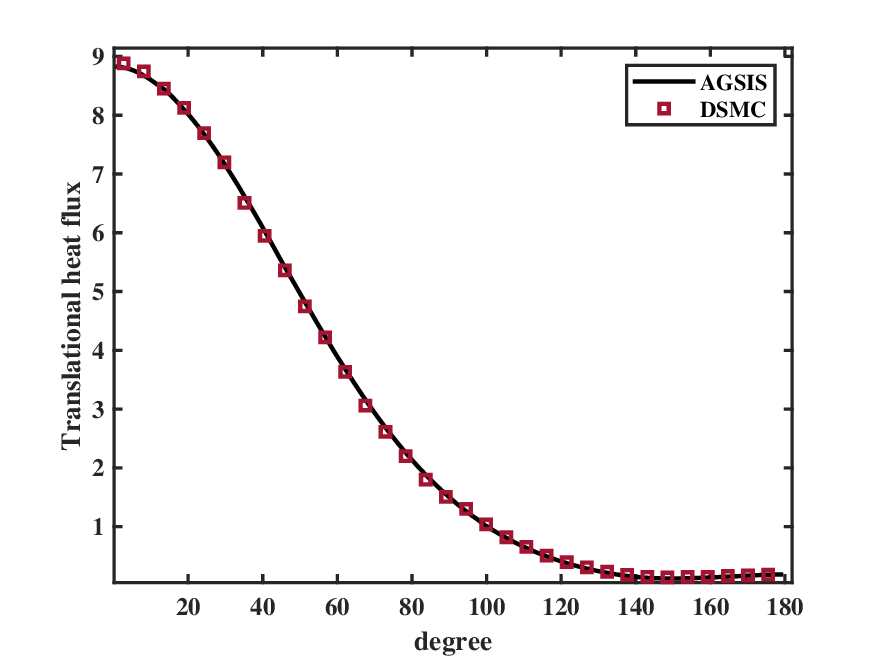}} 
    {\includegraphics[width=0.32\textwidth,trim={0 0 20 20},clip = true]{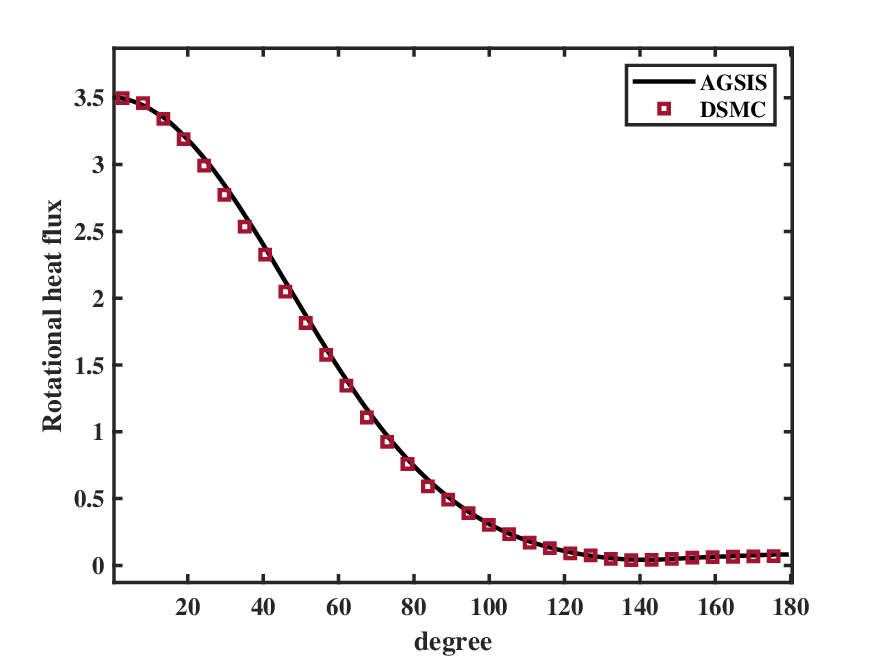}} \\
     {\includegraphics[width=0.32\textwidth,trim={0 0 20 20},clip = true]{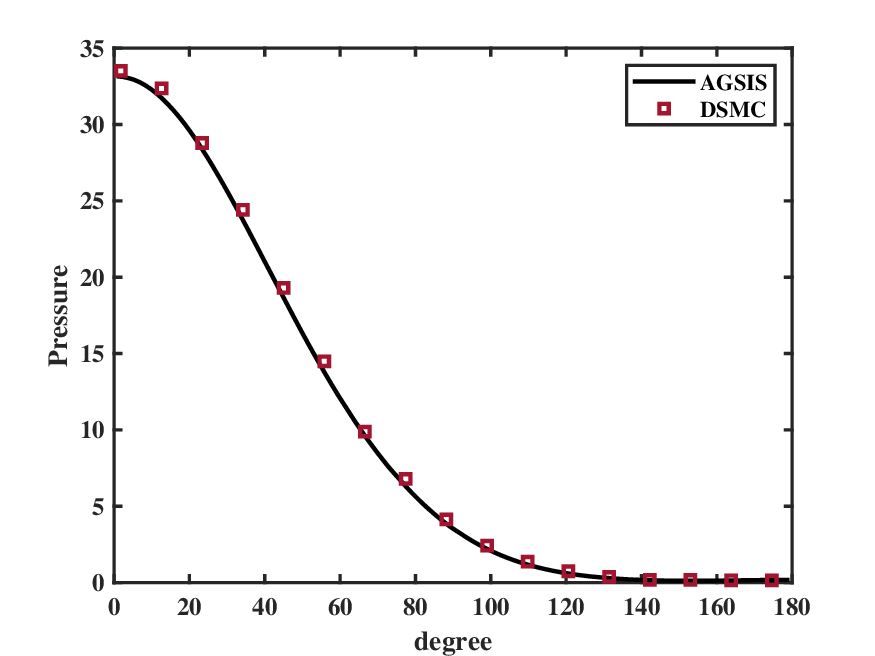}} 
    {\includegraphics[width=0.32\textwidth,trim={0 0 20 20},clip = true]{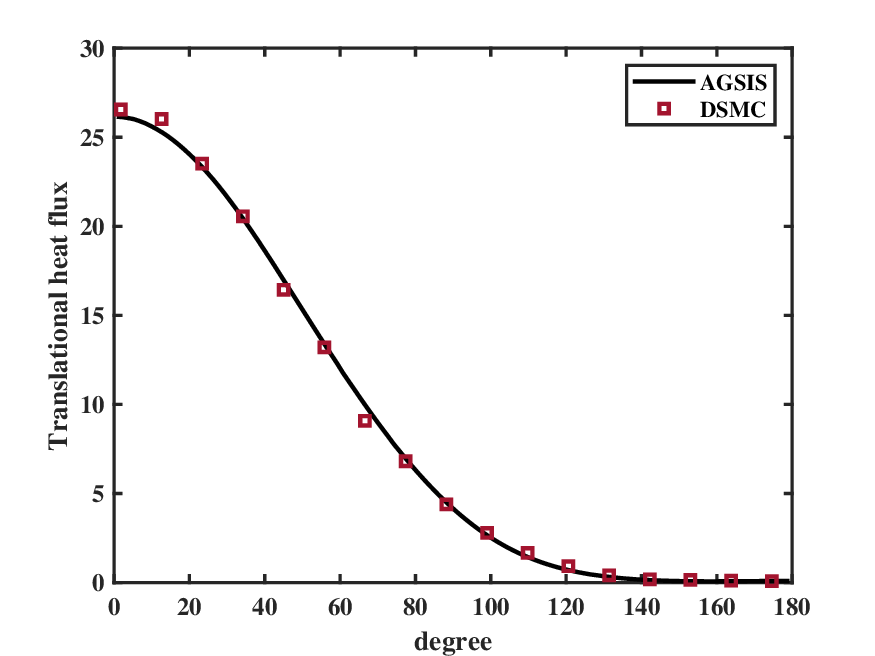}} 
    {\includegraphics[width=0.32\textwidth,trim={0 0 20 20},clip = true]{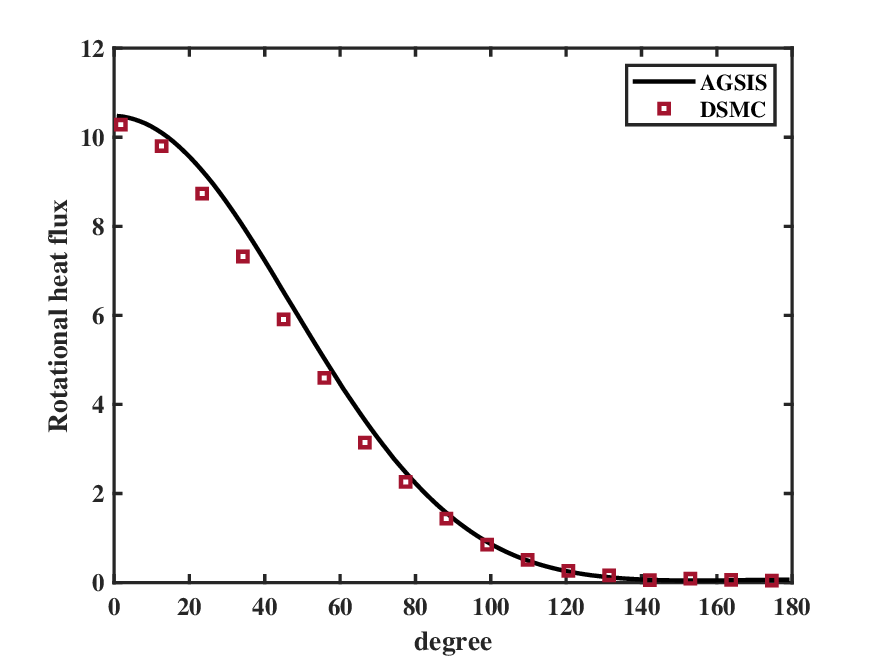}} 
    \caption{Comparisons of  macroscopic quantities along the cylinder surface in hypersonic flows. 
    	(First row) $\text{Kn} = 0.01$ and $\text{Ma} = 5$.
    	(Second row) $\text{Kn} = 0.1$ and $\text{Ma} = 5$.
    }
    \label{fig:2DCylinder_Ma5_Kn0d01_Wall_SQ_DSMC_AGSIS_cmp}
\end{figure}

\begin{figure}[!t]
    \centering
    {\includegraphics[width=0.4\textwidth,trim={20 20 100 90},clip = true]{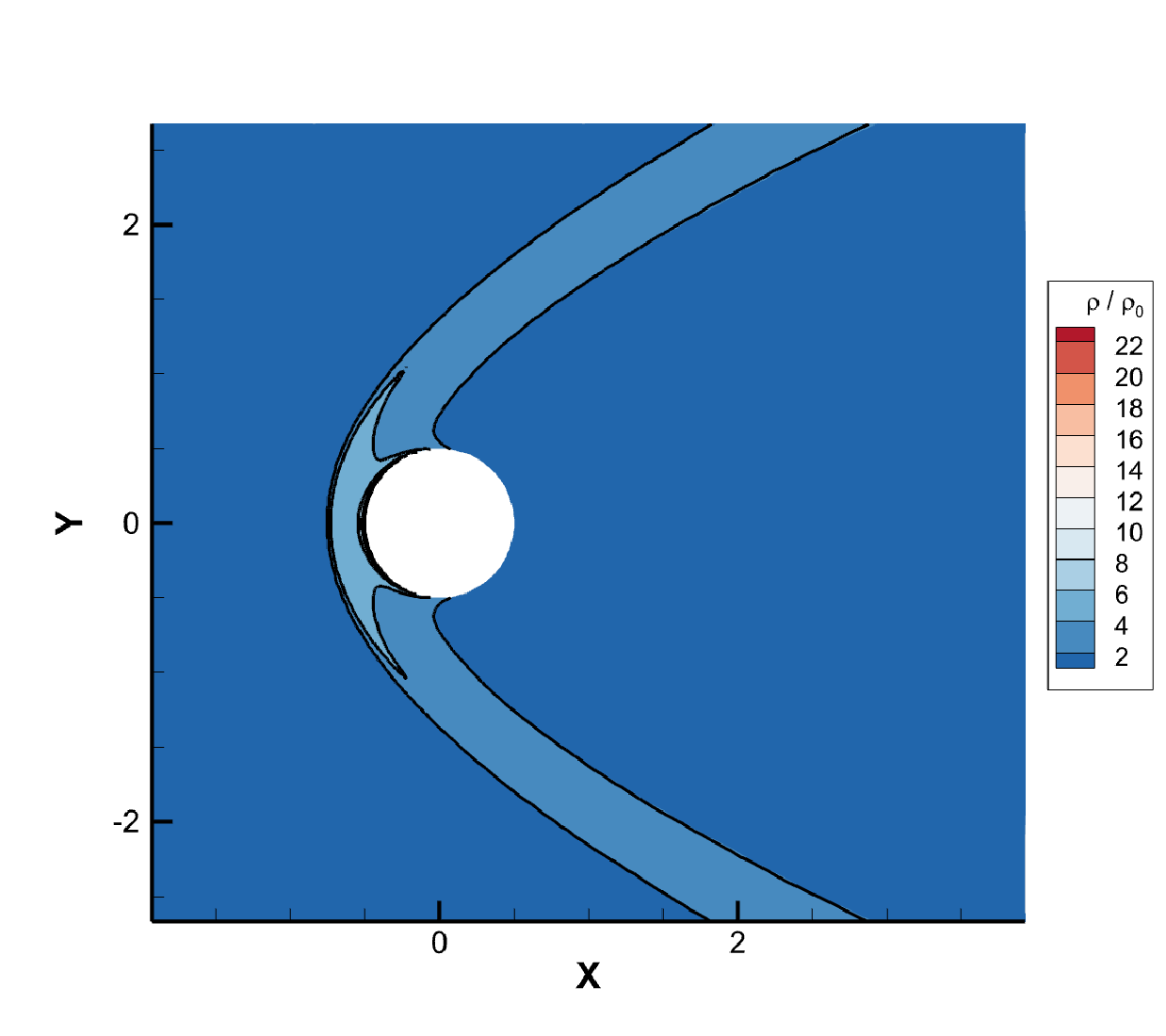}} 
    {\includegraphics[width=0.4\textwidth,trim={20 20 100 90},clip = true]{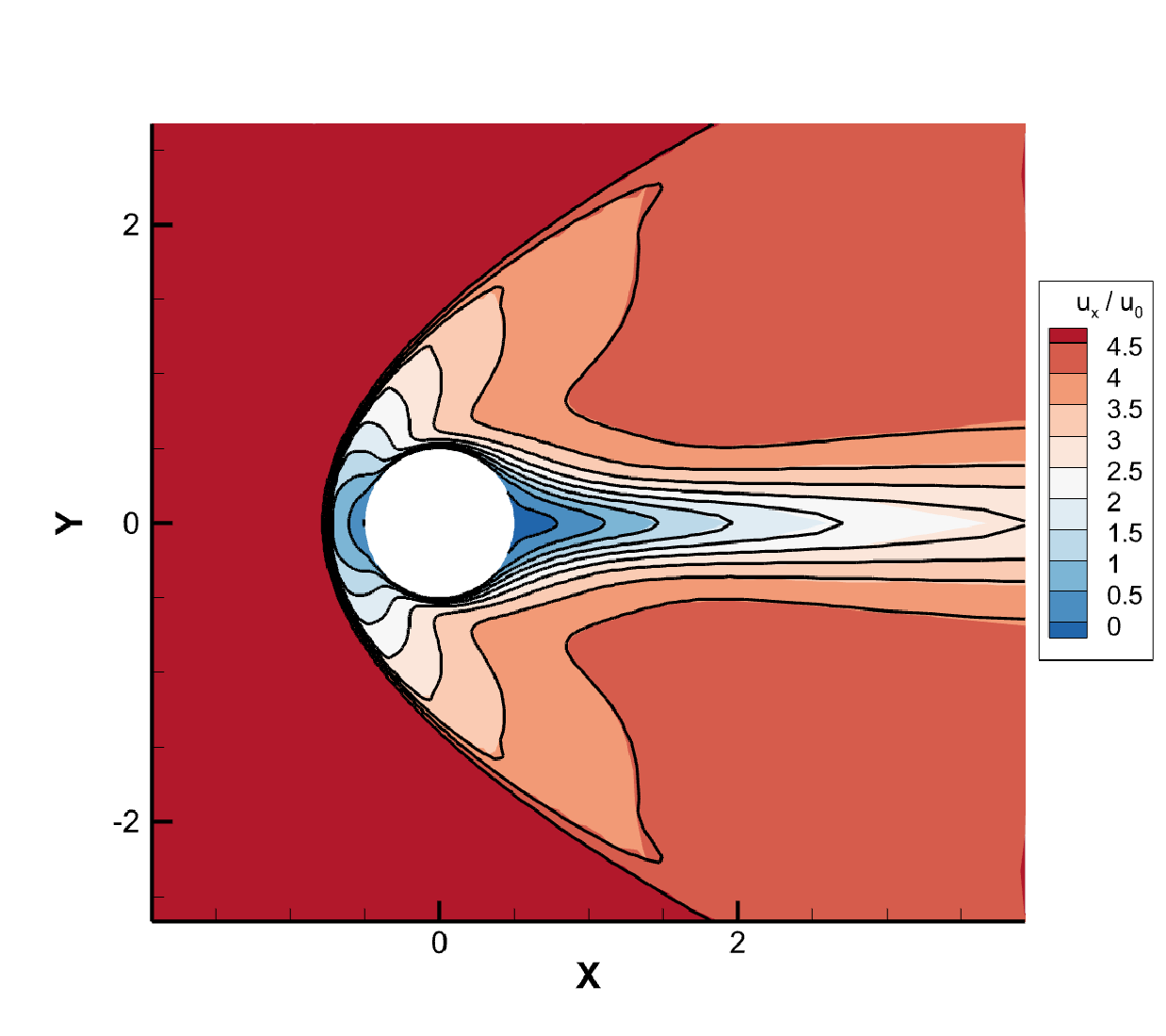}}
    \\
    {\includegraphics[width=0.4\textwidth,trim={20 20 100 90},clip = true]{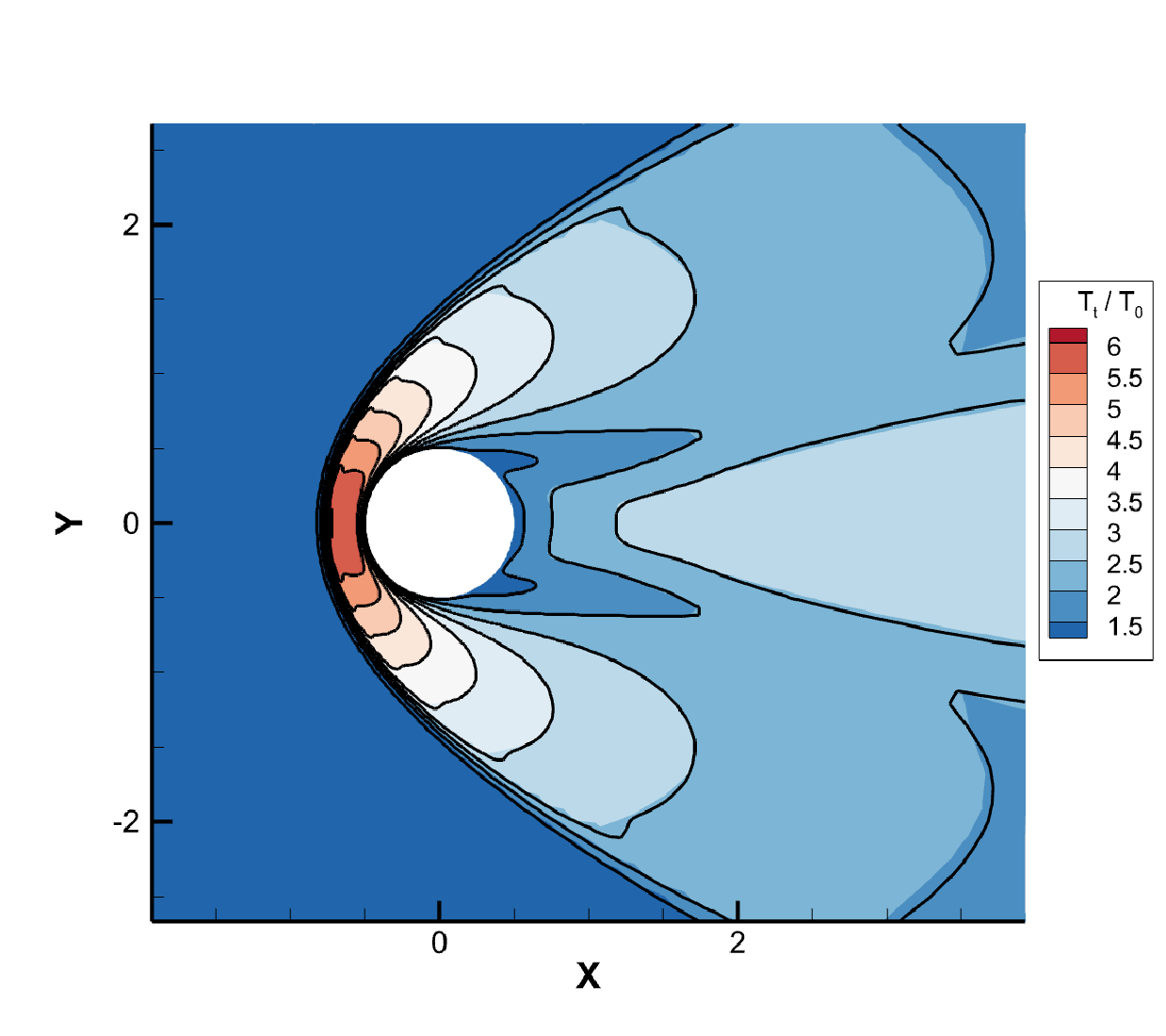}} 
    {\includegraphics[width=0.4\textwidth,trim={20 20 100 90},clip = true]{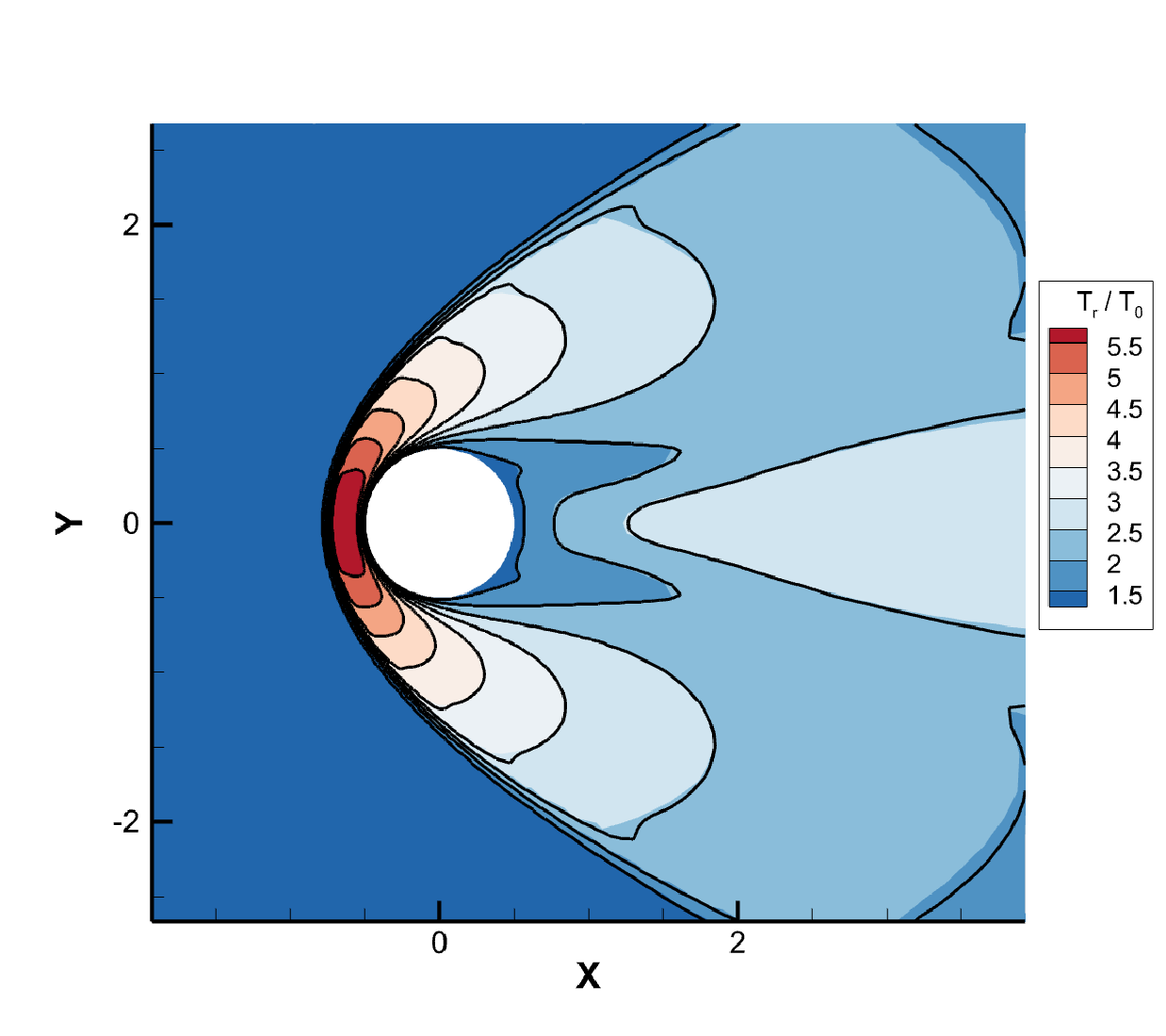}}
    
    \caption{Comparisons of macroscopic variables between the DSMC (contours) and aGSIS (black lines), when $\text{Kn} = 0.01$ and $\text{Ma} = 5$.}
    \label{fig:2DCylinder21120_Ma5_Kn0d01_DSMC_AGSIS_cmp}
\end{figure}

Figure~\ref{fig:2DCylinder_Ma5_Kn0d01_Wall_SQ_DSMC_AGSIS_cmp} compares the surface quantities (pressure, translational heat flux, and rotational heat flux) between aGSIS and DSMC. Good agreements are seen. 
Figure~\ref{fig:2DCylinder21120_Ma5_Kn0d01_DSMC_AGSIS_cmp} shows the density, velocity, and temperature (both translational and rotational) at $\text{Kn} = 0.01$. Agian, 
good consistency is seen between aGSIS and DSMC. 

Tables~\ref{tab:2DCylinder_gsis_Ma5_Effency}  compares the computational costs of original GSIS and aGSIS at different Knudsen numbers. The simulations of GSIS are conducted using 8 cores of an AMD EPYC 7763(2.45GHz). Note that the computation time includes the pre-processing computation time and the GSIS computation time, so the wall times in the table is not proportional to the iteration steps. Additionally, the pre-processing computation time will increase as the Knudsen number increases. Compared with the original GSIS method, the aGSIS shows a smaller number of iteration steps, faster computation time, and less memory consumption.
Note that the DSMC method at $\text{Kn} = 0.1$, the  implemented by the open-source software SPARTA~\cite{plimpton2015sparta}, requires 120 cores to compute for 0.2 hours.

\begin{table}[!t]
	\centering
    \caption{The comparison of computational costs around a cylinder in hypersonic flow at $\text{Ma} = 5$ for different Knudsen numbers using 8 cores with the initial number of mesh cells is 21,120.
    }
	\begin{threeparttable}
		\begin{tabular}{ccp{2cm}p{1.5cm}p{0.05cm}p{3cm}cp{2cm}p{1.5cm}}\hline
			\multirow{2}{*}{Kn}   &\multicolumn{3}{c}{Original GSIS} & ~ & \multicolumn{4}{c}{aGSIS}  \\
			\cline{2-4}  \cline{6-9}
			~               &  Steps & Wall times (s) & Memory (GB) & ~   & DVM cells \centering & Steps & Wall times (s) & Memory (GB)  \\ \hline
            0.1     & 70 & 239\centering  & \multirow{3}{*}{5.6} \centering     & ~ & 17,274\centering     & 39     & 173\centering   &  4.7    \\ 
            0.05    & 48 & 179\centering  & ~     & ~ & 16,138\centering     & 35     & 134\centering   &  4.3  \\ 
            0.01    & 28 & 120\centering  & ~     & ~ & 10,752\centering     & 19     & 68\centering    &  2.9  \\ 
            0.001   & 14 & 80\centering  & ~     & ~ & 3,180\centering     & 14     & 38\centering    &  1.0  \\ \hline
		\end{tabular}
	\end{threeparttable}
	\label{tab:2DCylinder_gsis_Ma5_Effency}
\end{table}

\subsection{Hypersonic flow passing Apollo}\label{3DApollo}

For the test of three-dimensional cases, the hypersonic flow passing Apollo at $\text{Ma}=5$ and angle of attack $\text{AoA}=30^\circ$ is simulated for $\text{Kn}=0.0012$. The reference length is $L_0=3.912$ m and temperature is $T_0=T_{\infty}=142.2$ K with $T_{\infty}$ being the free stream temperature. The isothermal surface with $T_w=300$ K and fully diffuse gas-wall interaction is adopted. The spatial domain is discretized by 372,500 hexahedral cells. The velocity space is discretized into a total of 8,166 grid cells, in a hybrid structured-unstructured manner~\cite{zhang2024efficient}.

\begin{figure}
    \centering
    \begin{minipage}{0.23\textwidth}
        \includegraphics[width=\textwidth,clip=true,keepaspectratio]{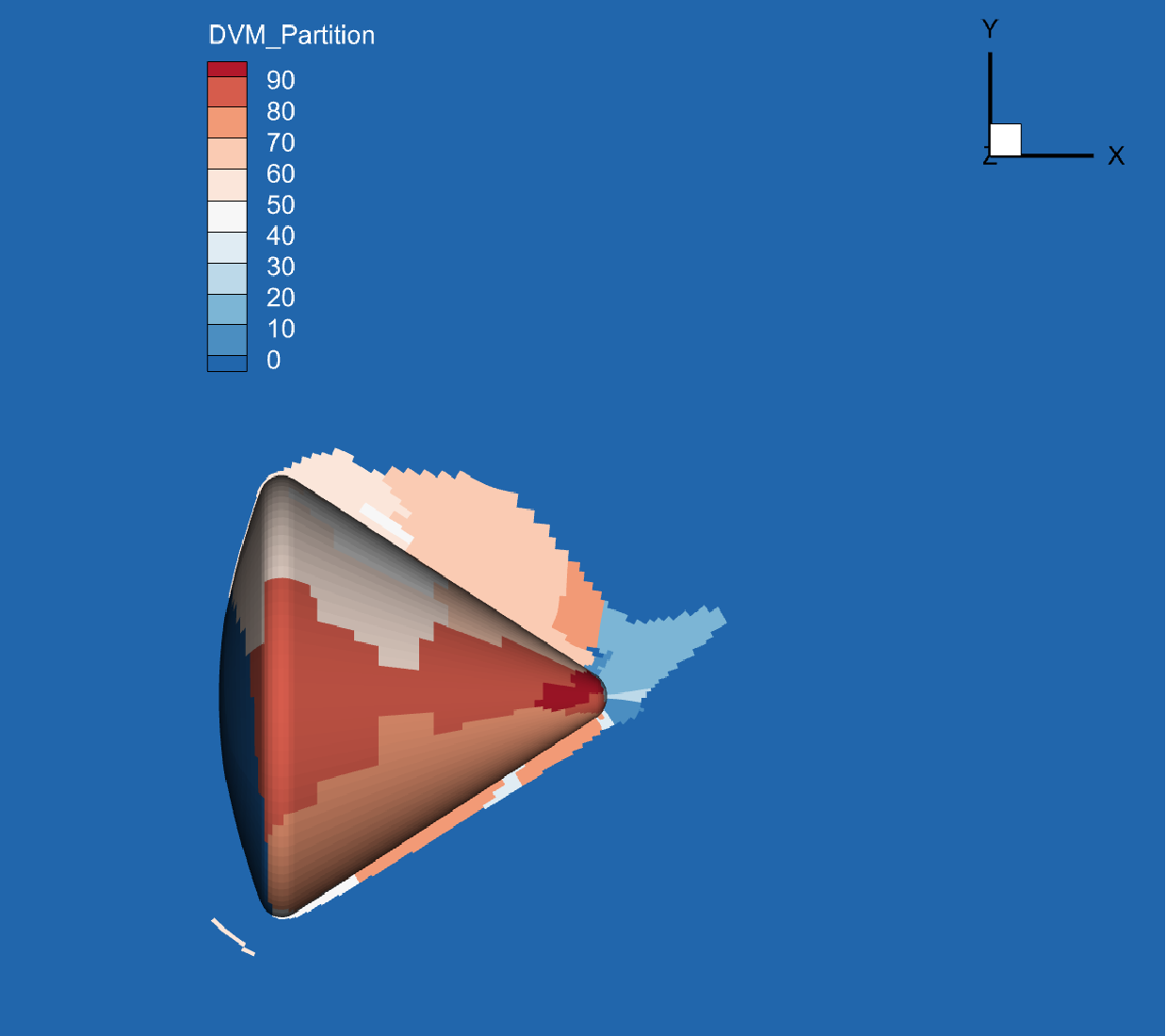}
    \end{minipage}
    \begin{minipage}{0.3\textwidth}
        \includegraphics[width=\textwidth,clip=true,keepaspectratio]{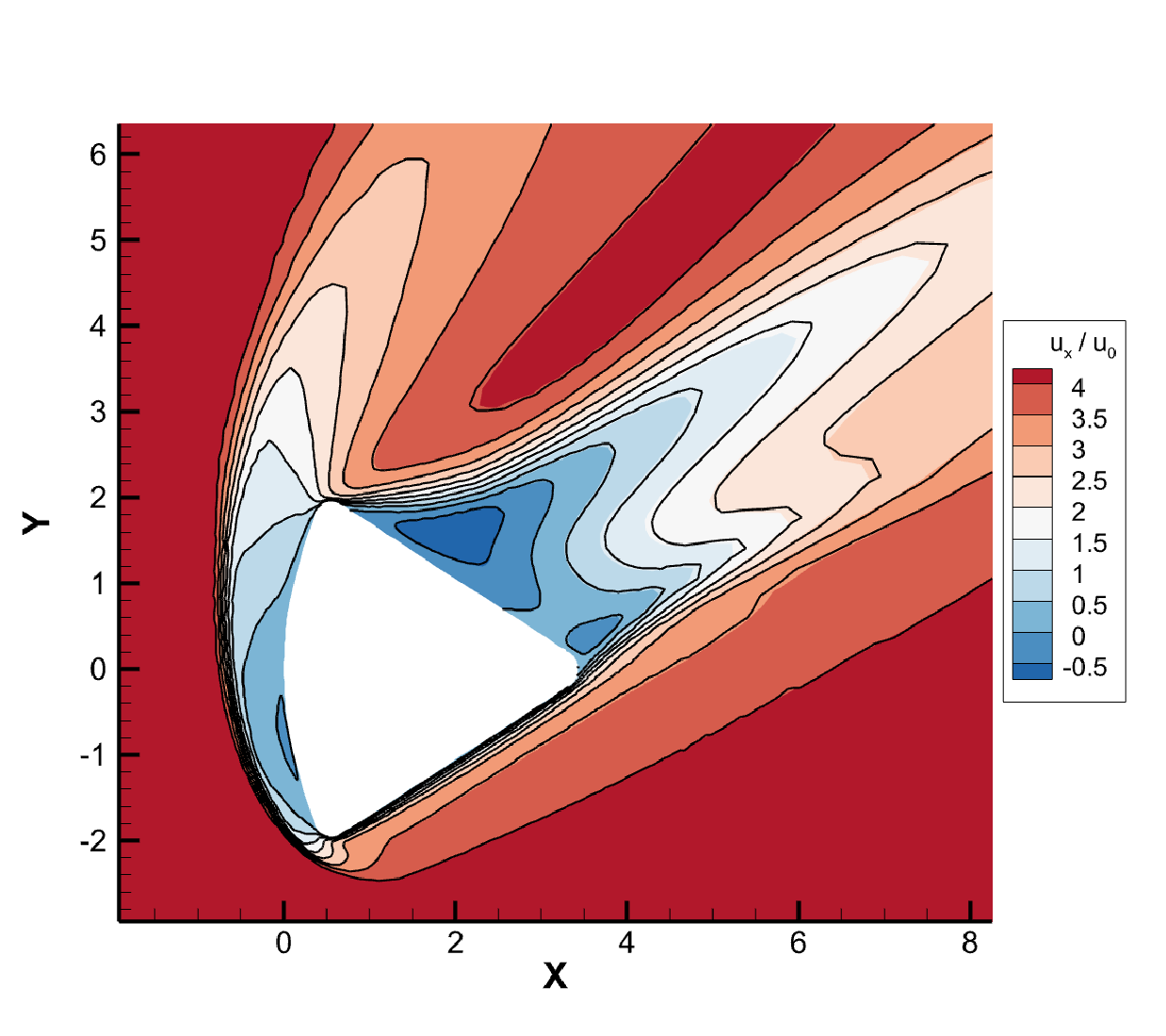}
    \end{minipage}
    \begin{minipage}{0.3\textwidth}
        \includegraphics[width=\textwidth,clip=true,keepaspectratio]{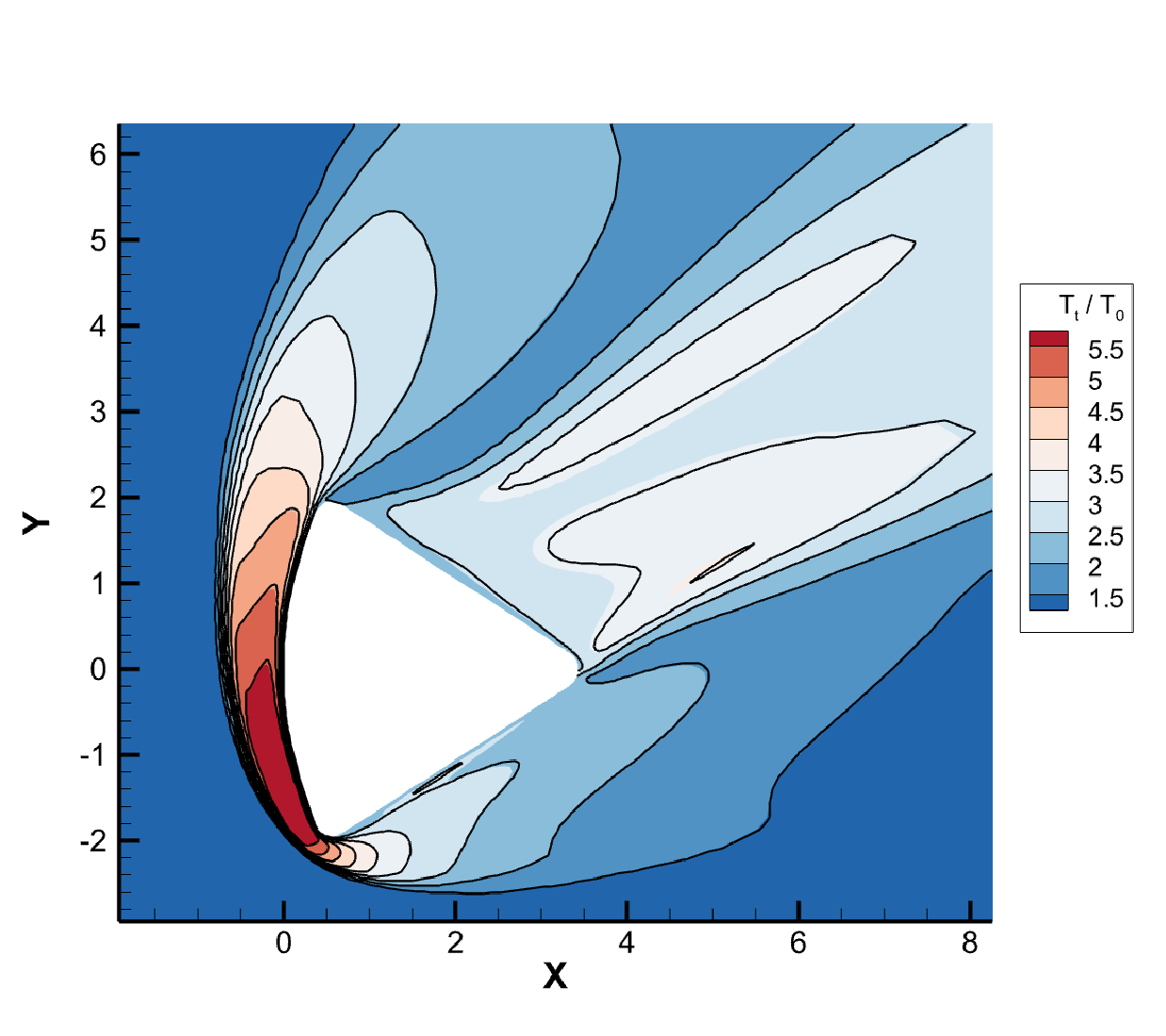}
    \end{minipage}
    \\
    \begin{minipage}{0.23\textwidth}
        \includegraphics[width=\textwidth,clip=true,keepaspectratio]{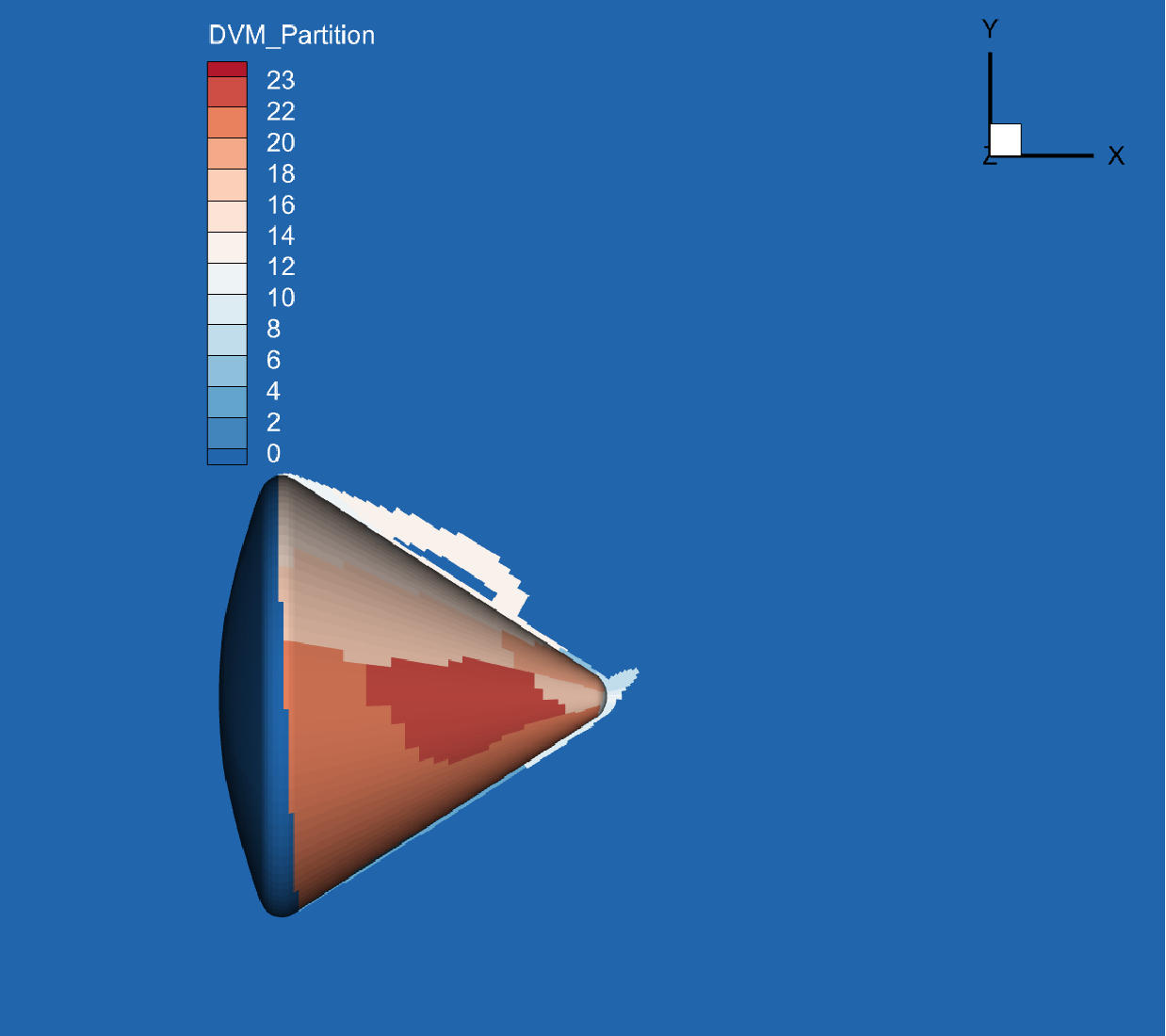}
    \end{minipage}
    \begin{minipage}{0.3\textwidth}
        \includegraphics[width=\textwidth,clip=true,keepaspectratio]{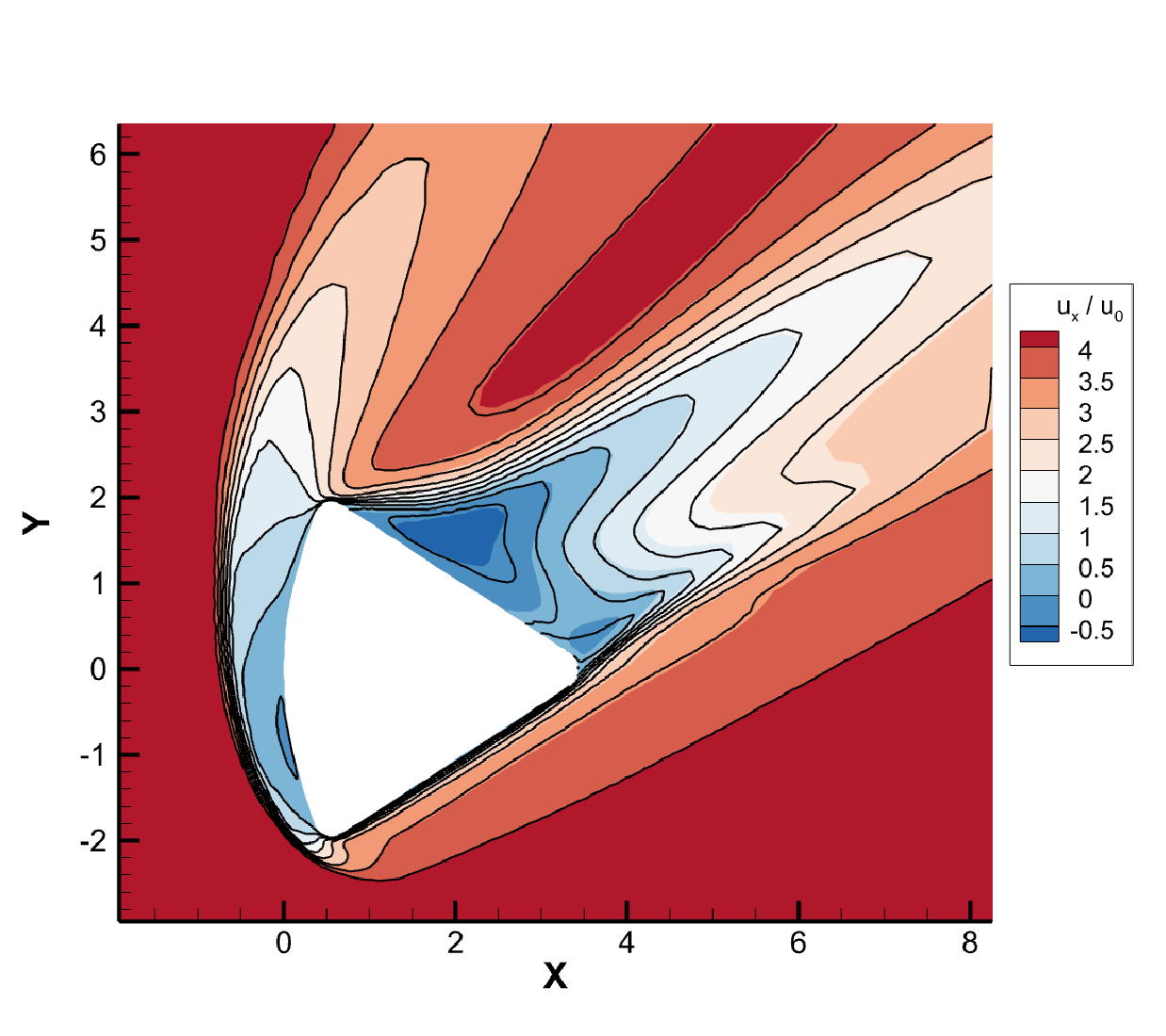}
    \end{minipage}
    \begin{minipage}{0.3\textwidth}
        \includegraphics[width=\textwidth,clip=true,keepaspectratio]{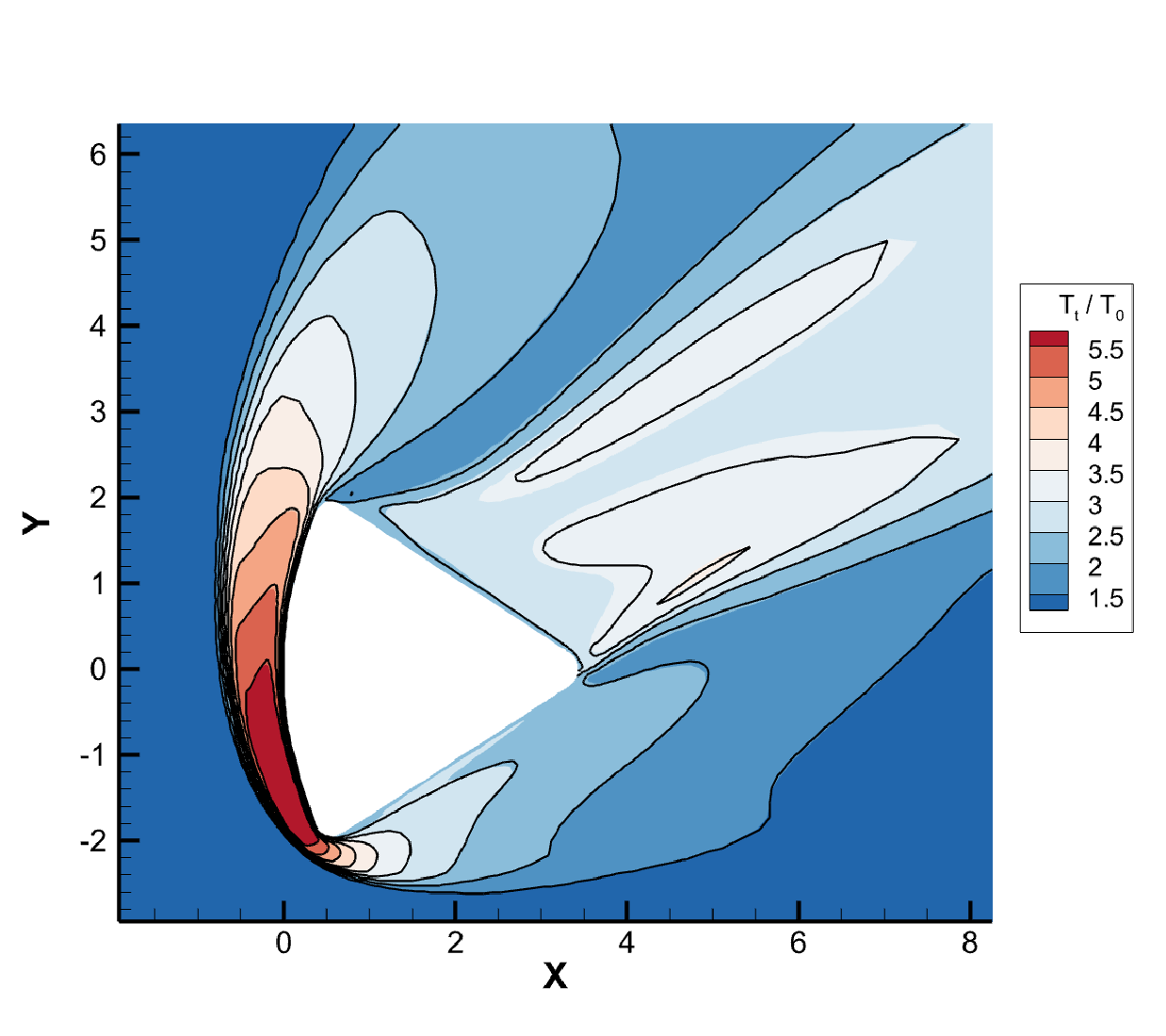}
    \end{minipage}
    \\
    \begin{minipage}{0.23\textwidth}
        \includegraphics[width=\textwidth,clip=true,keepaspectratio]{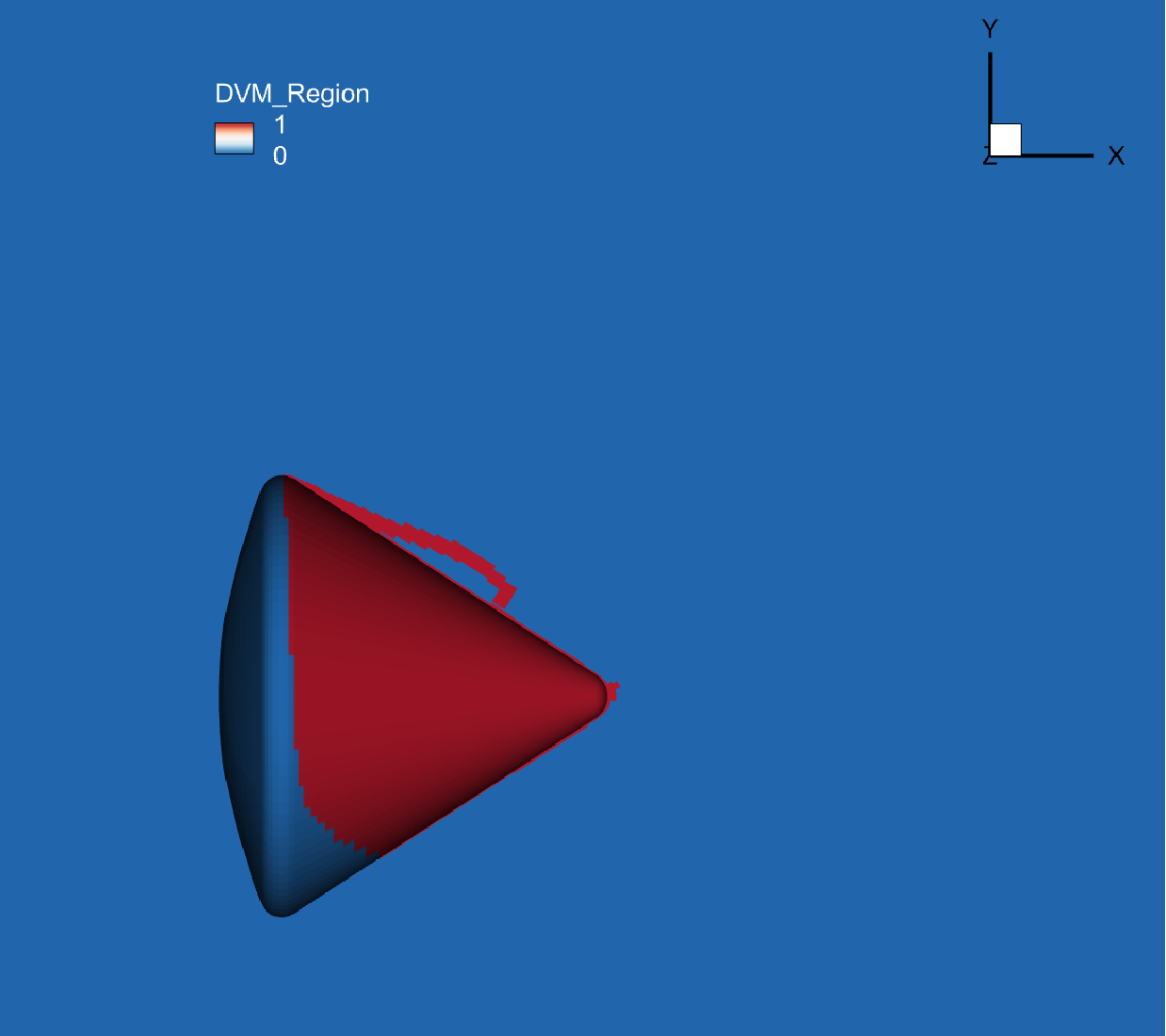}
    \end{minipage}
    \begin{minipage}{0.3\textwidth}
        \includegraphics[width=\textwidth,clip=true,keepaspectratio]{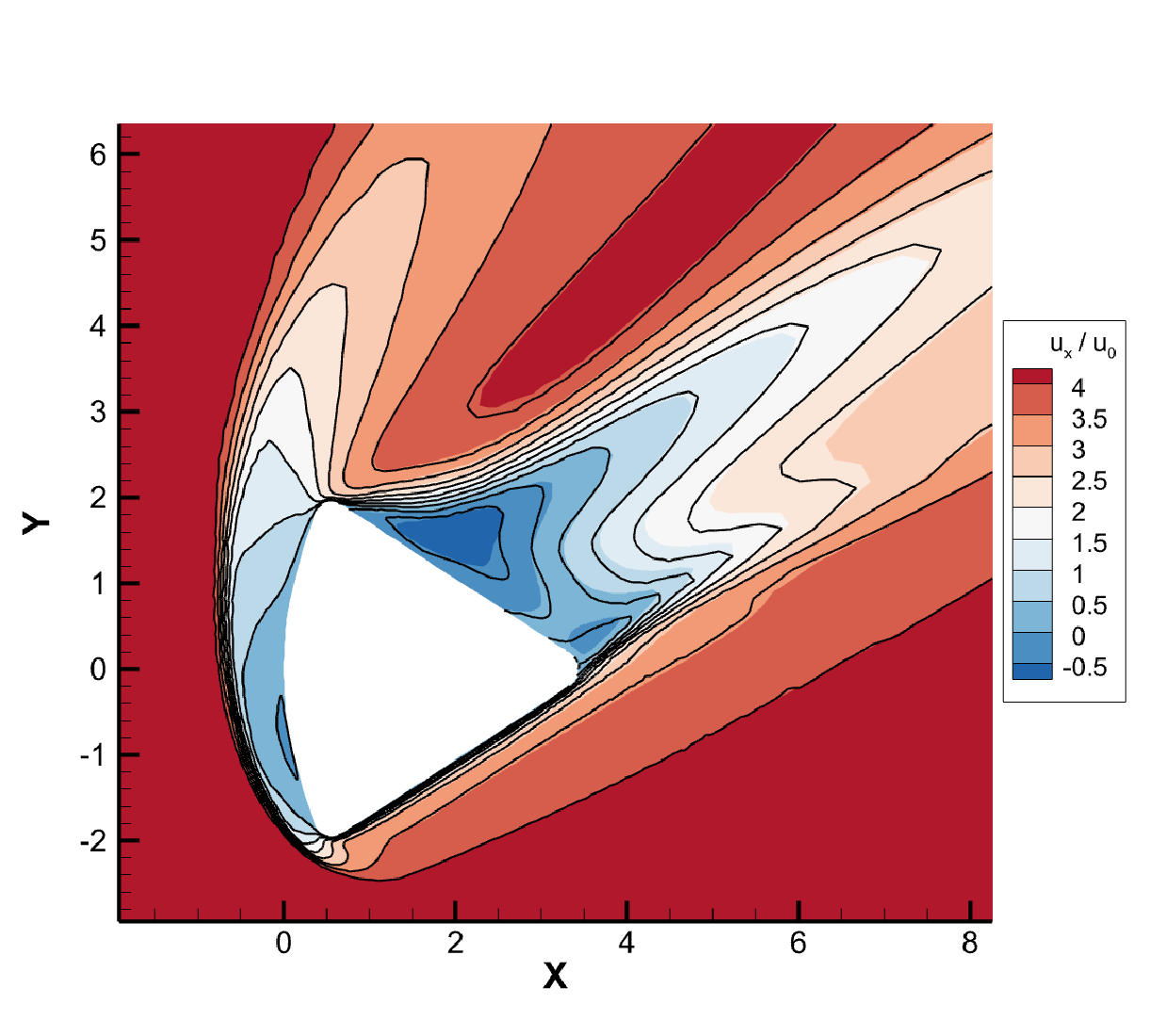}
    \end{minipage}
    \begin{minipage}{0.3\textwidth}
        \includegraphics[width=\textwidth,clip=true,keepaspectratio]{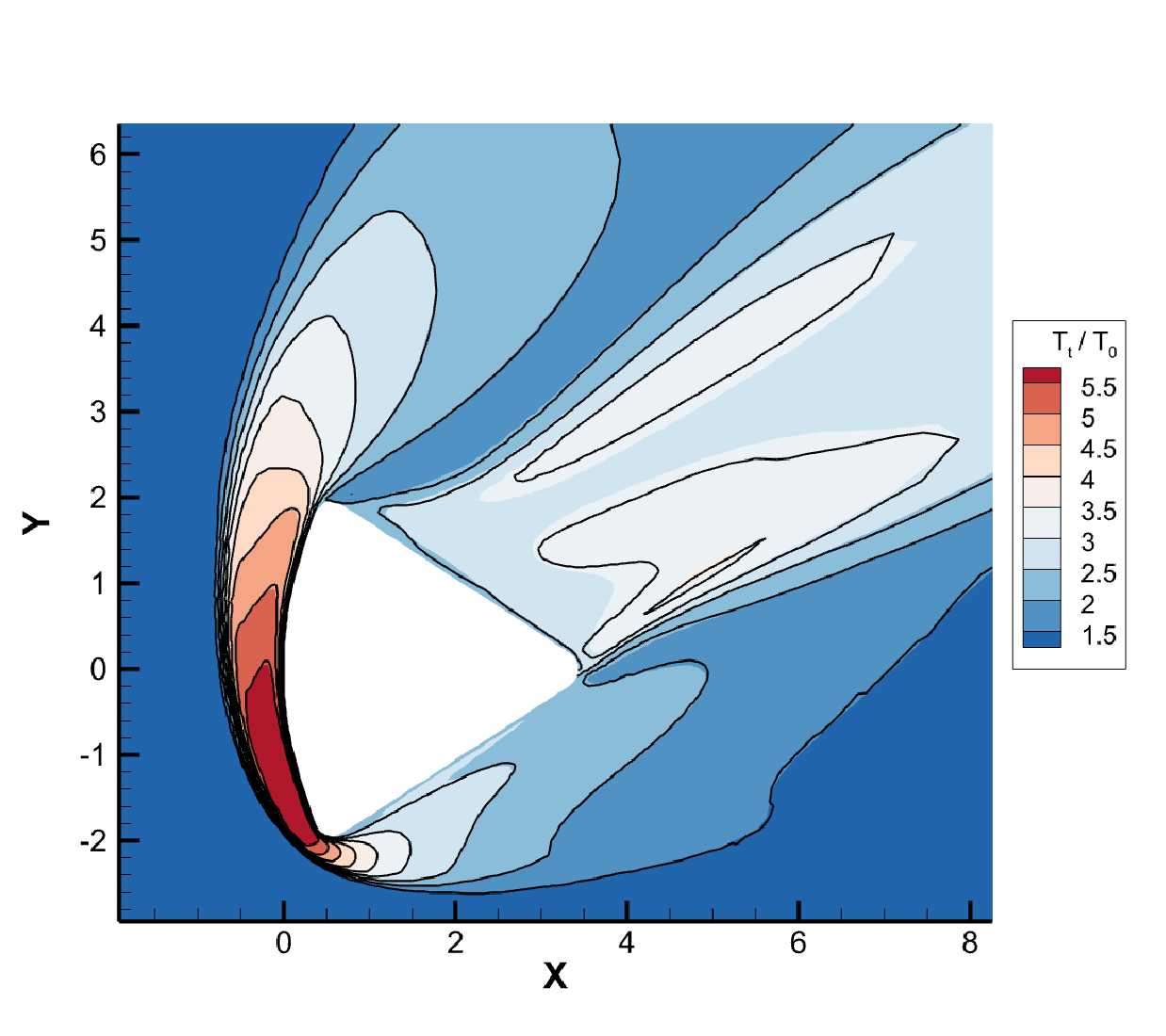}
    \end{minipage}

    \caption{The re-partitioning in non-equilibrium regions, as well as the contours of velocity and translational temperature, are compared between the DSMC (contours) and aGSIS (black lines), when $\text{Ma} = 5$, $\text{Kn} = 0.0012$ and $\text{AoA}=30^\circ$. From the top row to bottom, $\text{Kn}_{ref}=$0.01, 0.05, and 0.1, respectively. }
    \label{fig:3DApollo_Ma5_Kn0d0012_ux_tt_cmp}
\end{figure}

\begin{figure}
    \centering
    \begin{minipage}{0.23\textwidth}
        \includegraphics[width=\textwidth,clip=true,keepaspectratio]{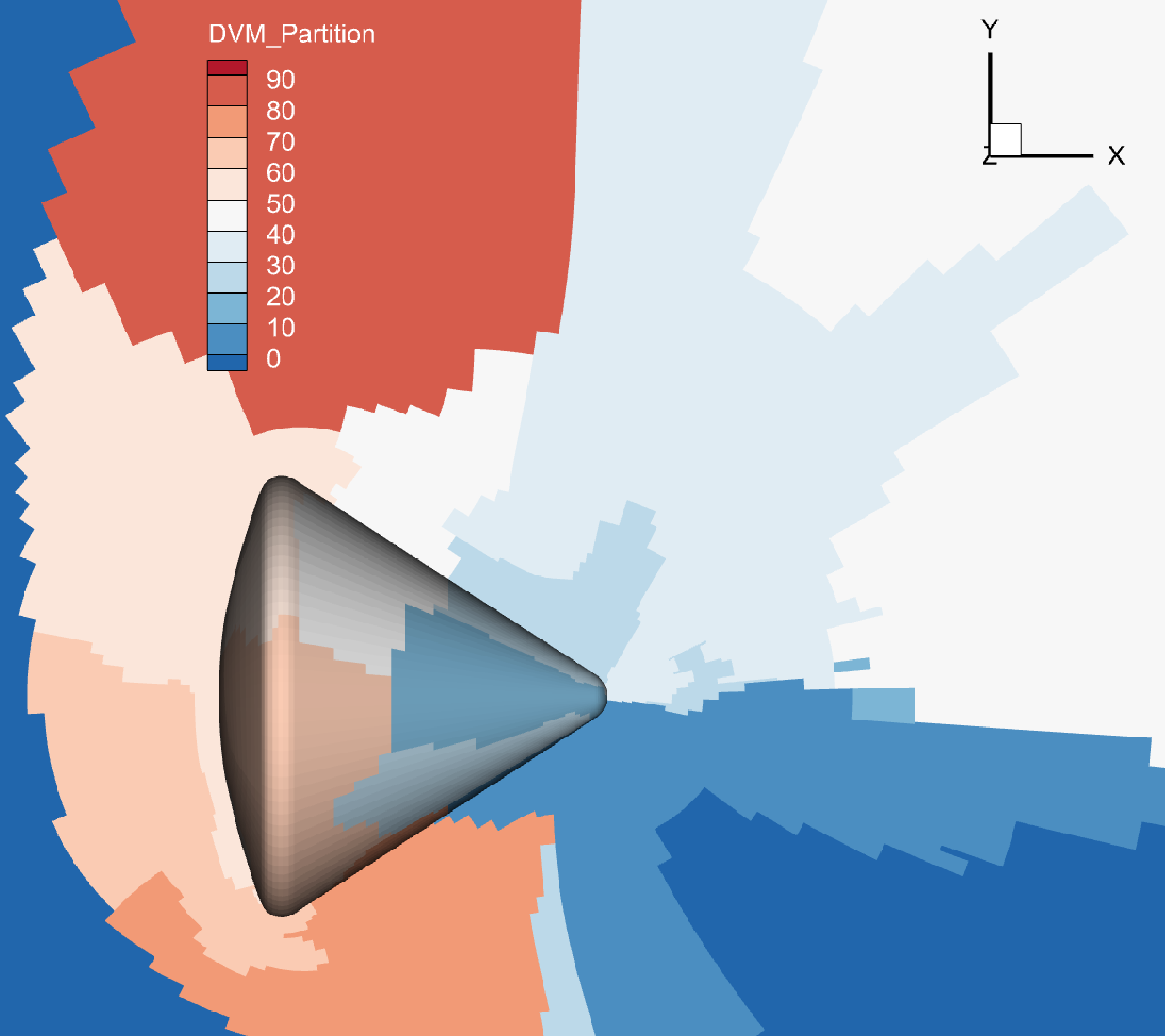}
    \end{minipage}
    \begin{minipage}{0.3\textwidth}
        \includegraphics[width=\textwidth,clip=true,keepaspectratio]{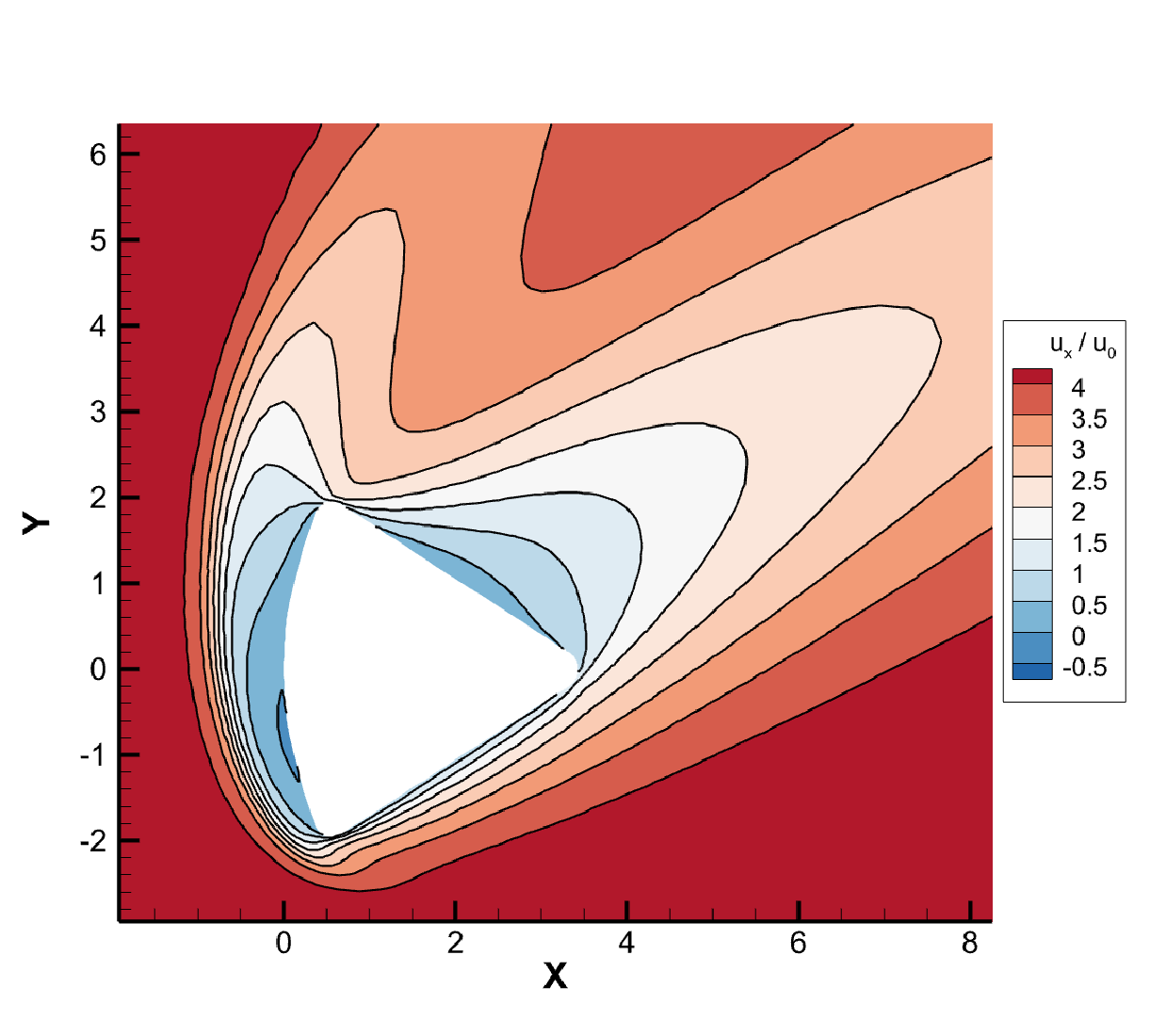}
    \end{minipage}
    \begin{minipage}{0.3\textwidth}
        \includegraphics[width=\textwidth,clip=true,keepaspectratio]{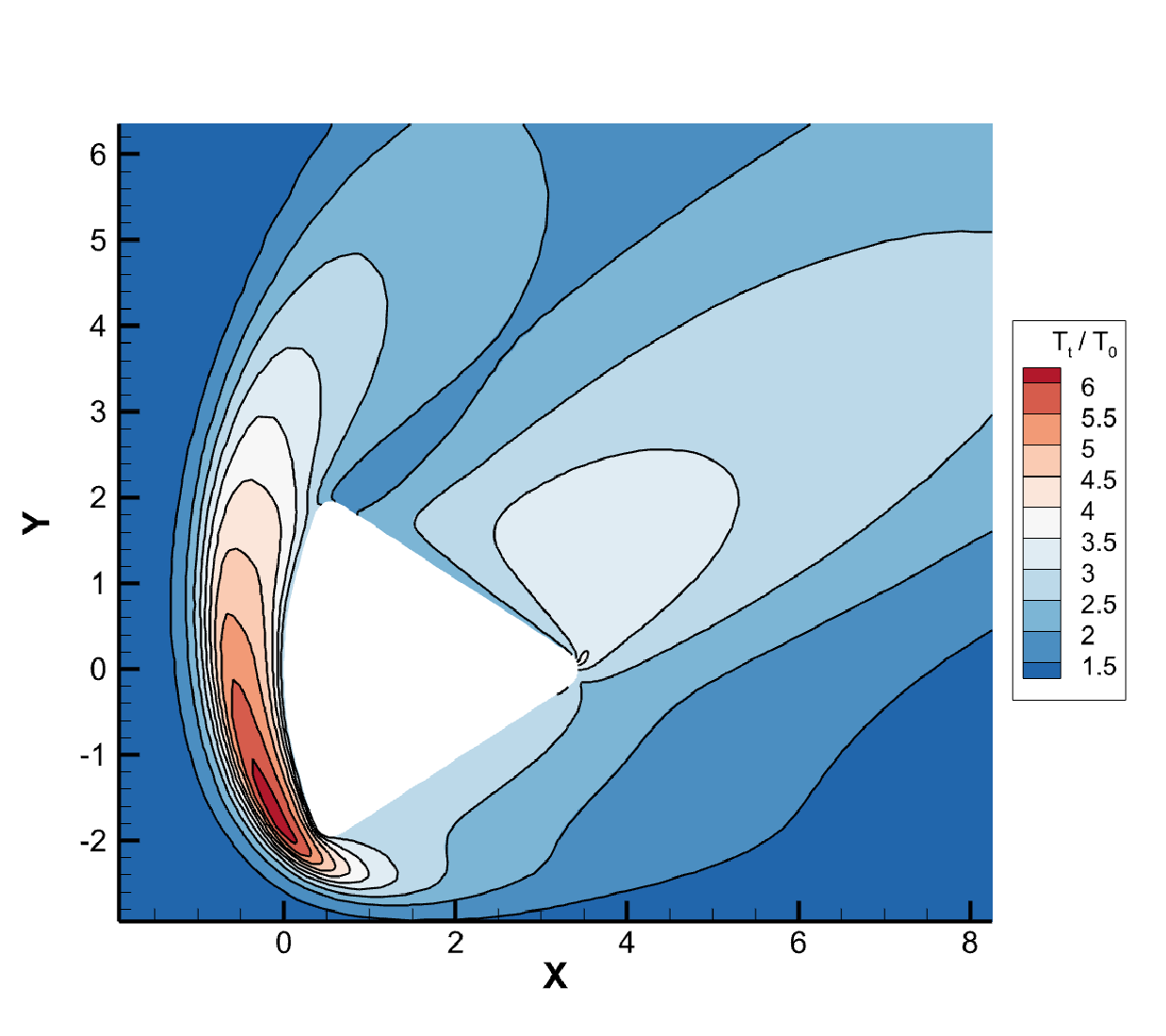}
    \end{minipage}
    \\
    \begin{minipage}{0.23\textwidth}
        \includegraphics[width=\textwidth,clip=true,keepaspectratio]{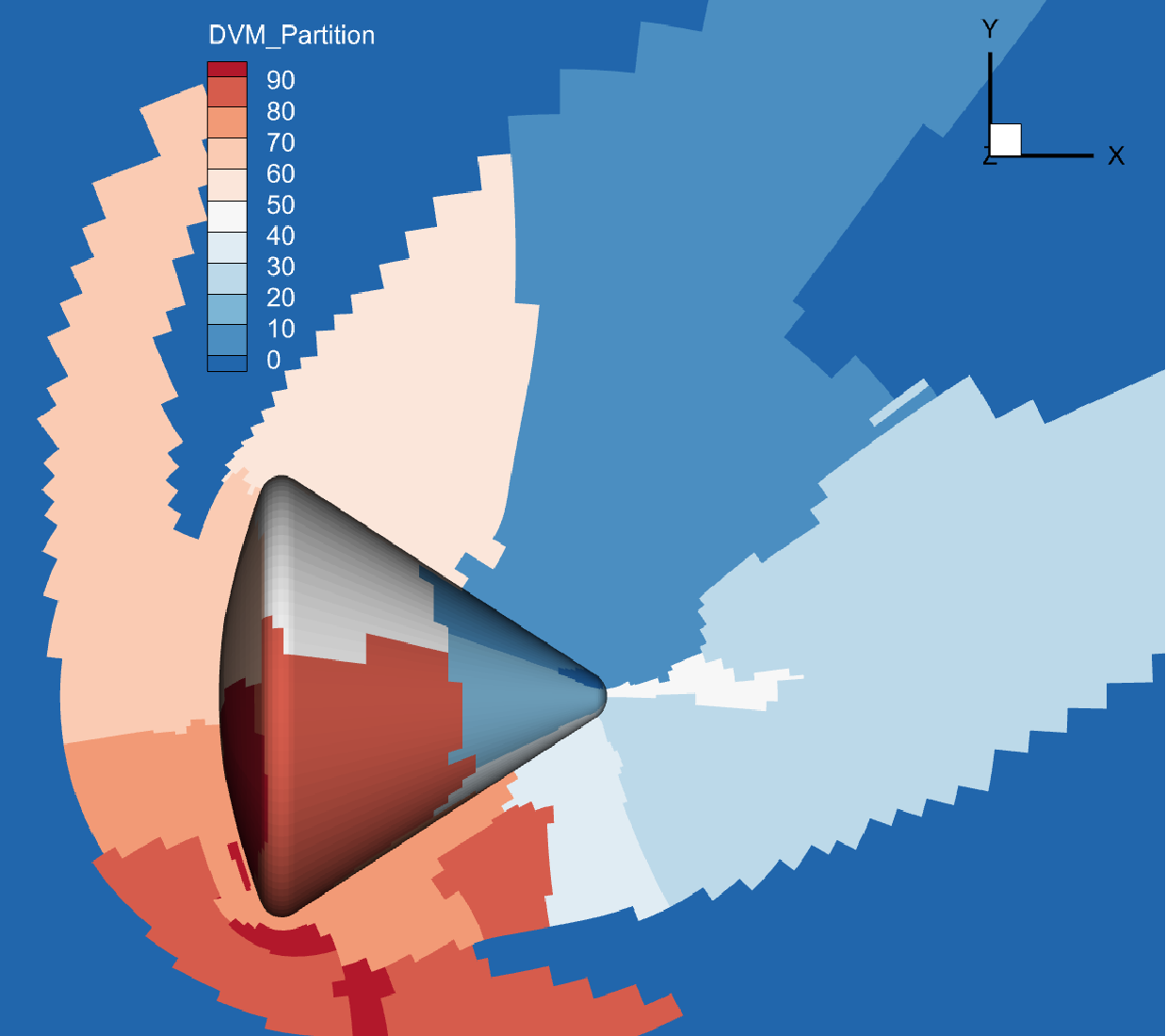}
    \end{minipage}
    \begin{minipage}{0.3\textwidth}
        \includegraphics[width=\textwidth,clip=true,keepaspectratio]{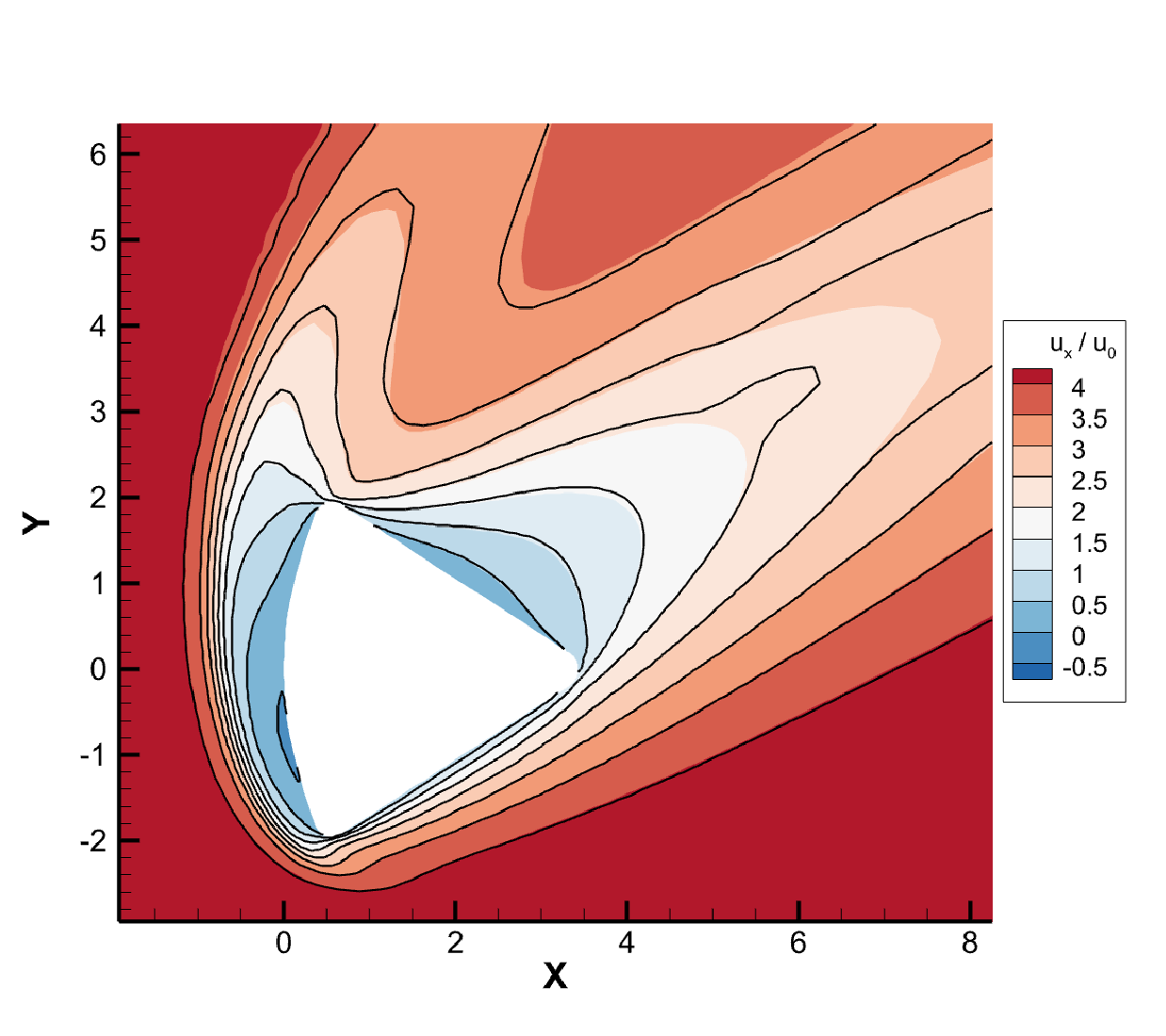}
    \end{minipage}
    \begin{minipage}{0.3\textwidth}
        \includegraphics[width=\textwidth,clip=true,keepaspectratio]{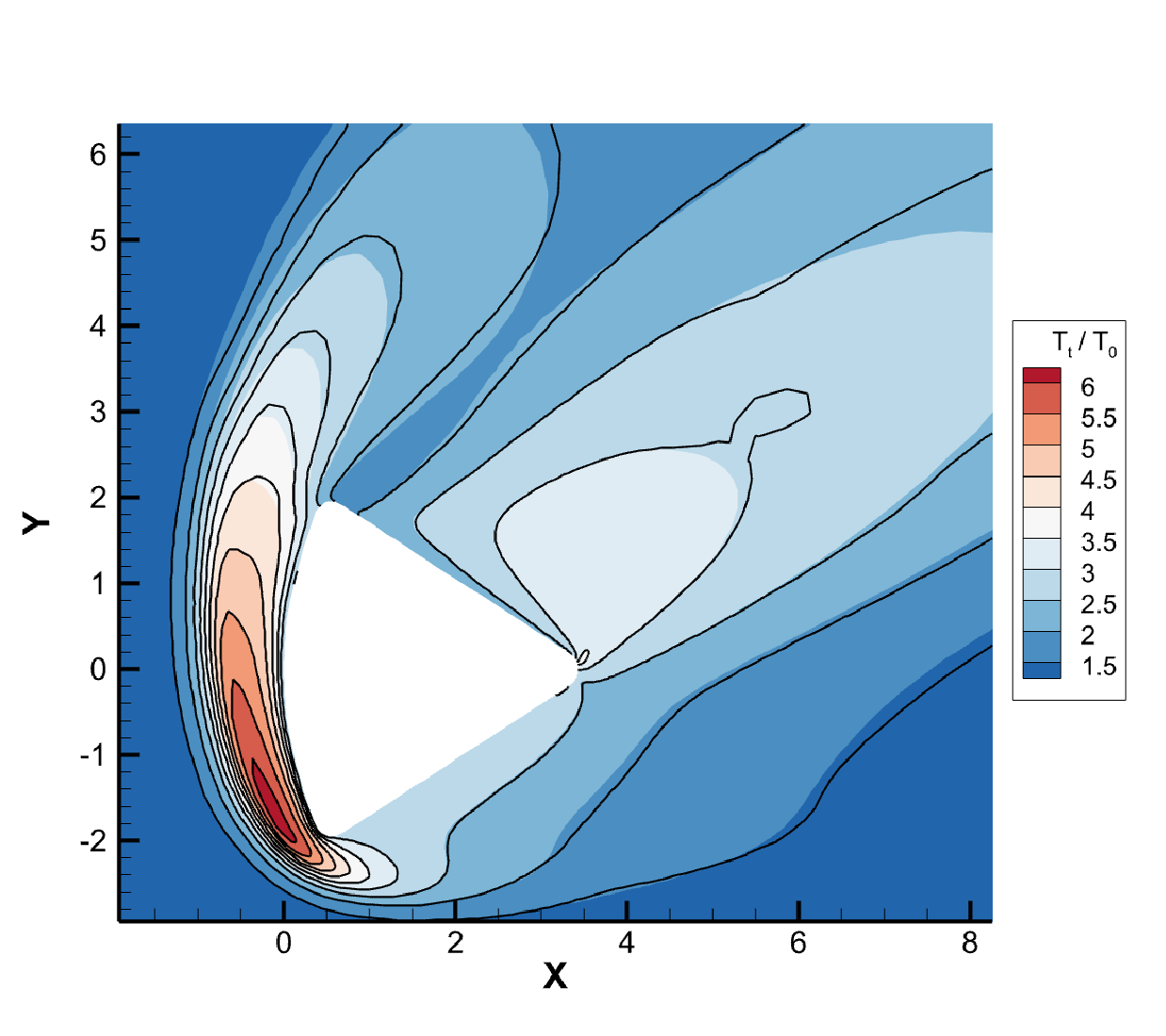}
    \end{minipage}
    \\
    \begin{minipage}{0.23\textwidth}
        \includegraphics[width=\textwidth,clip=true,keepaspectratio]{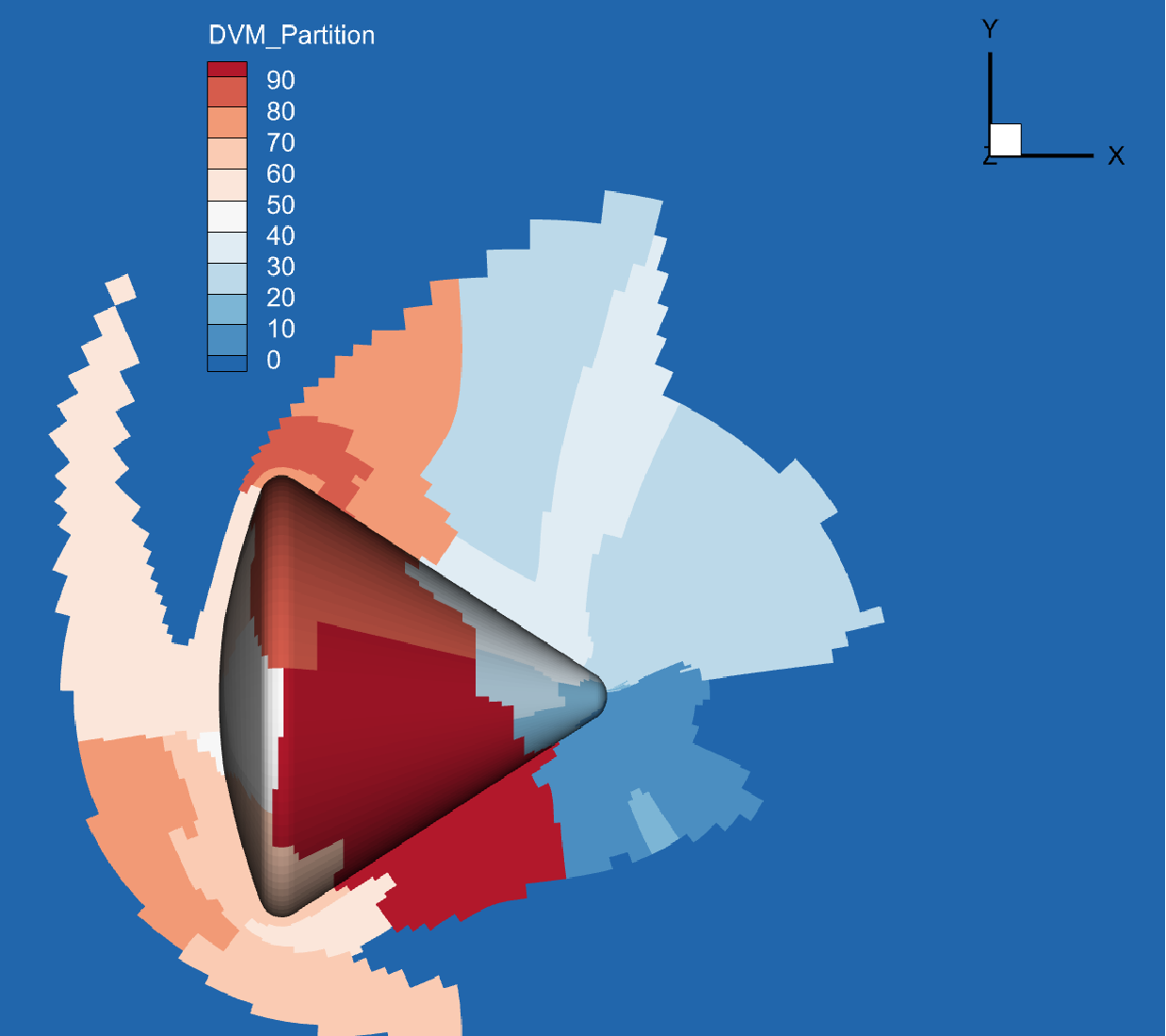}
    \end{minipage}
    \begin{minipage}{0.3\textwidth}
        \includegraphics[width=\textwidth,clip=true,keepaspectratio]{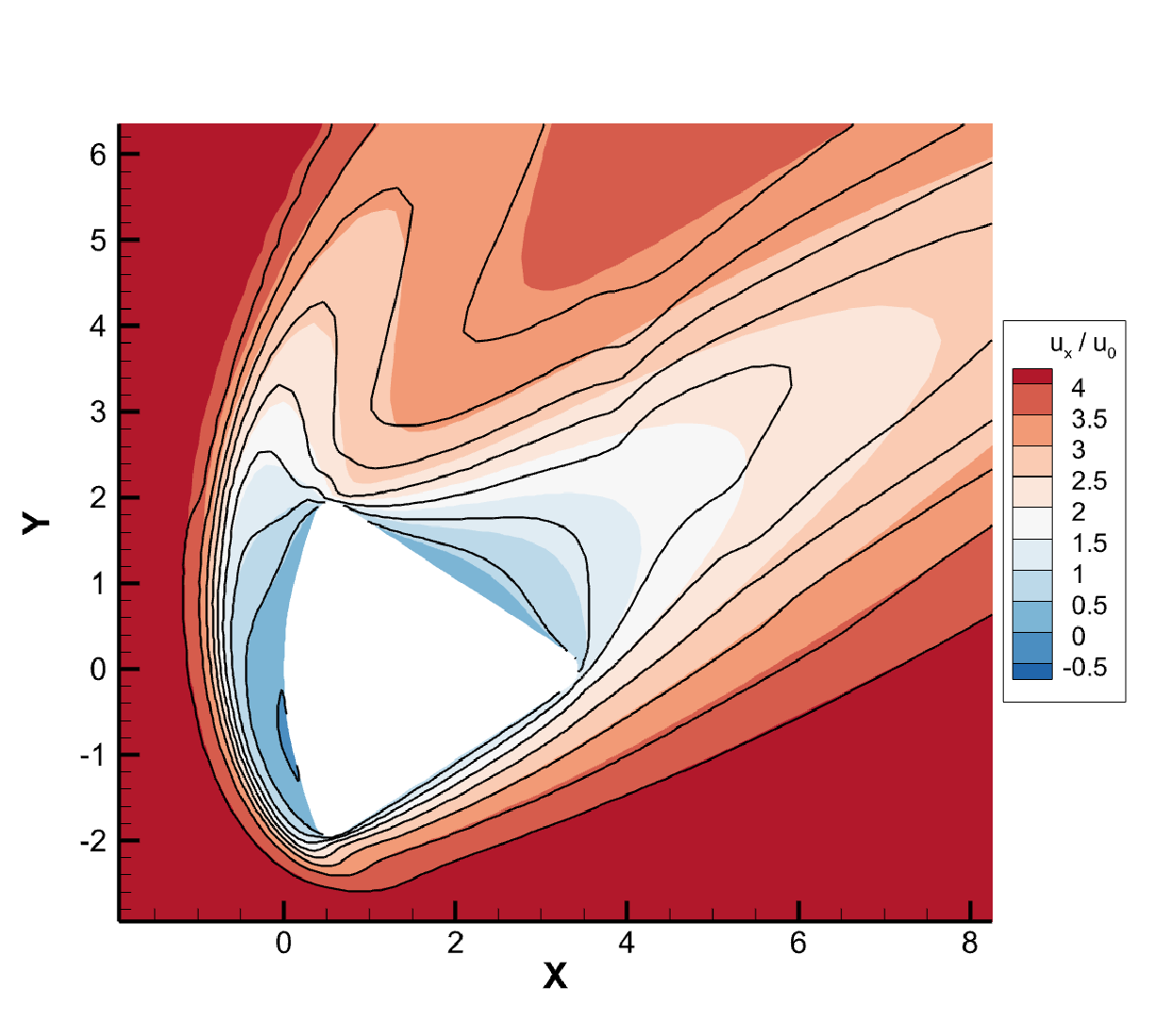}
    \end{minipage}
    \begin{minipage}{0.3\textwidth}
        \includegraphics[width=\textwidth,clip=true,keepaspectratio]{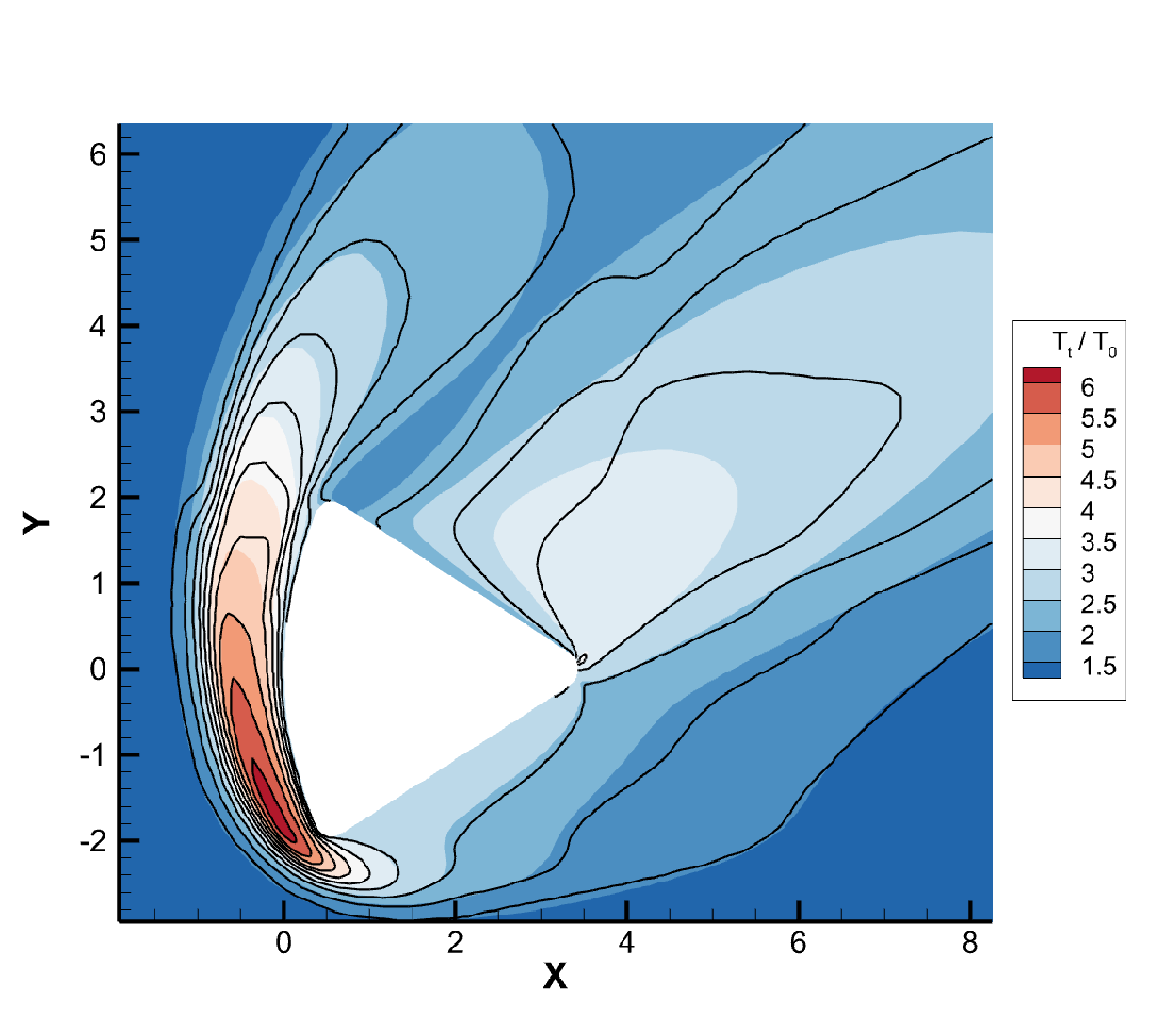}
    \end{minipage}

    \caption{The re-partitioning in non-equilibrium regions, as well as the contours of velocity and translational temperature, are compared between the DSMC (contours) and aGSIS (black lines), when $\text{Ma} = 5$, $\text{Kn} = 0.1$ and $\text{AoA}=30^\circ$. From the top row to bottom, $\text{Kn}_{ref}=$ 0.01, 0.05, and 0.1, respectively. 
    }
    \label{fig:3DApollo_Ma5_Kn0d1_ux_tt_cmp}
\end{figure}

Here, we discuss the impact of the reference Knudsen number $\text{Kn}_{ref}$ on the results of the aGSIS method. According to Eq.~\eqref{eq:Kn_Gll}, three scenarios with $\text{Kn}_{ref}=0.01$, 0.05 and 0.1, are simulated. The adaptive DVM grid number, number of iterations, computation time, and memory usage of aGSIS are summarized in Table~\ref{tab:3DApollo_Ma5_AGSIS_Effency}. Clearly, as $\text{Kn}_{ref}$ increases, the memory cost reduces. For example, while the original GSIS consumes 314.4 GB of memory, the aGSIS with $\text{Kn}_{ref}=0.1$ consumes about 20 times less memory.

Figure~\ref{fig:3DApollo_Ma5_Kn0d0012_ux_tt_cmp} shows that the non-equilibrium regions are primarily concentrated in the leeward area of the Apollo. For $\text{Kn}_{ref} \le 0.01$,  aGSIS results align well with the reference solutions from the original GSIS. However, for $\text{Kn}_{ref} \ge 0.05$, the aGSIS fails to accurately capture the non-equilibrium region in the leeward area, leading to noticeable discrepancies in the velocity field at the rear of the Apollo. Nevertheless, the lift and drag coefficients, which are important quantities, rarely change in Table~\ref{tab:3DApollo_Ma5_AGSIS_Effency}.

Furthermore, we test the case of $\text{Kn}=0.1$ with $\text{Kn}_{ref}$ values of 0.01, 0.05, and 0.1, respectively. The numerical results are illustrated in Fig.~\ref{fig:3DApollo_Ma5_Kn0d1_ux_tt_cmp}, and the associated computational cost  is given in Table~\ref{tab:3DApollo_Ma5_AGSIS_Effency}. Likewise, when  $\text{Kn}_{ref}$ is greater than or equal to 0.05, aGSIS exhibits a significant deviation in the flow field with the original GSIS, but the drag and lift coefficients rarely change. In this specific scenario, setting $\text{Kn}_{ref}$ to 0.01 proves to be an optimal choice.

\begin{table}[!t]
	\centering
    \caption{The comparison of computational costs passing Apollo in hypersonic flow at $\text{Ma} = 5$ and $\text{AoA}=30^\circ$ for different $\text{Kn}_{ref}$ using 96 cores with the initial number of mesh cells is 372,500. The reference area for the lift and drag coefficients is set to $13.26\ \text{m}^2$. Each GSIS iteration encompasses a single kinetic solver step followed by 400 macroscopic solver steps, and this process was conducted using 96 cores. 
    }
	\begin{threeparttable}
		\begin{tabular}{ccccccccc}\hline
			$\text{Kn}$ & $\text{Kn}_{ref}$  & DVM cells & Steps & Wall times (s) & Memory (GB) & $C_d$ & $C_l$  \\ \hline
            \multirow{3}{*}{0.0012} & 0.1     & 13,487\centering   & 30     & 357 & 15.5   & 1.0125 & 0.4186 \\ 
            ~ & 0.05    & 24,798\centering   & 26     & 339  & 21.5  & 1.0124 & 0.4183 \\ 
            ~ & 0.01    & 64,356\centering   & 21     & 310  & 50.6  & 1.0134 & 0.4176 \\    \hline
            \multirow{3}{*}{0.1} & 0.1     & 212,094\centering    & 72     & 1498  & 181.5  & 1.1250 & 0.3682 \\
            ~ & 0.05    & 249,756 \centering   & 74     & 1656  & 210.6  & 1.1295 & 0.3706 \\ 
            ~ & 0.01    & 292,920\centering    & 56     & 1371  & 249.2  & 1.1288 & 0.3702 \\ \hline
		\end{tabular}
	\end{threeparttable}
	\label{tab:3DApollo_Ma5_AGSIS_Effency}
\end{table}

\begin{table}[!t]
	\centering
    \caption{The comparison of computational costs passing Apollo in hypersonic flow at $\text{Ma} = 5$, $\text{Kn}_{ref} = 0.01$ and $\text{AoA}=30^\circ$ for different Knudsen numbers using 96 cores with the initial number of mesh cells is 372,500. 
    }
	\begin{threeparttable}
		\begin{tabular}{ccp{1.7cm}p{1.5cm}p{0.05cm}p{2.7cm}cp{1.7cm}p{1.5cm}}\hline
			\multirow{2}{*}{Kn}   &\multicolumn{3}{c}{Original GSIS} & ~ & \multicolumn{4}{c}{aGSIS}  \\
			\cline{2-4}  \cline{6-9}
			~               &  Steps & Wall times (s) & Memory (GB) & ~   & DVM cells \centering & Steps & Wall times (s) & Memory (GB)  \\ \hline
            $0.1$      &  57   &  1803\centering     & \multirow{4}{*}{314.4} \centering   & ~ & 292,920\centering     & 56     & 1371\centering   &249.2    \\ 
            $0.01$     &  32   &  1006\centering     & ~                                   & ~ & 210,459\centering     & 32     & 628\centering    & 181.2   \\ 
            $0.0012$   & 21    & 659 \centering      & ~                                   & ~ & 64,356\centering      & 21     & 310\centering    & 50.6  \\ 
            $10^{-4}$   & 13 & 495 \centering      & ~                                   & ~ & 35,277\centering      & 13     & 207\centering    & 27.4  \\ \hline
		\end{tabular}
	\end{threeparttable}
	\label{tab:3DApollo_Ma5_Knref0d01_AGSIS_Effency}
\end{table}

To test the capabilities of aGSIS across the entire flow field, Table~\ref{tab:3DApollo_Ma5_Knref0d01_AGSIS_Effency} summarizes the computational overhead of the aGSIS at different Knudsen numbers, when $\text{Kn}_{ref}=0.01$. All cases are computed on an Intel Xeon Platinum 9242(2.3GHz) machine using 96 cores. It is observed that in near-continuum flow regions, the aGSIS exhibits significant improvements over the original GSIS in terms of computation time and memory consumption: the smaller the Knudsen number, the greater the reduction in memory and computational cost.

\subsection{Hypersonic flow passing a space station}

\begin{figure}[p]
    \centering
    {\includegraphics[width=0.32\textwidth,clip = true]{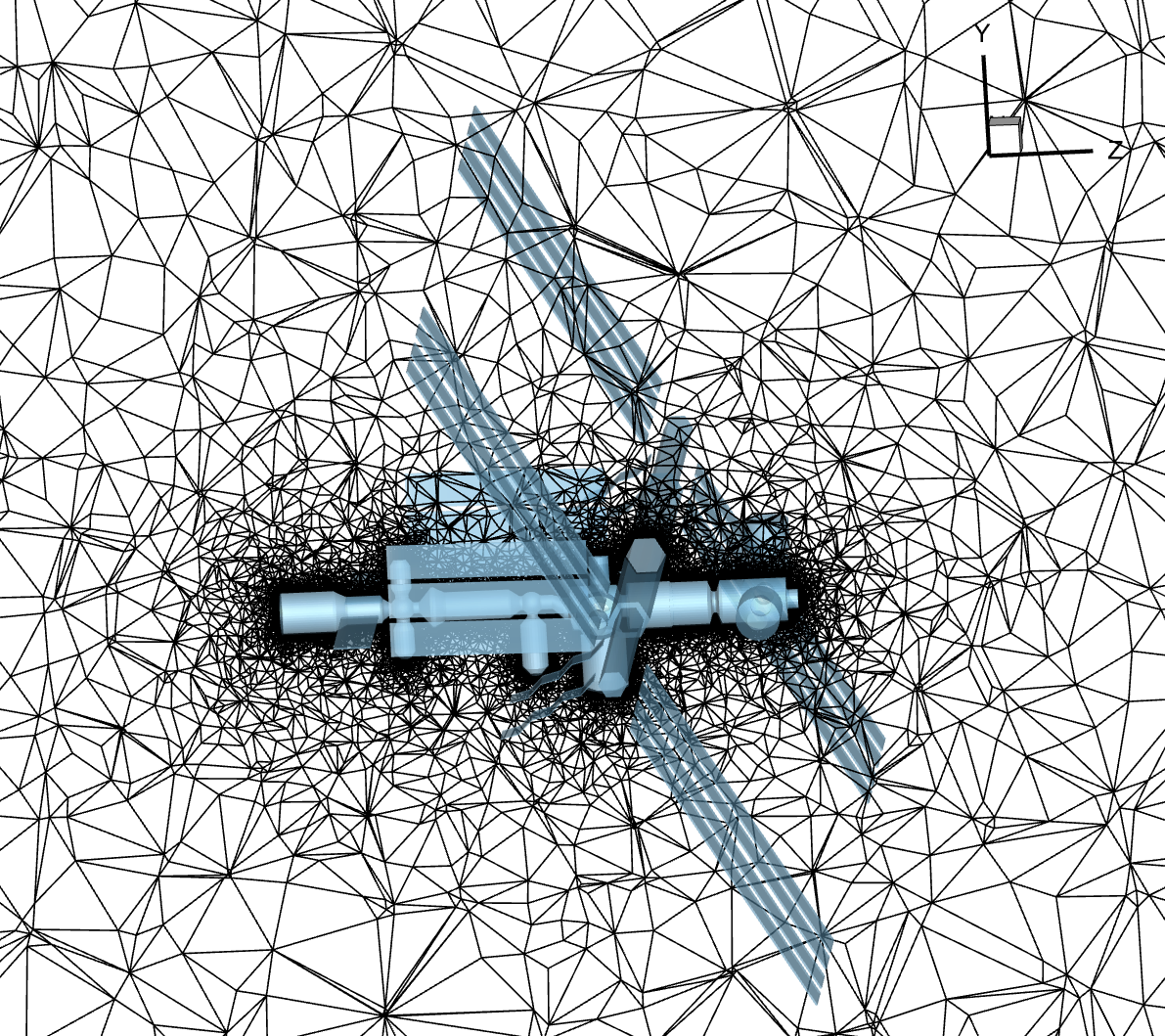}} 
    {\includegraphics[width=0.32\textwidth,clip = true]{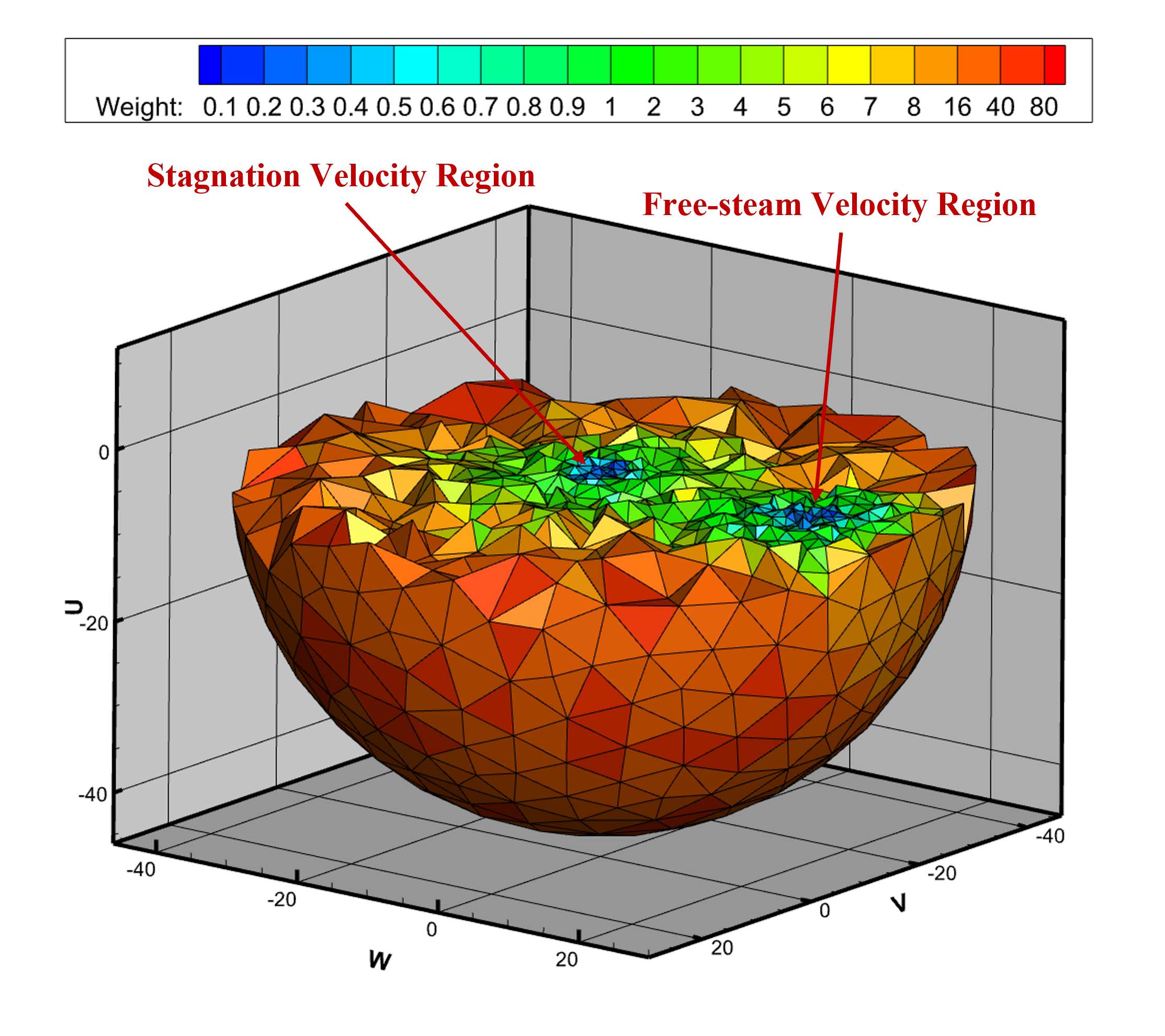}}
    {\includegraphics[width=0.32\textwidth,clip = true]{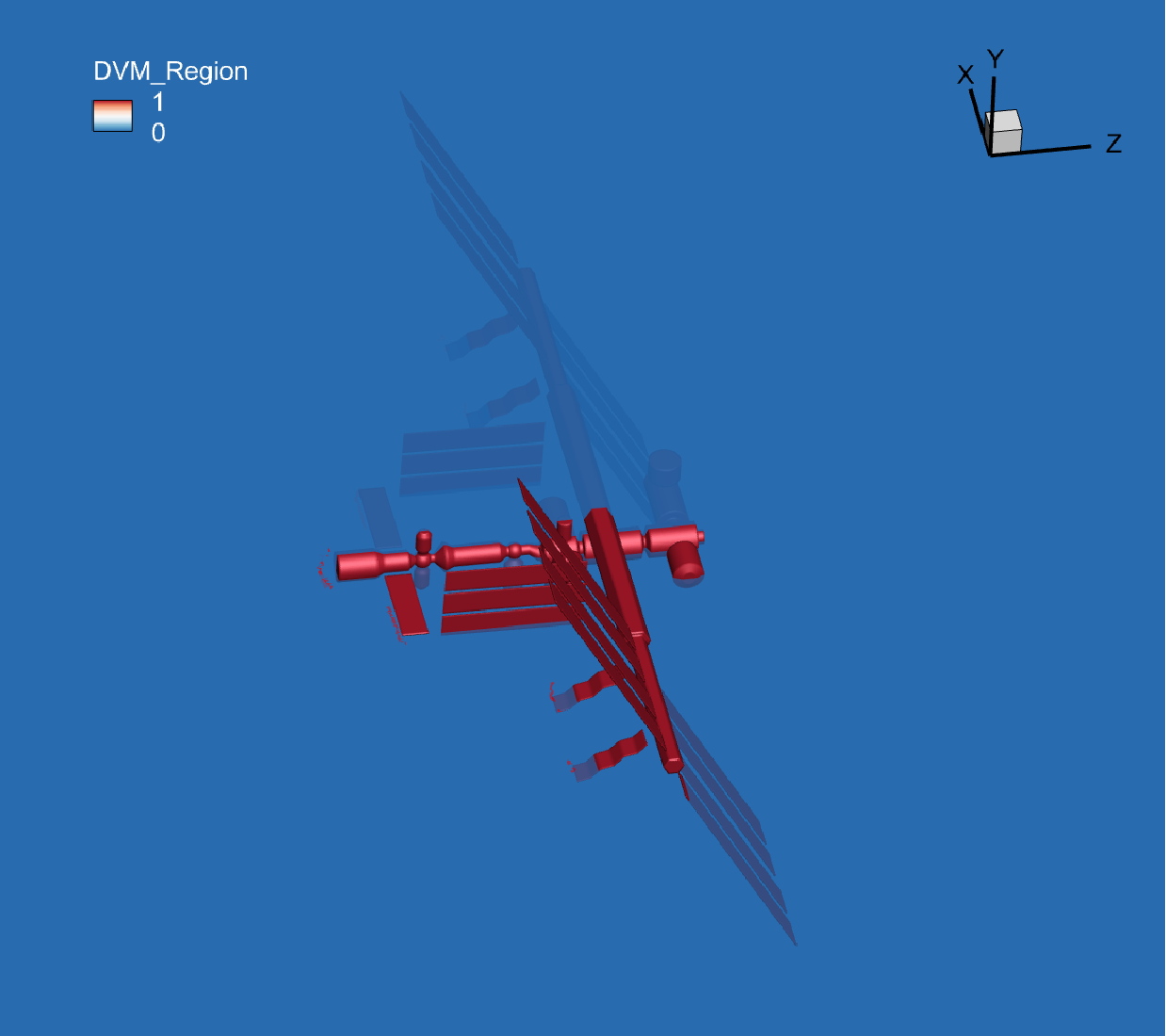}}
    \vspace{0.8cm}
    \\
    {\includegraphics[width=0.34\textwidth,clip = true]{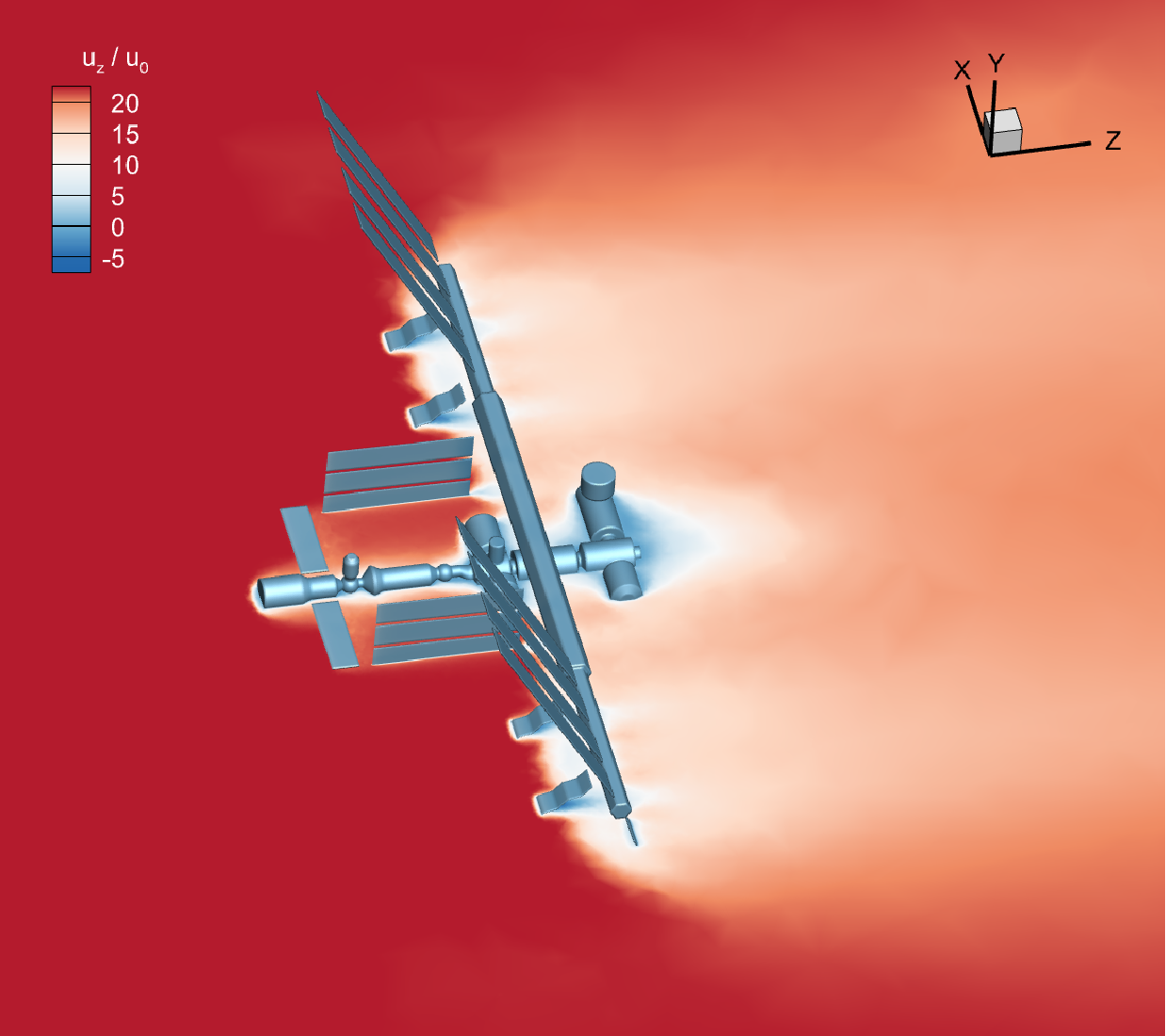}} 
    \hspace{1.8cm}
    {\includegraphics[width=0.34\textwidth,clip = true]{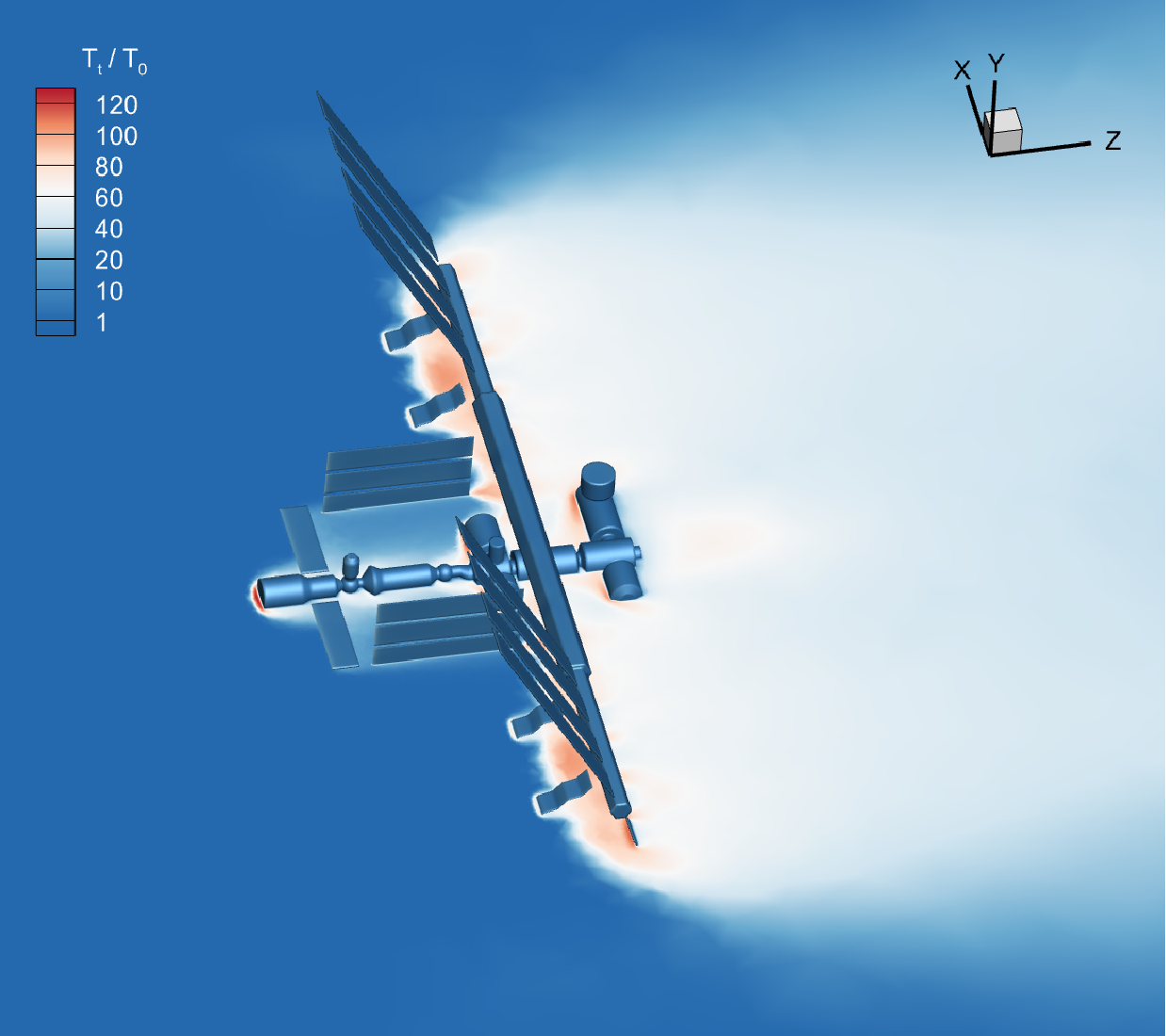}}
    \\
    {\includegraphics[width=0.45\textwidth,clip = true]{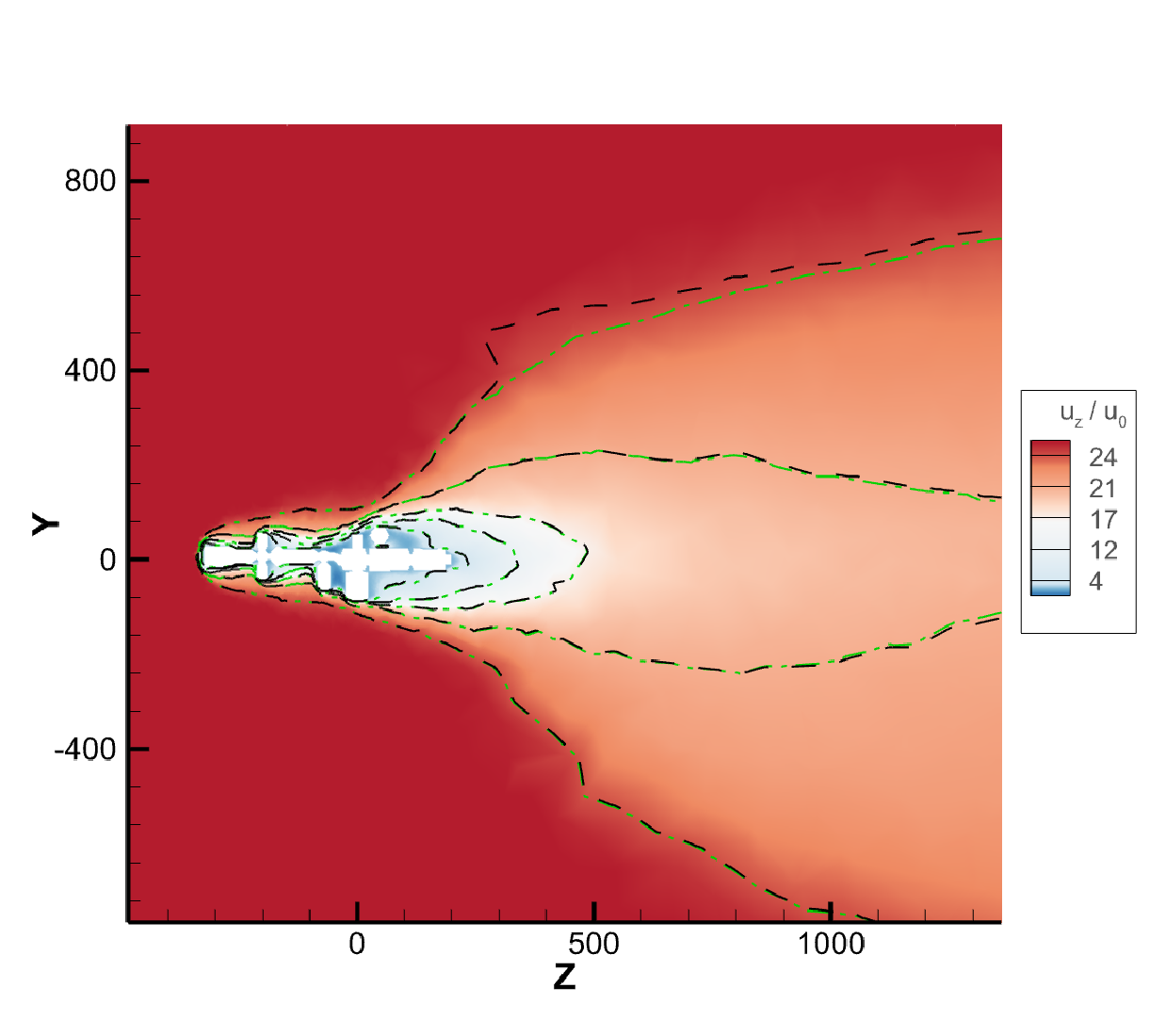}} 
    {\includegraphics[width=0.45\textwidth,clip = true]{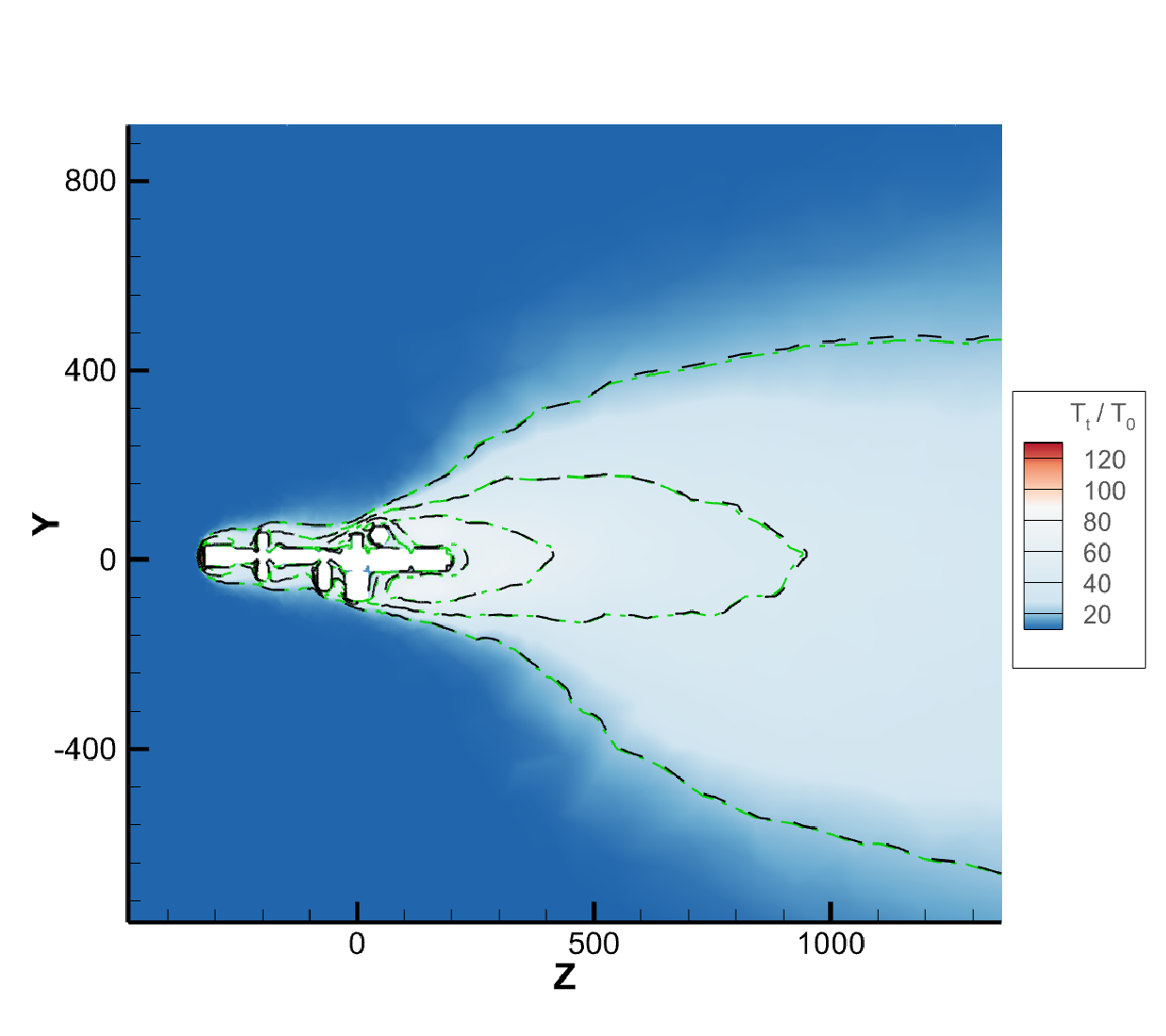}}
    
    \caption{ (First row) Spatial domain discretization, velocity space discretization and spatial adaptation in a hypersonic flow passing a space station at $\text{Kn}=0.01$ and $\text{Ma}=25$. When $\text{Kn}_{ref}=0.01$, the number of grid cells in the red DVM region is 251,198. 
    (Second row) The local Mach number and  translational temperature calculated by aGSIS.
   (Third row) Comparison of local Mach number and translational temperature distributions between the original GSIS (contours and green dash-dot) and adaptive GSIS (black dash) in the $x=0$ plane .}
    \label{fig:3DStation_Ma25_Kn0d01_Knref0d01}
\end{figure}

To further test the capabilities of aGSIS, we simulate the aerodynamic flow around a space station with an incoming velocity of 25 Ma and a Knudsen number of 0.01 when the reference length is $0.01 \text{m}$. The free stream temperature is 142.2 K, and the space station surface is an isothermal wall with a temperature of $300 \text{K}$, employing fully diffuse gas-wall interaction boundary conditions. Details on the mesh configuration can be found in  Ref.~\cite{zhang2024efficient}. The physical space is discretized into a total of 5,640,776 cells, and the velocity space is divided using an unstructured grid, resulting in 31,440 grid cells, see the first row in Fig.~\ref{fig:3DStation_Ma25_Kn0d01_Knref0d01}.


The reference Knudsen number, $\text{Kn}_{ref}$, is again set to 0.01. As depicted in Fig.~\ref{fig:3DStation_Ma25_Kn0d01_Knref0d01}, the red areas indicate the non-equilibrium computational regions, with the two layers of grids adjacent to the wall being forcibly designated as such. Overall, the  DVM region comprises 251,198 spatial grids, which represent 4.5\% of the total grid count.
Figure~\ref{fig:3DStation_Ma25_Kn0d01_Knref0d01} presents a comparison of the local Mach number and translational temperature between the original GSIS and the aGSIS, demonstrating that the results from both methods are essentially consistent. In this computation, 640 cores are employed, yielding a computation time of 0.92 hours and memory usage of 0.96 TB. Compared with the original GSIS~\cite{zhang2024efficient}, the memory usage is reduced by a factor of approximately 24, and the computation time is  decreased by a factor of approximately 7.

\section{Conclusions}\label{sec:conclusion}

To minimize computational memory usage in the discrete velocity method, we have devised an adaptive strategy within the GSIS framework. Specifically, the DVM is employed solely in regions where the local Knudsen number is substantial, while the Navier-Stokes equations, augmented with high-order corrections to the constitutive relations, are applied across the entire computational domain. To enhance parallel processing efficiency, we have introduced specialized parallel computing strategies for both two-dimensional and three-dimensional scenarios, with a particular emphasis on achieving load balance.

Consequently, the aGSIS has not only significantly reduced the computational memory demands of the program but also shortened the computation time. In particular, within near-continuum flow regions, the aGSIS has demonstrated substantial improvements in memory overhead and computation time compared to the original GSIS. Additionally, through simulations of supersonic flow in micro-nozzles, supersonic flow around circular cylinders, Apollo capsules, and International Space Station, the aGSIS has been validated to maintain rapid convergence and the capability to capture the non-equilibrium physics inherent in GSIS. 

This approach can be seamlessly extended to problems involving vibration and radiation, delivering accurate and efficient simulation results across various practical application fields.


\section*{Acknowledgments} 
This work is supported by the National Natural Science Foundation of China (12172162) and the Stable Support Plan (80000900019910072348). Special thanks are given to the Center for Computational Science and Engineering at the Southern University of Science and Technology.


\bibliographystyle{elsarticle-num}

\bibliography{ref}

\end{document}